\documentclass[%
 reprint,
superscriptaddress,
 amsmath,amssymb,
 aps,
]{revtex4-2}

\usepackage{graphicx}% Include figure files
\graphicspath{{Main-figs/}}
\usepackage{dcolumn}% Align table columns on decimal point
\usepackage{bm}% bold math
\usepackage{siunitx}% a fun way to get nicely formatted units
    \DeclareSIUnit \counts{Cts}
    \DeclareSIUnit \electron{e}
    \DeclareSIUnit \molar{M}
\usepackage{braket}
\usepackage[colorlinks=true, allcolors=black]{hyperref}% add hypertext capabilities
\def\GdConc {{[\mathrm{Gd^{3+}}]}}

\def \ToneS {T_1^\mathrm{S}}

\def\FigureOne {\begin{figure*}[]
\centering
\includegraphics[scale=1]{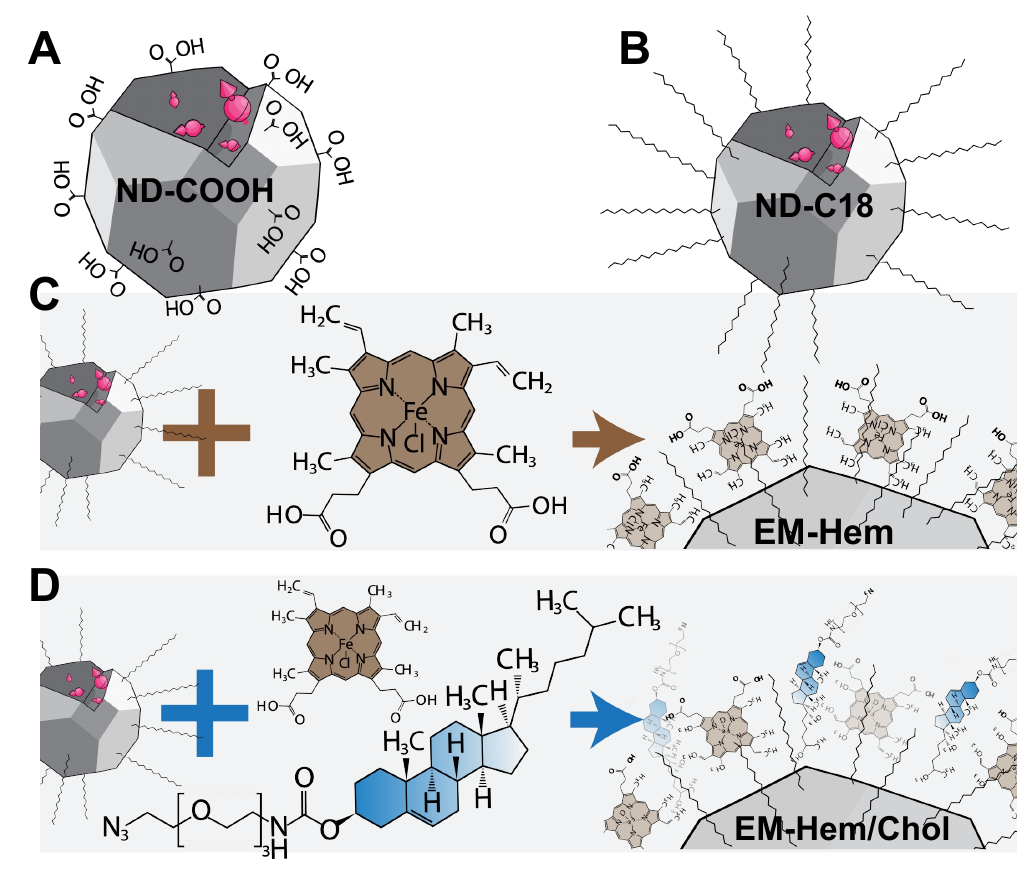}
\caption{\label{fig:Emlusion Overview}
 \textbf{Nanodiamond Samples} 
 Schematics of (A) a carboxyl-terminated, hydrophilic nanodiamond containing NV centers (ND-COOH);
 (B) an octadecane-terminated, hydrophobic nanodiamond containing NV centers (ND-C18); 
 (C) an ND-C18 nanodiamond coated with amphiphilic hemin, formed as an emulsion  (EM-Hem); and   
 (D) an ND-C18 nanodiamond coated with hemin and cholesteryl-TEG azide, formed as an emulsion (EM-Hem/Chol). 
}\end{figure*}}

\def \FigureTwo {\begin{figure}[]
\includegraphics[scale=1]{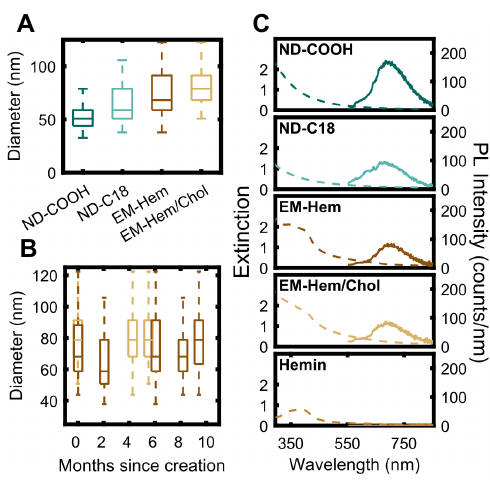}
\caption{\label{fig:Emlusion Properties}
\textbf{Nanodiamond Emulsion Properties}
 (A) Box plots of the particle size distribution measured by dynamic light scattering (DLS) for the four nanodiamond samples in Figure~\ref{fig:Emlusion Overview}.  
 (B) Box plots of the particle size distribution measured by DLS  for EM-Hem (brown) and EM-Hem/Chol (yellow) monitored over 10 months.   
 (C) UV-Vis extinction (left axis; dashed curves) and photoluminescence (right axis; solid curves) spectra under 532~nm excitation for ND-COOH, ND-C18, EM-Hem, EM-HChol and hemin micelle aqueous dispersions.
}\end{figure}}

\def \FigureThree{\begin{figure}[]
\centering
\includegraphics[scale=1]{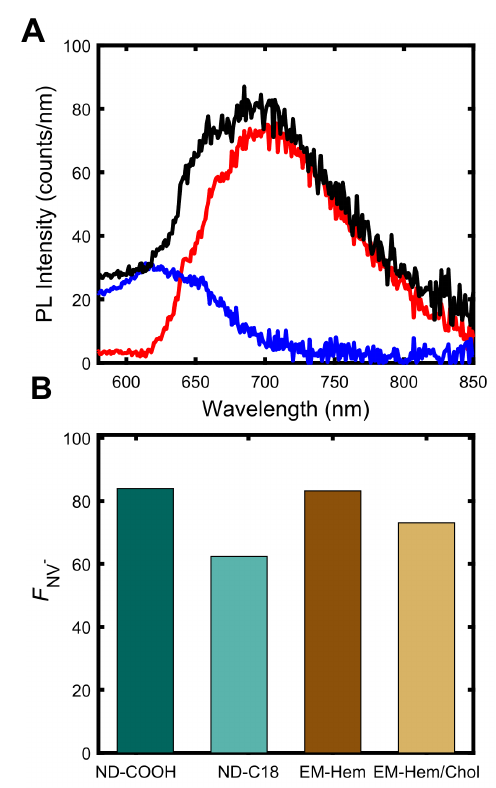}
\centering
\caption{\label{fig:Charge Properties of Emulsions}
\textbf{NV-center Charge State} 
(A) PL spectrum of a representative, EM-Hem/Chol emulsion (black) and its decomposition into  NV$^{-}$ emission (red) and NV$^{0}$ emission (blue) spectra \textit{via} nonnegative matrix factorization. 
(B) $F_\mathrm{NV^{-}}$ for the nanodiamond samples.  
}\end{figure}}

\def \FigureFour{\begin{figure}[]
\centering
\includegraphics[scale=1]{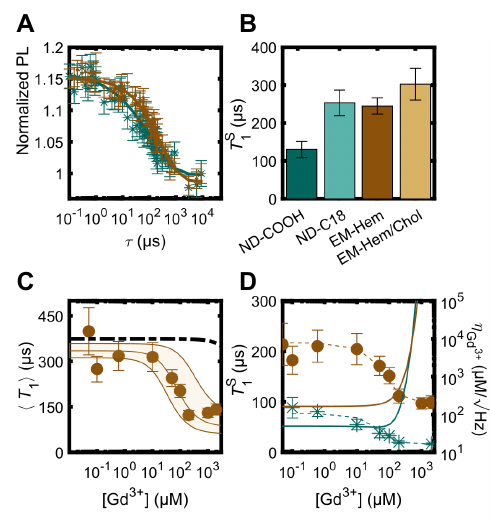}
\caption{\label{fig:T1 Overview}
\textbf{Spin Lifetime} 
(A) PL as a function of delay time, $\tau$, between an optical initialization pulse and subsequent readout pulse, for ND-COOH (dark green stars) and EM-Hem (brown circles). 
Solid curves are fits using Equation~(\ref{Eq: Stretch Fit}) of the main text.
The PL signal is normalized to the intensity at $\tau = 9$~ms, and error bars represent the uncertainty from photon shot noise. 
(B) Best-fit $\ToneS$ for the four nanodiamond samples. Error bars represent the standard deviation of $\ToneS$ from measurements of multiple samples (for ND-COOH, ND-C18, and EM-Hem) or the fit confidence interval from a single measurement for EM-Hem/Chol (Supporting Information Figure~S12). 
(C) $\left\langle {T}_1 \right\rangle$ as a function of $\GdConc$ added to EM-Hem (brown circles). Error bars represent the uncertainty in $\langle T_1\rangle$ through propagation of the best-fit confidence intervals for $\ToneS$ and $\beta$ through Equation~(\ref{Eq:avgT1}).
The curves represent models for the effect of $\GdConc$ on $\langle T_1\rangle$ that respectively assume no surface adsorption (black dashed curve) or adsorption according to a Langmuir model (solid curve for best fit and shaded region for 68 $\%$ confidence intervals). 
(D) Best-fit $\ToneS$ \textit{vs.} $\GdConc$ for ND-COOH (green stars) and EM-Hem (brown circles) (left axis).
Error bars represent 68\% confidence intervals.
Dotted curves are fits using Equation~(\ref{Eq: T1s(Gd)}).
The right axis shows $\eta_\GdConc$ for ND-COOH (green solid curve) and EM-Hem (brown solid curve) as a function of $\GdConc$ calculated according to Equation~(\ref{Eq:ChemicalSensitivity}). 
}\end{figure}}

\def\FigureFive{\begin{figure}[]
\centering
\includegraphics[scale=1]{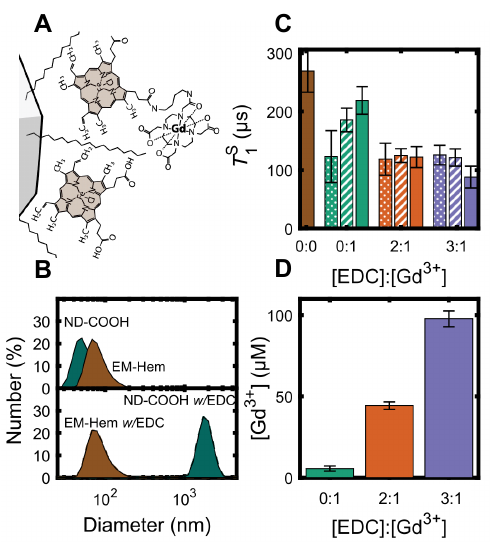}
\centering
\caption{\label{fig:EDC Conjugation}
\textbf{EDC Conjugation} 
(A) Model of EDC-enabled crosslinking of amine-terminated Gd$^{3+}$ chelate to EM-Hem. 
(B) DLS size distribution for ND-COOH (dark green) and EM-Hem (brown) before (upper) and after (lower) addition of crosslinkers. 
(C) $\ToneS$ values as a function of [EDC]:[Gd$^{3+}$] ratio following the EDC-Gd$^{3+}$ conjugation experiment; after conjugation (dotted bars), initial after centrifugation washing (striped bars),  and after a two-day dialysis step (solid bars).
Error bars represent the best-fit 68 $\%$ confidence intervals for $\ToneS$ through Equation~(\ref{Eq: Stretch Fit}).
(D) ICP-OES measurements of remaining $\GdConc$ after dialysis. 
Error bars represent propagated uncertainty from the best-fit confidence intervals for ICP-OES Gd$^{3+}$ calibration curve.
}\end{figure}}

\def \FigureSix{\begin{figure}[]
\centering
\includegraphics[scale=1]{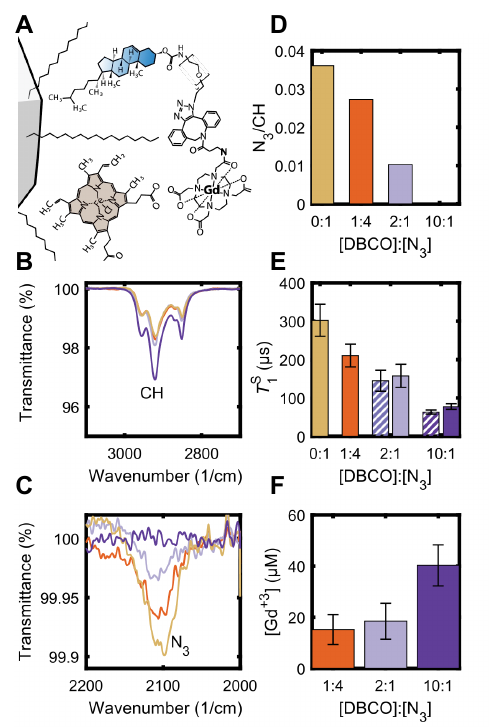}
\centering
\caption{\label{fig:Click Conjugation}
\textbf{Click Conjugation}  (A) Model of click conjugation of a DBCO-terminated Gd$^{+3}$ spin label to EM-Hem/Chol. Data shown for conjugations at  0:1 (yellow) 1:4 (orange), 2:1 (lavender) and 10:1 (purple) molar ratios of dibenzocyclooctyne (DBCO) to azide (N$_{3}$). 
FTIR transmittance curves of (B) the CH and (C) the N$_{3}$ stretching regions of the four samples.
(D) Ratio of the integrated areas of N$_{3}$  to CH stretches for each sample. 
(E) $\ToneS$ for the  samples. Striped bars refer to  measurements taken before dialysis. 
Error bars represent the best-fit  68~$\%$ confidence intervals for $\ToneS$ through Equation~(\ref{Eq: Stretch Fit}).
(F) ICP-OES measurements of remaining $\GdConc$ in the click-conjugated samples after dialysis. 
Error bars represent propagated uncertainty from the best-fit confidence intervals for ICP-OES gadolinium calibration curve.
}\end{figure}}

\begin{document}

%\preprint{APS/123-QED}

\title{Nanodiamond emulsions for enhanced quantum sensing \\ and click-chemistry conjugation}% Force line breaks with \\
%\thanks{A footnote to the article title}%

\author{Henry J. Shulevitz}
\affiliation{Department of Electrical and Systems Engineering, University of Pennsylvania, Philadelphia Pennsylvania 19104, USA}

\author{Ahmad Amirshaghaghi}
\affiliation{Department of Bioengineering, University of Pennsylvania, Philadelphia Pennsylvania 19104, USA}

\author{Mathieu Ouellet}
\affiliation{Department of Electrical and Systems Engineering, University of Pennsylvania, Philadelphia Pennsylvania 19104, USA}

\author{Caroline Brustoloni}
\affiliation{Department of Electrical and Systems Engineering, University of Pennsylvania, Philadelphia Pennsylvania 19104, USA}

\author{Shengsong Yang}
\affiliation{Department of Chemistry, University of Pennsylvania, Philadelphia Pennsylvania 19104, USA}

\author{Jonah J. Ng}
\affiliation{Department of Electrical and Systems Engineering, University of Pennsylvania, Philadelphia Pennsylvania 19104, USA}

\author{Tzu-Yung Huang}
\altaffiliation[Present address:]{ Nokia Bell Labs, 600 Mountain Ave., Murray Hill, NJ 07974, USA
}
\affiliation{Department of Electrical and Systems Engineering, University of Pennsylvania, Philadelphia Pennsylvania 19104, USA}

\author{Davit Jishkariani}
\affiliation{Department of Bioengineering, University of Pennsylvania, Philadelphia Pennsylvania 19104, USA}
\affiliation{Chemical and Nanoparticle Synthesis Core, University of Pennsylvania,  Philadelphia, 19104, USA}

\author{Christopher B. Murray}
\affiliation{Department of Chemistry, University of Pennsylvania, Philadelphia Pennsylvania 19104, USA}
\affiliation{Department of Materials Science and Engineering, University of Pennsylvania, Philadelphia Pennsylvania 19104, USA}

\author{Andrew Tsourkas}
\affiliation{Department of Bioengineering, University of Pennsylvania, Philadelphia Pennsylvania 19104, USA}

\author{Cherie R. Kagan}
\email[Email: ]{kagan@seas.upenn.edu \& lbassett@seas.upenn.edu}
\affiliation{Department of Electrical and Systems Engineering, University of Pennsylvania, Philadelphia Pennsylvania 19104, USA}
\affiliation{Department of Materials Science and Engineering, University of Pennsylvania, Philadelphia Pennsylvania 19104, USA}
\affiliation{Department of Chemistry, University of Pennsylvania, Philadelphia Pennsylvania 19104, USA}

\author{Lee C. Bassett}
\email[Email: ]{kagan@seas.upenn.edu \& lbassett@seas.upenn.edu}
\affiliation{Department of Electrical and Systems Engineering, University of Pennsylvania, Philadelphia Pennsylvania 19104, USA}

\date{\today}% It is always \today, today,
             %  but any date may be explicitly specified

\begin{abstract}
Nanodiamonds containing nitrogen-vacancy (NV) centers can serve as colloidal quantum sensors of local fields in biological and chemical environments. 
However, nanodiamond surfaces are challenging to modify without degrading their colloidal stability or the NV center's optical and spin properties. 
Here, we report a simple and general method to coat nanodiamonds with a thin emulsion layer that preserves their quantum features, enhances their colloidal stability, and provides functional groups for subsequent crosslinking and click-chemistry conjugation reactions. 
To demonstrate this technique, we decorate the nanodiamonds with combinations of carboxyl- and azide-terminated amphiphiles that enable conjugation using two different strategies. 
We study the effect of the emulsion layer on the NV center's spin lifetime, and we quantify the nanodiamonds' chemical sensitivity to paramagnetic ions using $T_1$  relaxometry. 
This general approach to nanodiamond surface functionalization will enable advances in quantum nanomedicine and biological sensing.
\end{abstract}

%\keywords{Suggested keywords}%Use showkeys class option if keyword
                              %display desired
\maketitle

%\tableofcontents

\section{\label{sec:level1}Introduction}
Nanodiamonds containing nitrogen-vacancy (NV) centers represent a promising material for quantum and biological sensing. 
The NV center is a nonbleaching emitter in the near-infrared biological window, allowing for fluorescence imaging within living biological systems \cite{wu2021nanodiamond, igarashi2020tracking,  mohan2010vivo}.
The carbon nanoparticles are nontoxic while being robust and unlikely to degrade within the body \cite{yu2005bright}. 
And most importantly, the NV center hosts an optically-addressable spin state, allowing for nanoscale quantum measurement of a variety of external fields and perturbations \textit{in vitro} or \textit{in vivo} \cite{wu2022recent, schirhagl2014nitrogen, rondin2014magnetometry, petrakova2015charge}.
Through proper functionalization, nanodiamonds could act as platforms for both sensing and drug delivery \cite{chow2011nanodiamond, say2011luminescent}.
Realizing such a general theranostics platform, however, requires stable colloidal nanoparticles with multiple distinct and controllable surface terminations.    

Unfortunately nanodiamonds, and in particular the milled nanodiamonds that are of interest to quantum sensing, are not easily functionalized \cite{rondin2010surface, kaviani2014proper, say2011luminescent}. 
Although several surface terminations are possible \cite{reineck2019not, petrakova2015charge, kaviani2014proper, shenderova2017commercial}, most work has focused on carboxylated nanodiamonds that are  stable in aqueous environments \cite{ say2011luminescent,ando1996chemical, yamano2017charge, schirhagl2014nitrogen, wu2022recent}.
However, the coverage of carboxyl groups on the surface of milled nanodiamonds is patchy as they are reported to occur predominantly at undercoordinated carbon sites at the edges between (111) and (100) crystalline facets \cite{nguyen2007adsorption, datta2011surface}.
With a limited number of surface carboxyl groups to stabilize the particles, the nanodiamonds tend to agglomerate when added to non-neutral solutions or biological fluids \cite{tzeng2011superresolution, yuan2009biodistribution, say2011luminescent}.
While  techniques exist for coating nanodiamonds with materials such as silica, polyethylene glycol, polymers, or lipids \cite{neburkova2017coating, takimoto2010preparation, von2013core, sukhova2019preparation, rendler2017optical, vavra2018supported, zhang2014use, zvi2023engineering}, 
these methods are specific to the type of coating and typically require complex synthetic processes that can degrade the quantum and optical properties of the NV center. 
%\FigureOne

Here we present a one-pot method for functionalizing milled nanodiamonds through the formation of emulsions. 
By sonicating hydrophobic nanodiamonds with amphiphilic small molecules, we form stable aqueous nanoparticle dispersions \cite{amirshaghaghi2018site}. 
By selecting the composition and tailoring the concentration of the amphiphile, we control the ligand chemistry and density. 
This method maintains nanodiamonds' unique material, optical, and spin properties while enhancing colloidal stability  and enabling carbodiimide crosslinker and click-chemistry conjugation reactions.

\FigureOne

\section{Results and Discussion}

\subsection{Emulsion Synthesis and Characterization}
%The emulsion process involves the rapid mixing of a hydrophobic nanodiamonds and an amphiphilic compound in water, resulting in aqueous dispersions of nanodiamond cores with organic molecular coatings. 
%The emulsions are created from commercially available, hydrophobic, octadecane-terminated milled, fluorescent nanodiamond material from Ad\'amas Nanotechnologies.%
In this study, we examine four types of colloidal nanodiamond samples. %including two emulsions that provide tailored surface chemistry%.
These include commercial, aqueous dispersions of carboxyl-terminated, milled fluorescent nanodiamonds (ND-COOH, shown in Figure~\ref{fig:Emlusion Overview}A) from Ad\'amas Nanotechnologies, which have an average diameter of 53 nm (Supporting Information Figure~S1) and contain less than 1 ppm of NV centers.
Based on previous statistical studies of the  heterogeneous distribution of NV centers in this material \cite{shulevitz2022template}, we estimate that each nanodiamond contains between 0-20 NV centers.
We also study commercial hydrophobic, octadecane-terminated nanodiamonds (ND-C18, shown in Figure 1B), fabricated by Ad\'amas Nanotechnologies using the same conditions as those for the ND-COOH samples. 
The ND-C18 have the same NV center density, but a larger average diameter of 68 nm, due to variability in the milling and size-purification process.
We create two emulsions from commercial ND-C18 material, one  with hemin (EM-Hem, shown in Figure 1C), 
and another with equal weight of hemin and cholesteryl-TEG azide (EM-Hem/Chol, shown in Figure 1D). 

To form emulsions, we adapt a previously reported procedure to coat superparamangetic iron oxide nanoparticles with protoporphyrin IX \cite{yan2018protoporphyrin}.
We first mix 1 mg of dry ND-C18 nanodiamonds with the amphiphilic compound(s) in 100--200 \si{\micro\liter} of organic solvent.  
We then add this mixture to 4 mL of water by pipette, while sonicating and using the pipette tip to vigorously stir until a homogeneous mixture is  observed (see Methods). 
%We first mix non-aqueous, dispersions of ND-C18 nanodiamonds in toluene with 10:1 toluene:dimethyl sulfoxide (DMSO) solutions of the amphiphilic compound(s) at a high concentration in a small volume of organic solvent (See Methods). 
%We then add this mixture to a larger volume of water and sonicate vigorously for five minutes. 
The mixture is left overnight, uncovered, to allow the most of organic solvent to evaporate, and finally, the samples are dialyzed to remove any remaining organic solvent or free amphiphile. 
For EM-Hem, we utilize hemin as the amphiphilic material at a weight ratio of 10:1 nanodiamond to hemin.
We used nanodiamond to hemin ratios ranging from 20:1 to 5:1 (Supporting Information Figures~S2-S8), and we found that 10:1 results in the most stable and consistent single-particle emulsions.
For EM-Hem/Chol, we modify the EM-Hem synthesis procedure by replacing half of the weight of hemin (MW = 651.94 g/mol) with cholesteryl-TEG azide (MW = 630.90 g/mol) to provide an azide termination for subsequent click-chemistry conjugation.
We are unable to form stable emulsions using cholesteryl-TEG azide as the sole amphiphilic material, likely due to the lower solubility of cholesteryl-TEG azide, compared to hemin, in water. 

\FigureTwo
 
Dynamic light scattering (DLS) is used to measure the particle size distribution for the four types of samples (Figure~\ref{fig:Emlusion Properties}A and Supporting Information Figure~S1).
The uncoated nanodiamond samples, ND-COOH ($53\pm15$~nm) and ND-C18 ($68\pm23$~nm), exhibit average diameters and standard deviations consistent with the vendors specifications.
For EM-Hem ($75\pm23$~nm)  and EM-Hem/Chol ($87\pm26$~nm), compared to the ND-C18 sample, the average diameters are larger by 7 nm and 19 nm, respectively. The increased diameters of the emulsions exceed twice the 1.5~nm and 3~nm molecular lengths of hemin and cholesteryl-TEG azide, consistent with coatings comprised of multiple amphiphilic layers. 
While we consider the possibility that multiple nanodiamonds might also cluster within a single coating layer, the  marginal increase in average particle size, with no significant change in standard deviation, suggests that clustering does not play a significant role.
Moreover, van der Waals interactions would cause larger particles to preferentially agglomerate, which is not consistent with our observations.
Transmission electron microscopy (TEM) images also show the particles remain separated (Supporting Information Figures~S9-S11). Figure~\ref{fig:Emlusion Properties}B shows the size distributions for EM-Hem and EM-Hem/Chol dispersions monitored over 9 months, demonstrating their long-term stability.
 
We measure the optical extinction and photoluminescence (PL) spectra of the emulsions to confirm that we can still optically address and measure NV centers in EM-Hem and EM-Hem/Chol dispersions (Figure~\ref{fig:Emlusion Properties}C). 
%Apart from stability, it is essential to consider the optical characteristics of the emulsions. 
Figure~\ref{fig:Emlusion Properties}C also displays, for comparison, the spectra of ND-COOH and ND-C18 dispersions and the spectra of free hemin micelles (see Methods). 
ND-COOH and ND-C18 exhibit monotonically decreasing extinction, arising from a combination of absorption and scattering at shorter wavelengths, while EM-Hem and EM-Hem/Chol feature an additional absorption peak around 325~nm associated with hemin.
As expected, EM-Hem shows a more pronounced absorption peak than EM-Hem/Chol arising from the larger amount of hemin present. 
Using these extinction spectra, we can estimate the average coating density of hemin molecules per nanodiamond (see Supporting Information section III). 
In the red and near-infrared regions under 532~nm excitation, the hemin control sample exhibits almost no emission, while all nanodiamond samples feature the characteristic NV-center PL emission spectrum.

\subsection{Effects on the NV-Center Charge State}

\FigureThree

Under optical excitation at 532~nm, the NV center fluctuates between the neutral (NV$^{0}$) and negative (NV$^{-}$) charge-state configurations through a process of ionization and recombination \cite{aslam2013photo}. 
Moreover, the time-averaged charge configuration of individual NV centers depends on the local chemical potential, which varies spatially due to inhomogeneities in the nanodiamonds' impurity concentrations and their surface chemistry \cite{shulevitz2022template, ong2017shape, reineck2019not, karaveli2016modulation}.
The nanodiamond ensemble emission spectra are therefore a linear combination of the characteristic spectra for NV$^{0}$ and NV$^{-}$, due to both temporal and spatial averaging.
Since the spin properties required for quantum sensing are only accessible in the NV$^{-}$ state, it is important to maximize the charge ratio, $F_\mathrm{NV^{-}} = S_\mathrm{NV^{-}}/(S_\mathrm{NV^{0}} +S_\mathrm{NV^{-}})$, where $S_\mathrm{NV^{0}}$ and $S_\mathrm{NV^{-}}$ are the integrated PL intensities of the neutral and negative charge states.
The quantity $F_\mathrm{NV^{-}}$ is extracted directly from measured ensemble emission spectra using nonnegative matrix factorization\cite{berry2007algorithms} (see Supporting Information section I.1), which decomposes each spectrum into its $\mathrm{NV^{0}}$ and $\mathrm{NV^{-}}$ components, as shown in Figure~\ref{fig:Charge Properties of Emulsions}A.

Figure~\ref{fig:Charge Properties of Emulsions}B shows $F_\mathrm{NV^{-}}$ for the four nanodiamond samples.
ND-C18 exhibits a significantly lower $F_\mathrm{NV^{-}}$ than ND-COOH, which is consistent with prior works showing that surface carboxyl groups donate electrons and raise the chemical potential \cite{meara2020computational, rondin2010surface, reineck2019not, petrakova2015charge, karaveli2016modulation}.
Interestingly, addition of the organic coatings in the emulsion samples acts to increase $F_\mathrm{NV^{-}}$, compared to the ND-C18 source material.
This observation suggests that the coating of hemin molecules, which also feature carboxyl groups, could similarly increase the chemical potential.
It is also possible that the hemin coating modifies the nanodiamond emission spectrum through preferential absorption or energy transfer of the shorter wavelengths associated with NV$^{0}$.
As shown in Fig~\ref{fig:Emlusion Properties}C, hemin features a broad absorption peak between 300-450~nm, with a tail extending to $\approx$600 nm.  
In either case, the slightly higher value of $F_\mathrm{NV^{-}}$ for EM-Hem compared to EM-Hem/Chol is consistent with the larger hemin concentration.
Regardless of the mechanism, the increased $F_\mathrm{NV^{-}}$ is desirable, since it improves the signal-to-noise ratio and sensitivity for quantum sensing experiments.
%Interestingly, through the proper chemical functionalization of the coating materials, the emulsion process could allow $F_\mathrm{NV^{-}}$ to be maximized.

\subsection{Effects on the NV-Center Spin Lifetime}

The NV center's spin lifetime ($T_1$) is an important figure of merit for nanodiamond quantum sensors. 
Variations in $T_1$ are used to detect broadband magnetic noise associated with free radicals, and $T_1$ represents an upper limit on the spin coherence time, $T_2$, which is relevant for detecting ac fields.
For NV centers in nanodiamonds, $T_1$ is typically limited by fluctuating spins on the nanodiamond surface; the strong $\approx 1/r^6$ scaling of this coupling, where $r$ is the distance between the fluctuating spin and an NV center, means that $T_1$ can vary over orders of magnitude for NV centers in nanodiamonds depending on their placement relative to the nanodiamond surface.
As a consequence, the ensemble average spin lifetime, $\langle T_1\rangle$, depends strongly on nanodiamond size as well as on surface termination \cite{shulevitz2022template, tetienne2013spin, peng2020reduction, ong2017shape,sangtawesin2019origins,tsukahara2019removing}.
Because the measurement can be performed optically, without the need for microwave infrastructure, $T_1$ is the preferred spin measurement for \textit{in vivo} sensing. 
Larger $T_1$ typically correlates with increased spin coherence times.
Moreover, increased $T_1$ lowers the threshold for detecting magnetic field noise, potentially improving the sensitivity to free radicals.
%Conversely, decreased $T_1$  lowers measurement times, potentially improving measurement speed and sensitivity. 

\FigureFour

We measure $\left\langle T_1 \right\rangle$ for nanodiamond dipsersions using an all-optical protocol.
In each measurement sequence, we initialize the NV centers into their $m_s=0$ spin-triplet sublevel using a
3~\si{\micro\second}, 532~nm laser excitation pulse and then count PL photons during a 350~ns readout laser pulse at delay times ($\tau$) ranging from 50~ns to 9~ms.
Based on the optical configuration and nanodiamond concentration, we estimate that each measurement probes approximately 200 nanodiamonds, each containing 0-20 NV centers. 
Figure~\ref{fig:T1 Overview}A shows the results of this measurement for ND-COOH and EM-Hem.
The PL signal is initially bright due to polarization in the $m_s=0$ state, and it decays as a function of $\tau$ due to relaxation into the $m_s=\pm 1$ dark states.
For an ensemble measurement, the shape of the observed  PL signal, $I(\tau)$, results from the summed contribution of several thousand NV centers, 
  \begin{equation}\label{Eq: Sum of PL}
 I(\tau)  = \sum_{i} I(\infty)  \left[1 + C \exp\left(-\frac{\tau}{T_{1,i}}\right)\right], 
 \end{equation}
where $T_{1,i}$ is the spin lifetime of the $i^\mathrm{th}$ NV center, $C$ is the spin-dependent PL contrast, and $I(\infty)$ is the brightness associated with a spin-unpolarized state.
To account for the ensemble average over many spin lifetimes, we model the data using a stretched exponential function:
 \begin{equation}\label{Eq: Stretch Fit}
  S(\tau)  = S(\infty)  \left[1 + C_\mathrm{S}{\exp\left(-\frac{\tau}{\ToneS}\right)^\beta}\right].
 \end{equation}
Here, $S(\infty)$ and $C_\mathrm{S}$ represent the asymptotic long-delay signal and PL contrast, respectively.
The stretched-exponential spin lifetime, $\ToneS$, describes the effective ensemble decay rate, whereas the stretching exponent, $\beta$, accounts for averaging over the distribution of lifetimes. 
The variables $\ToneS$ and $\left\langle T_1 \right\rangle$ are related by the formula\cite{johnston2006stretched}:
 \begin{equation}\label{Eq:avgT1}
  \left\langle {T}_1 \right\rangle  =  \ToneS \frac{\Gamma(\frac{1}{\beta})}{\beta}, 
 \end{equation}
where $\Gamma$ represents the gamma function. 
For $\beta$ in the range 0.5--1, as we typically observe, $\langle T_1\rangle$ is larger than $\ToneS$ by a factor between one and two.

Figure~\ref{fig:T1 Overview}B shows $\ToneS$ for the four samples.
The 87$\%$ increase in $\ToneS$ of ND-C18 compared to ND-COOH may reflect the larger average size of ND-C18 (see Fig.~\ref{fig:Emlusion Properties}A), since there will be a greater average distance between NV centers and the surface of a larger particle, or the difference in surface chemistry, which may result in a lower surface noise density \cite{tetienne2013spin}.
The difference in $\ToneS$ for EM-Hem and EM-Hem/Chol, in comparison to  ND-C18, are  within the measurement uncertainty.
This observation is surprising, since hemin contains paramagnetic Fe$^{3+}$ ions that would be expected to reduce $T_1$;  however, we find that this effect is mitigated by Fe$^{3+}$'s low intrinsic spin relaxation rate and low spin number compared to other paramagnetic ions. 
In the Supporting Information, we extract the hemin concentrations from the extinction measurements (Figure~\ref{fig:Emlusion Properties}C) and utilize a model for $T_1$, described below, to  quantify the impact of the hemin coating on $T_1$ (Supporting Information Figures~S13 and~S14). 
%The effect of the the hemin coating on $T_1$ is discussed in the Supporting Information. 

In order to examine the emulsions' sensitivity to free radicals, we prepare $\approx5$~nM Em-Hem dispersions and track $T_1$ as a function of increasing concentration of a gadolinium  (Gd$^{3+}$) chelate with butylamine termination, with $\GdConc$ varying from 0.05~\si{\micro\molar} to 2000~\si{\micro\molar} (see Figure~\ref{fig:T1 Overview}C).
We observe that $\langle T_1\rangle$ begins to decrease for $\GdConc\gtrsim 10$~\si{\micro\molar} before saturating when $\GdConc\gtrsim 1000$~\si{\micro\molar}.
To interpret these observations, we adapt a model by Tetienne \textit{et al}.,\cite{tetienne2013spin} to calculate the effect of fluctuating spins on $T_1$ for NV centers inside nanodiamonds.
The drop in $T_1$ will depend on both [Gd$^{3+}$] as well as the distance of the NV centers from the outside spin bath. 
We model the nanodiamonds as 68~nm diameter spheres, each containing a random spatial distribution of NV centers.
%We first model the each nanodiamond as a perfect sphere with a fixed diameter of 60nm, a random distribution of NV centers, and an initial  $\left\langle T_1 \right\rangle$  of 145$\mu$s.
We initially assume that no NV centers exist within 3~nm of the nanodiamond surface \cite{bluvstein2019identifying}. 
%and that the hemin coating blocks Gd$^{3+}$  from being within  nm of the nanodiamond surface. 
By integrating over possible locations for the NV centers and assuming a uniform [Gd$^{3+}$] in the surrounding medium, we calculate $\left\langle T_1 \right\rangle$ as a function of $\GdConc$. 
The result of this calculation is shown as a dotted line in Fig.~\ref{fig:T1 Overview}C.
This model significantly underestimates the impact of $\GdConc$ on $\left\langle T_1 \right\rangle$.
As shown in Supporting Information Figure~S15, we can improve the model's agreement with the data in the range of $\GdConc\approx10$--100~\si{\micro\molar} by  allowing for NV centers to reside within 0.1~nm of the nanodiamond surface, however, this is inconsistent with the proposed instability of near-surface NV centers and, moreover, this explanation does not reproduce the saturation of $\langle T_1\rangle$ for larger $\GdConc$.
To resolve this discrepancy, we hypothesize that some of the Gd$^{3+}$ chelates adsorb to the particle surface, resulting in a Gd$^{3+}$ surface density $\sigma(\GdConc)$ that follows the Langmuir model:
 \begin{equation}\label{Eq: rho to sigma}
\sigma(\GdConc) = \sigma_{\infty} \frac{ K  \GdConc}{1 + K   \GdConc },
 \end{equation}
where $\sigma_\infty$ and $K$ are empirical constants.
The solid curve and shaded region in Figure~\ref{fig:T1 Overview}C is the result of a fit using this model, with $\sigma_\infty=$ 0.10  $\pm$ 0.04~nm$^{-2}$ and $K=$6000 $\pm$ 4000~\si{\molar}$^{-1}$.
These values correspond to a maximum of 1500 $\pm$ 600~Gd$^{3+}$ adsorbed to a 68~nm diameter nanodiamond. 
This maximum is $<10$\% of the $\approx$ 20000 surface sites for a perfectly packed Gd$^{3+}$-chelate monolayer on a 68~nm diameter sphere.  
Although this model makes many simplifying assumptions, the results distinctly show how NV centers can serve as sensitive reporters of variations in their nanoscale local environment, in this case distinguishing the effects of adsorbed free radicals from those in solution.

\subsection{Chemical sensitivity}

A typical sensing experiment with NV centers involves tracking the change in PL as a function of variations in the environment.
The chemical sensitivity, $\eta_{\GdConc}$, quantifies the minimum detectable change in $\GdConc$ that can be identified in a given measurement bandwidth.
Here, we consider the case of $T_1$ relaxometry, where the PL signal is given by Equation~(\ref{Eq: Stretch Fit}) for a fixed delay time, $\tau_0$.
The variables $\ToneS$, $\beta$, and $C_\mathrm{S}$ are, in principle, all functions of $\GdConc$. 
In practice, however, we empirically find that the predominant effect is on $\ToneS$ (Supporting Information Figures~S16-S20).
Figure~\ref{fig:T1 Overview}D shows the variation in $\ToneS$ as a function of $\GdConc$ for ND-COOH and EM-Hem.
Based on the discussion in the previous section, and similar to previous works \cite{vavra2018supported}, we utilize an empirical model:
 \begin{equation}\label{Eq: T1s(Gd)}
  \ToneS(\GdConc) = T_{1,\mathrm{Sat}} + \Delta T_1 \exp\left({\frac{-\GdConc}{\GdConc_0}}\right)
 \end{equation} 
Assuming the measurement uncertainty is dominated by photon shot noise and that the duration of each measurement cycle is dominated by $\tau_0$, the chemical sensitivity is given by:
 \begin{equation}\label{Eq:ChemicalSensitivity}
\eta_{\GdConc}= \sqrt{\frac{\tau_0 S(\tau_0)}{R t_\mathrm{m}}}
\left(\frac{\partial S}{\partial \ToneS}\cdot \frac{d\ToneS}{d\GdConc} \right)^{-1},
 \end{equation}
where $R$ is the steady-state photon count rate, and $t_\mathrm{m}$ is the photon integration time per cycle.

Figure~\ref{fig:T1 Overview}D shows $\eta_{\GdConc}$ as a function of $\GdConc$, for the ND-COOH and EM-Hem samples. 
In evaluating Equation~(\ref{Eq:ChemicalSensitivity}), we assume a fixed $\tau_0=T_1^{\mathrm{S}}/2$, and $R=500$~kCts/s.
This count rate represents the typical brightness of a single nanodiamond containing $\approx$20 NV centers
measured using a high-NA optical collection system\cite{shulevitz2022template}.
It is also similar to the signal levels in our ensemble measurements, where we probed $\approx$200 nanodiamonds and attenuated the signal in order to remain within the dynamic range of our single-photon detector. 
%Here we assume $R = 500$kcts/s as the typical count rate for a single nanodiamond containing $\approx$ 15 NV centers measured with a high-NA optical collection system. 
In an optimized sensing experiment, orders-of-magnitude improvements in sensitivity can be achieved by collecting the fluorescence of $N$ nanodiamonds, with a scaling according to $\eta_\GdConc\propto 1/\sqrt{N}$.
%Therefore, in our measurement setup which probes ensembles of nanodiamonds, we expect orders of magnitude increases in sensitivity according to a $\sqrt(N)$ scaling. 

For small $\GdConc$, we observe that $\eta_{Gd^{3+}}$ for ND-COOH is approximately three times smaller than $\eta_{Gd^{3+}}$ for EM-Hem, however this advantage rapidly declines for $\GdConc>10$~\si{\micro\molar} and reaches parity at $\GdConc\approx500$~\si{\micro\molar}.
The initially smaller $\eta_\GdConc$ results from ND-COOH's shorter $T_1^{S}$, which improves sensitivity by reducing the required delay time in each measurement.
%The difference in $\eta_\GdConc$ for small $\GdConc$ results from the samples' different $T_1^{S}$; in this case, a shorter $\ToneS$ improves sensitivity by reducing the required delay time in each measurement.
%5Just transition to next setction
The spin lifetime can be tuned for the needs of particular applications. 
Since past work \cite{tetienne2013spin, ong2017shape, shulevitz2022template} and the previous section shows that  $\ToneS$ increases with particle size, the emlusions' $\eta_\GdConc$ could be improved by employing smaller initial ND-C18 nanodiamonds.
On the other hand, ac-field sensing experiments benefit from longer $\ToneS$.
% Not sure we need this transition 
The interaction of nanodiamonds with the Gd$^{3+}$ spins could also be improved by employing conjugation techniques instead of relying indirectly on adsorption, as we explore in the next section.

% LCB location of editing, 5/31 3pm

\subsection{Conjugation}

The application of nanodiamonds as sensor platforms for nanomedicine requires chemical conjugation of various biomolecules to the nanodiamonds.
Examples include DNA oligos or antibodies for selective binding and targeting, or labels and transducers such as dyes or paramagnetic species.
Carboxyl-terminated nanoparticles (ND-COOH, EM-Hem, and EM-Hem/Chol) can be conjugated to amine-terminated compounds through carbodiimide crosslinking chemistry \cite{hermanson2013bioconjugate}.
In this method, a combination of N-(3-dimethylaminopropyl)-N'-ethylcarbodiimide hydrochloride (EDC) and N-hydroxysuccinimide (Sulfo-NHS) is utilized to mediate the conjugation reaction. 
To crosslink the carboxyl-terminated particles to amine-terminated compounds, we first add equal amounts of EDC and Sulfo-NHS to the carboxy-terminated nanoparticles in a $5\%$ N-(2-Hydroxyethyl)piperazine-N'-(2-ethanesulfonic acid) (HEPES) buffer, used to maintain a neutral pH. 
We add excess crosslinker (EDC and Sulfo-NHS) and target conjugates, utilizing concentrations $\approx$100X higher than the number of surface -COOH groups, in order to compensate for EDC hydrolysis and increase the likelihood conjugation. 
We shake the mixture for 25 min before adding a chosen conjugate and then continue to shake the mixture overnight. 
Although previous studies have demonstrated  EDC-based crosslinking to nanodiamonds\cite{manus2010gd, cordina2018reduced, chang2013highly, garg2019synthesis}, we find it challenging to reproduce these results for ND-COOH without observing significant agglomeration  (Figure~\ref{fig:EDC Conjugation}B, Supporting Information Figure~S21).
In contrast, EM-Hem exhibits greater stability toward the crosslinking reaction.
The samples are cleaned using centrifugation or dialysis to remove any unconjugated target particles and crosslinker by-products.  

\FigureFive

The crosslinking reaction is general, allowing the addition of amine-terminated molecules of interest. 
We considered amine-terminated dyes (Supporting Information Figures~S22), and as we focus here on spin-based sensing, conjugation of the butylamine-terminated, Gd$^{3+}$-chelate spin label (Figure~\ref{fig:EDC Conjugation}A).
The Gd$^{3+}$ conjugation is especially useful since the spin label is expected to decrease ${T}_1$ without altering other nanodiamond optical properties.
We performed two conjugations using 1.5 mL samples of 0.25 mg/mL EM-Hem ($\approx$ 1~\si{\nano\molar} of 68~nm spherical nanodiamonds) and 3~mg of Gd$^{3+}$ (3~\si{m\molar}), with different molar ratios of EDC to Gd$^{3+}$, i.e., [EDC]:[Gd$^{3+}$] of 2:1 and 3:1.
A control sample (0:1) is also made in parallel to probe the extent of Gd$^{3+}$ chelate adsorption in the absence of the EDC crosslinker.
Figure~\ref{fig:EDC Conjugation}C shows $\ToneS$ of each sample immediately after conjugation (dotted) and after washing \textit{via} either centrifugation (striped) or dialysis (solid). 
As expected, after the initial mixing with Gd$^{3+}$, all samples exhibit a significant drop in $\ToneS$. 
The washing results in an increase in $\ToneS$  for the control sample as the free Gd$^{3+}$ chelate is removed from the solution. 
Conversely, regardless of the cleaning method, the two conjugated samples retain a shortened $\ToneS$, indicating successful chemical bonding.

Inductively coupled plasma-optical emission spectroscopy (ICP-OES) measurements are used to quantify the amount of  Gd$^{3+}$ chelate in each sample after dialysis (Figure~\ref{fig:EDC Conjugation}D).
We observe significantly more Gd$^{3+}$ in the two conjugated samples than in the control, and the amount of retained Gd$^{3+}$ is proportional to the amount of EDC used. 
Since the amount of EDC and Gd$^{3+}$ spin label are added in significant excess of the number of hemin -COOH groups, this trend implies that the conjugation is limited by the efficiency of the EDC reaction. %or the available conjugates (Gd$^{3+}$). 
By once again assuming spherical 68~nm diameter nanodiamonds, the ICP-OES measurements imply 1600 $\pm$ 1000, 12000 $\pm$ 2000
 and 27000 $\pm$ 5000 Gd$^{+3}$ chelates per nanodiamond for the control and two conjugated samples, respectively. 
The residual Gd$^{3+}$ in the control sample likely reflects an incomplete dialysis of the samples. 
Some of the remaining Gd$^{3+}$ may be adsorbed to the nanodiamonds' surfaces, but the control sample's long $\ToneS$ implies that the majority of Gd$^{3+}$ remains unbound.
The approximately 10-fold to 20-fold increase in retained Gd$^{3+}$ for the two conjugated samples over the control highlights the success of crosslinking conjugation. 
In fact, here the ICP-OES measurements reveal higher retained [Gd$^{3+}$] than the observed decrease in $\ToneS$ would imply. 
From the relatively small decrease in $\ToneS$, we infer that the conjugated Gd$^{3+}$ chelates are positioned farther from the nanodiamond surface than the adsorbed chelates responsible for the reduced $T_1$ in Figure~\ref{fig:T1 Overview}D.  
We did not notice agglomeration in DLS measurements from any of the samples immediately after the conjugation process. 
After dialysis, however, the conjugated samples showed signs of agglomeration (Supporting Information Figure~S23), which suggests that the conjugation process destabilizes the emulsions. 
%which likely interferes with cleaning via dialysis. 
%With an agglomerated sample, Gd$^{3+}$ may be adsorbing to the clusters which could limit chelate removal during dialysis, but would not significantly decreases the average distance between the NV centers and the spin labels. 
%While the instability of the EM-Hem-Gd$^{3+}$ complicates both our analysis and future applications, the results still indicate the feasibility of crosslinking conjugation. 

\FigureSix
The crosslinking reaction has several disadvantages, most notably the instability of EDC and the added complexity of adding Sulfo-NHS to increase stability and buffers to control the pH.
In contrast, copper-free click chemistry reactions are a widely adopted method for biological conjugation\cite{thorek2009comparative, meyer2016click, battigelli2022recent, devaraj2021introduction, jewett2010cu}.
Here, we utilize the emulsion process to enable click chemistry based on the reaction of  dibenzocyclooctyne (DBCO) compounds with terminal-azide ($N_{3}$) groups \cite{debets2011bioconjugation,agard2004strain, campbell2011strain}. %, which we successfully introduce at the nanodiamond surface using the emulsion technique. 
The emulsion technique provides a simple route to introducing $N_3$ groups around the nanodiamond surface, as seen in Figure~\ref{fig:Emlusion Overview}C for the example of EM-Hem/Chol. 
Again, this conjugation reaction is general, and allows EM-Hem/Chol to be conjugated with any DBCO-terminated particle including: dyes (Supporting Information Figure~S24) and DBCO-terminated Gd$^{+3}$ spin labels (Figure~\ref{fig:Click Conjugation}A). 
In contrast to the crosslinking reaction, which involves multiple steps where pH must be carefully controlled, the click-chemistry conjugation of Gd$^{3+}$-DBCO with EM-Hem/Chol simply consists of mixing the two materials together overnight.

To test the reaction efficiency, we carry out the conjugation reaction with increasing Gd$^{3+}$-DBCO to N$_{3}$ molar ratios (1:4, 2:1, 10:1), and compare it with a nonconjugated control sample (0:1).
Figure~\ref{fig:Click Conjugation}B,C shows Fourier transform infrared transmission (FTIR) measurements used to monitor the CH and N$_{3}$ stretching regions for the four samples.
Successful conjugation is evidenced by the increase in the CH stretching modes and a corresponding reduction in the N$_{3}$ modes, as quantified by the ratio of their integrated areas in Figure~\ref{fig:Click Conjugation}D.
Similar to the crosslinking conjugation, we require an excess of DBCO, here only 10X, to completely quench the N$_{3}$ signal and ensure complete conjugation. 
The greater conjugation as DBCO:N$_{3}$ increases is also reflected by a decrease in $\ToneS$ (Figure~\ref{fig:Click Conjugation}E), which remains low before (striped) and after dialysis (solid), and by an increase in  [Gd$^{3+}$] measured by ICP-OES after dialysis (Figure~\ref{fig:Click Conjugation}F). 
 The ICP-OES measurements imply concentrations of 4000
 $\pm$  2000, 
 5000
$\pm$ 3000 and 11000
 $\pm$ 4000 Gd$^{3+}$ per spherical 68~nm nanodiamond for the conjugated samples with increasing DBCO:N$_{3}$. 
The maximum coating is less than we observed for crosslinking, which we expected due to the smaller number of potential conjugations sites. 
Notably, $\ToneS$ for the click-conjugated reactions are lower than those of the crosslinked samples in Figure~\ref{fig:EDC Conjugation} and the unconjugated samples in Figure~\ref{fig:T1 Overview}, even though the distance between the conjugated chelate and the nanodiamond surface is larger and the amount of initially added conjugation material is smaller.
%This result demonstrates the efficiency of the click conjugation reaction. 
%due to the increased length of the cholesteryl-TEG azides and the only 10X excess of DBCO to N$_{3}$.  
To control for possible adsorption of Gd$^{3+}$-DBCO to the nanodiamonds, we repeated the experiments for ND-COOH and EM-Hem  samples (Supporting Information Figures~S26-S34), where we do not expect any click reactions to occur.
We observe longer $\ToneS$ times for these samples after dialysis and significantly lower concentrations of Gd$^{3+}$ detected by ICP-OES, consistent with the 0:1 control sample of Figure~\ref{fig:EDC Conjugation}.

We do not observe agglomeration of the samples after click-chemistry conjugation (Supporting Information Figure~S35).
We hypothesize that the hybrid EM-Hem/Chol surface coating allows the functional separation of stabilization, imparted by the COOH groups of hemin, with conjugation, achieved using the azide-terminated cholesteryl-TEG.  
Our results illustrate a simple and efficient nanodiamond emulsion click-chemistry procedure that could not be achieved with commercially available carboxyl-terminated nanodiamonds.

\section{Conclusion}
We report nanodiamond emulsions as versatile building blocks for quantum sensing and nanomedicine platforms that maintain the beneficial optical and quantum properties of the NV center, while allowing for generalized chemical conjugation.
The hybrid surface termination enables multiple simultaneous yet distinct conjugation reactions without jeopardizing the colloidal stability of the nanoparticles.
By separating the tasks of stabilizing and conjugating across two distinct surface groups, the EM-hemin/chol nanoparticles outperformed the singularly terminated alternatives. 
We also observed that the charge and spin properties of the NV center can be modified by altering the composition of the coating materials. 
These controls will enable higher signal-to-noise ratios and tunable sensitivity to external perturbations.  

The polydispersity of milled nanodiamonds currently hinders biological sensing applications. 
In our measurements and modeling, we averaged over a large number of nanodiamonds to account for the variations in size, NV-center density, and surface termination. 
A more precise quantification of these parameters, and their variations across individual nanodiamonds, would improve our understanding and ability to engineer the materials for specific applications.
Nevertheless, in their current form, EM-Hem and EM-Hem/Chol represent a major improvement over commercially available milled nanodiamonds (ND-COOH and ND-C18) and demonstrate the utility of forming emulsions to develop chemical quantum sensors.

While we have focused here on the imaging and sensing aspects of nanomedicine, our work lays the foundation for advances in targeting and drug delivery.
These functionalities could be achieved through the addition of amphiphiles with specifically chosen functional groups, such as folic-acid-terminated lipids for targeting, 
or the simultaneous conjugation to biomolecules, such as proteins, antibodies, or DNA oligos.  
In this way, a complete nanomedicine platform can be designed, combining targeting and drug delivery functionality with the quantum sensitivity of biocompatible nanodiamonds. 
%Although we have only provided an initial step, these quantum nanomedical devices could drastically change the way we detect and treat disease.  
%That the required control over the surface terminations of  nanodiamonds can be achieved through a simple one-pot method highlights the importance of the emulsions method. 

\section{Methods}\label{MethodsFor Emulsions}

\subsection{Synthesis Methods}
To form the emulsions, we first disperse 1 mg of ND-C18 in 60 $\mu$L of toluene.
Separately, we disperse  3 mg of  hemin (Sigma Aldrich) in  1 mL of 10:1 toluene to dimethyl sulfoxide (DMSO) and 1 mg of cholesteryl-TEG azide (Sigma Aldrich) in  1 mL of toluene. 
For EM-Hem we add 33 $\mu$L of the hemin dispersion to the nanodiamonds. 
For EM-Hem/Chol we add 17 $\mu$L of the hemin dispersion and 50 $\mu$L of the cholesteryl-TEG azide dispersion to the nanodiamonds. 
We then add toluene to the dispersion so that the final volume of the ND-C18/amphiphilic mixture is between 100$\mu$L- 200$\mu$L. 
Next, we pipette the nanodiamond-amphiphile mixture into a glass vial containing 4 mL of Milli-Q (18.2 M$\Omega$·cm) water and stir vigorously, during sonication, for at least 5 min until a homogeneous mixture is observed. 
The toluene is then allowed to evaporate  overnight uncovered at room temperature. 
Next, we perform dialysis with  4  L  of  Milli-Q (18.2 M$\Omega$·cm) water to  remove  DMSO and free amphiphilic compounds.  
Hemin micelles are formed in the same manner but without the addition of nanodiamonds. 

To synthesize the Gd$^{3+}$-DBCO chelates we first dissolve 15 mg of 1,4,7,10-tetraazacyclo\-dodecane-1,4,7\--tris\-(acetic acid)\--10\--(3\--oxo\--3\--(5\--azadibenzocyclootyne)\-acetamide) 
(DO3A-\-DBCO, Macrocyclics B-283) in a 0.5 mL  H2O/MeOH= 1/1 solution. 
We then add  2 equivalents of GdCl$_3$ (Sigma Aldrich) and 10 equivalents of N,N-Diisopropylethylamine (DIPEA) and stir at room temperature for 3 h to initially mix and then stir overnight at 40 $^{\circ}$C. 
We then evaporate to dryness under vacuum and purify the sample by high-performance liquid chromatography (HPLC).

\subsubsection{Conjugations}
For carbodiimide crosslinking conjugation we add equal amounts of EDC (Thermo Fisher) and Sulfo-NHS (Thermo Fisher) to 1.5 mL of EM-Hem at a 0.25 mg/mL concentration (nanodiamond mass) in a 5$\%$ HEPES solution. 
The crosslinkers are measured out in powder form and added directly to the EM-Hem solutions to minimize the impact of hydrolysis. 
The solution is then shaken for 25 min before adding the Gd$^{3+}$ chelate (Macrocyclics	
X-287) and then is left to shake overnight to complete the conjugation.

For click chemistry conjugation, GD$^{3+}$-DBCO chelates are dissolved in  Milli-Q water and then mixed with 0.6~mL samples of EM-Hem/Chol at a 0.25 mg/mL concentration, (nanodiamond mass) with the proper molar ratio and shaken overnight.

\subsubsection{Optical Measurements}
Spin relaxation measurements are taken using a custom-built confocal microscope described by Huang \textit{et al}.,\cite{huang2019monolithic} and Shulevitz \textit{et al}.,\cite{shulevitz2022template}
configured to preform measurements of nanodiamond dispersions.
A 532~nm (green) continuous-wave laser (Coherent, Compass 315M-150) is focused through a 5X objective (Olympus UMPlanFL) into the center of a 300 \si{\micro\liter} cuvette (Spectrocell).
To average over a sufficient number of nanodiamonds, we calculate the beam width of this laser and increase the concentration of nanodiamond dispersion by a factor of 5X, \textit{via} vacuum evaporation or centrifugation, to ensure at least 150 nanodiamonds are within the excitation volume, assuming uniform spherical  nanodiamonds.  
$T_1$ lifetime measurements are implemented by programming the required pulse sequences onto an arbitrary-waveform generator (AWG; AWG520 Tektronix).  
The AWG control signals are passed to the microscope's  acousto-optic modulator (AOM) for generating optical pulses and to three high isolation switches (ZASWA-250DR Mini-Circuits) for time-gating photon detection events recorded by counters in a data acquisition card (National Instruments PCIe-6323). 

Spectral emission measurements are taken on an Edinburgh Instruments FLS1000 spectrometer with a PMT-980 photodetector at  532~nm excitation using a  450~W Xe lamp.
Absorption spectra are measured with an Agilent Cary 5000 spectrophotometer.
ICP-OES measurements were performed on a SPECTRO GENESIS ICP-OES spectrometer.
DLS measurements are conducted on a Malvern Instruments Zetasizer  Nano-s and analyzed using Malvern's software. 
FTIR transmittance measurements are performed on a
Nicolet 6700 (Thermo Scientific) spectrometer with a mercury
cadmium telluride detector.
To perform the measurements, samples were drop cast on double-side polished Si chips and vacuum dried. 
Measurements are then taken at multiple spots on the film to account for  potentially uneven film thickness.

\begin{acknowledgments}
H.J.S., M.O., T-Y.H., L.C.B. and C.R.K. acknowledge primary support for this work by the NSF through the University of Pennsylvania Materials Research Science and Engineering Center (MRSEC) (DMR-1720530).
M.O. acknowledges support from the Natural Sciences and Engineering Research Council of Canada (NSERC).
C.B. acknowledges support from the NSF SUNFEST Research Experiences for Undergraduates (REU) program at the University of Pennsylvania (1950720).
S.Y. and C.B.M. acknowledge support from the NSF Center Integration of Modern Optoelectronic Materials on Demand (STC-IMOD) (DMR-2019444).
D.J. acknowledges support from the Chemical and Nanoparticle Synthesis  Core. 
A.T. acknowledges the support of the National Institutes of Health (NIH) (R01 EB029238 and R01 EB028858).
L.C.B. and A.T. also acknowledge support from the Grainger Foundation Frontiers of Engineering Grant under the National Academy of Sciences (Award Number: 2000010464).
The authors thank Peter Maurer for helpful comments on an early version of this manuscript.
\end{acknowledgments}

%\appendix

%\section{Appendixes}

\bibliography{Bib_merged}% Produces the bibliography via BibTeX.

\end{document}

%\preprint{APS/123-QED}
\def \ToneS {T_1^\mathrm{S}}

\title{ Supporting Information for \\ Nanodiamond Emulsions for Enhanced Quantum Sensing and Click-Chemistry Conjugation}% Force line breaks with \\

\author{Henry J. Shulevitz}
\affiliation{Department of Electrical and Systems Engineering, University of Pennsylvania, Philadelphia Pennsylvania 19104, USA}

\author{Ahmad Amirshaghaghi}
\affiliation{Department of Bioengineering, University of Pennsylvania, Philadelphia Pennsylvania 19104, USA}

\author{Mathieu Ouellet}
\affiliation{Department of Electrical and Systems Engineering, University of Pennsylvania, Philadelphia Pennsylvania 19104, USA}
\author{Caroline Brustoloni}
\affiliation{Department of Electrical and Systems Engineering, University of Pennsylvania, Philadelphia Pennsylvania 19104, USA}

\author{Shengsong Yang}
\affiliation{Department of Chemistry, University of Pennsylvania, Philadelphia Pennsylvania 19104, USA}

\author{Jonah J. Ng}
\affiliation{Department of Electrical and Systems Engineering, University of Pennsylvania, Philadelphia Pennsylvania 19104, USA}

\author{Tzu-Yung Huang}
\altaffiliation[Present address:]{ Nokia Bell Labs, 600 Mountain Ave., Murray Hill, NJ 07974, USA
}
\affiliation{Department of Electrical and Systems Engineering, University of Pennsylvania, Philadelphia Pennsylvania 19104, USA}
\author{Davit Jishkariani}
\affiliation{Department of Bioengineering, University of Pennsylvania, Philadelphia Pennsylvania 19104, USA}
\affiliation{Chemical and Nanoparticle Synthesis Core, University of Pennsylvania,  Philadelphia, 19104, USA}

\author{Christopher B. Murray}
\affiliation{Department of Chemistry, University of Pennsylvania, Philadelphia Pennsylvania 19104, USA}
\affiliation{Department of Materials Science and Engineering, University of Pennsylvania, Philadelphia Pennsylvania 19104, USA}

\author{Andrew Tsourkas}
\affiliation{Department of Bioengineering, University of Pennsylvania, Philadelphia Pennsylvania 19104, USA}

\author{Cherie R. Kagan}
\email[Email: ]{kagan@seas.upenn.edu \& lbassett@seas.upenn.edu}
\affiliation{Department of Electrical and Systems Engineering, University of Pennsylvania, Philadelphia Pennsylvania 19104, USA}
\affiliation{Department of Materials Science and Engineering, University of Pennsylvania, Philadelphia Pennsylvania 19104, USA}
\affiliation{Department of Chemistry, University of Pennsylvania, Philadelphia Pennsylvania 19104, USA}

\author{Lee C. Bassett}
\email[Email: ]{kagan@seas.upenn.edu \& lbassett@seas.upenn.edu}
\affiliation{Department of Electrical and Systems Engineering, University of Pennsylvania, Philadelphia Pennsylvania 19104, USA}

\def \ToneS {T_1^\mathrm{S}}
\maketitle

\def\SuppDLSRaw{\begin{figure}[]
\renewcommand\figurename{Figure}
\centering
\includegraphics[width = \textwidth ]{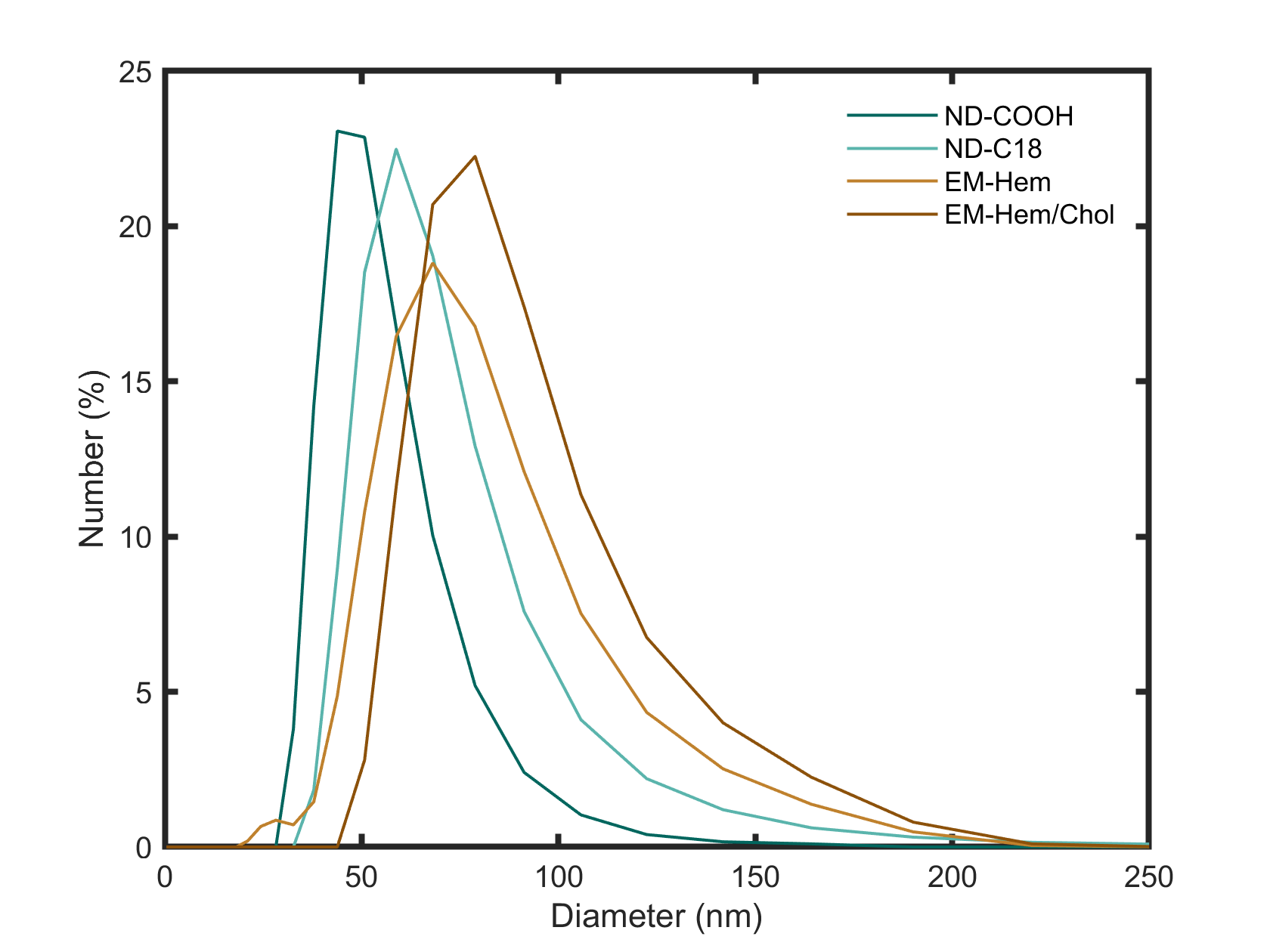}
%\justifying
\caption[DLS Data For Emulsions]{Dynamic light scattering (DLS) data for aqueous, carboxyl-terminated, hydrophilic nanodiamonds containing NV centers (ND-COOH), octadecane-terminated, hydrophobic nanodiamonds containing NV centers (ND-C18), ND-C18 nanodiamonds coated with amphiphilic hemin, formed as an emulsion (EM-Hem), and ND-C18 nanodiamonds coated with hemin and cholesteryl-TEG azide, formed as an emulsion (EM-Hem/Chol).}
\label{fig:DLS Raw}
\end{figure}}

 \def\TEMFiveHundred{\begin{figure}[t]
\renewcommand\figurename{Figure}
\centering
\includegraphics[width = \textwidth ]{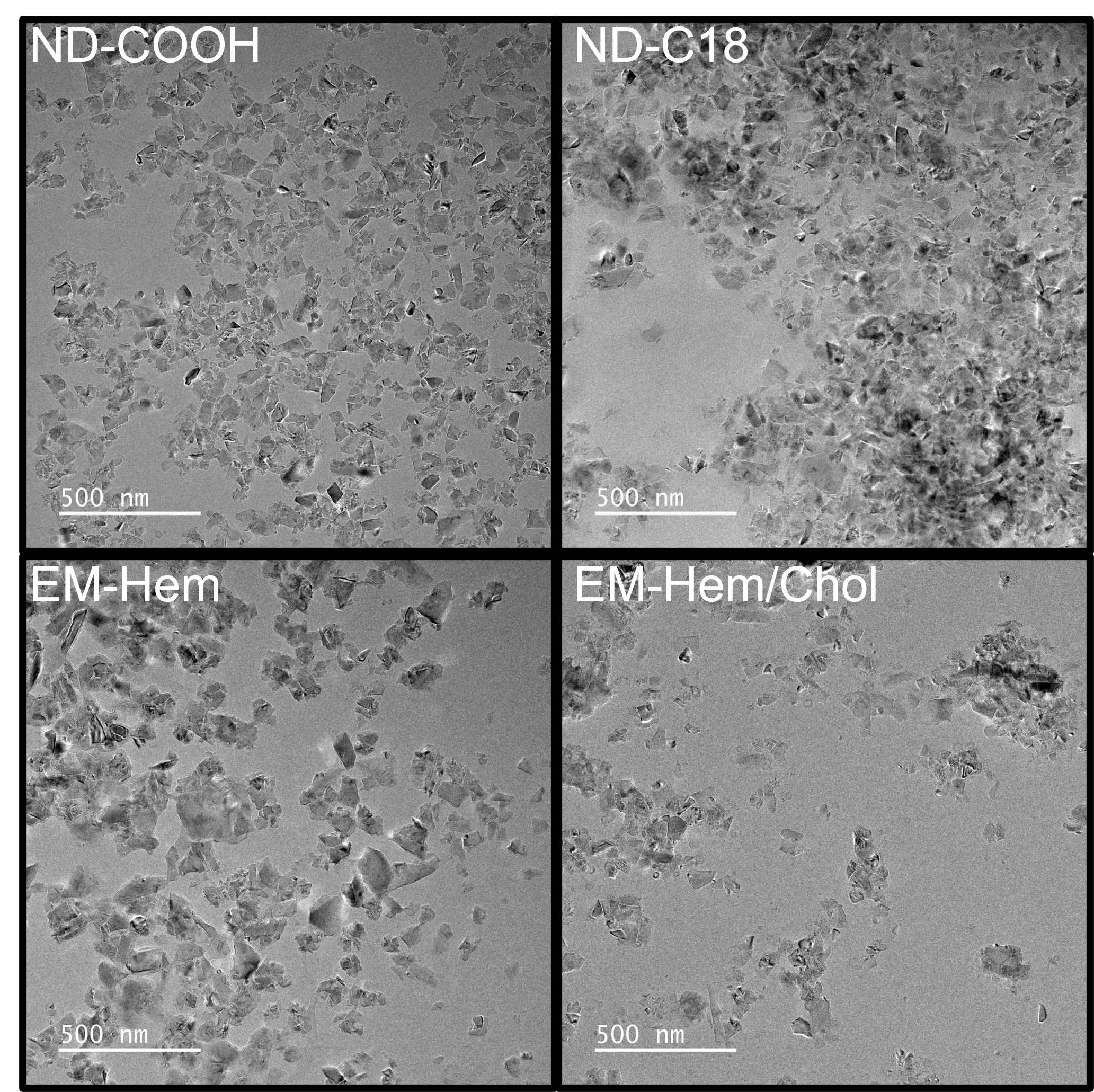}
%\justifying
\caption[Nanodiamond Emulsions TEM Wide-Area]{ 
Low-resolution transmission electron microscopy (TEM) images of ND-COOH, ND-C18, EM-Hem, and EM-Hem/Chol.}
\label{fig:TEM500}
  
 \end{figure}}

\def\TEMOneHundred{\begin{figure}[t]
\renewcommand\figurename{Figure}
\centering
\includegraphics[width = \textwidth ]{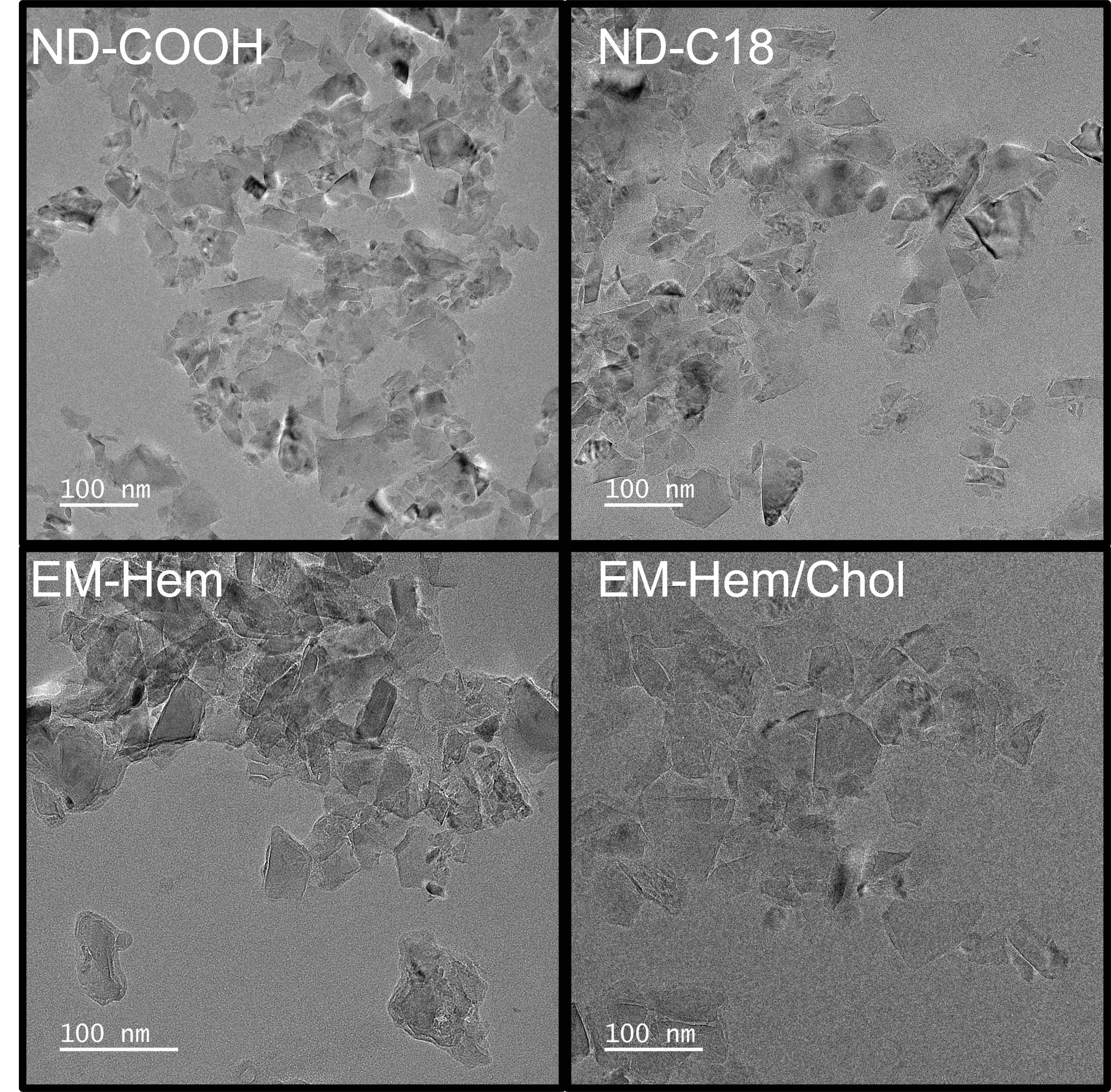}
%\justifying
\caption[Nanodiamond Emulsions TEM Mid-Area]{ 
Intermediate-resolution TEM images of ND-COOH, ND-C18, EM-Hem, and EM-Hem/Chol.}
\label{fig:TEM100}
  
 \end{figure}}

 \def\TEMTwent{\begin{figure}[t]
\renewcommand\figurename{Figure}
\centering
\includegraphics[width = \textwidth ]{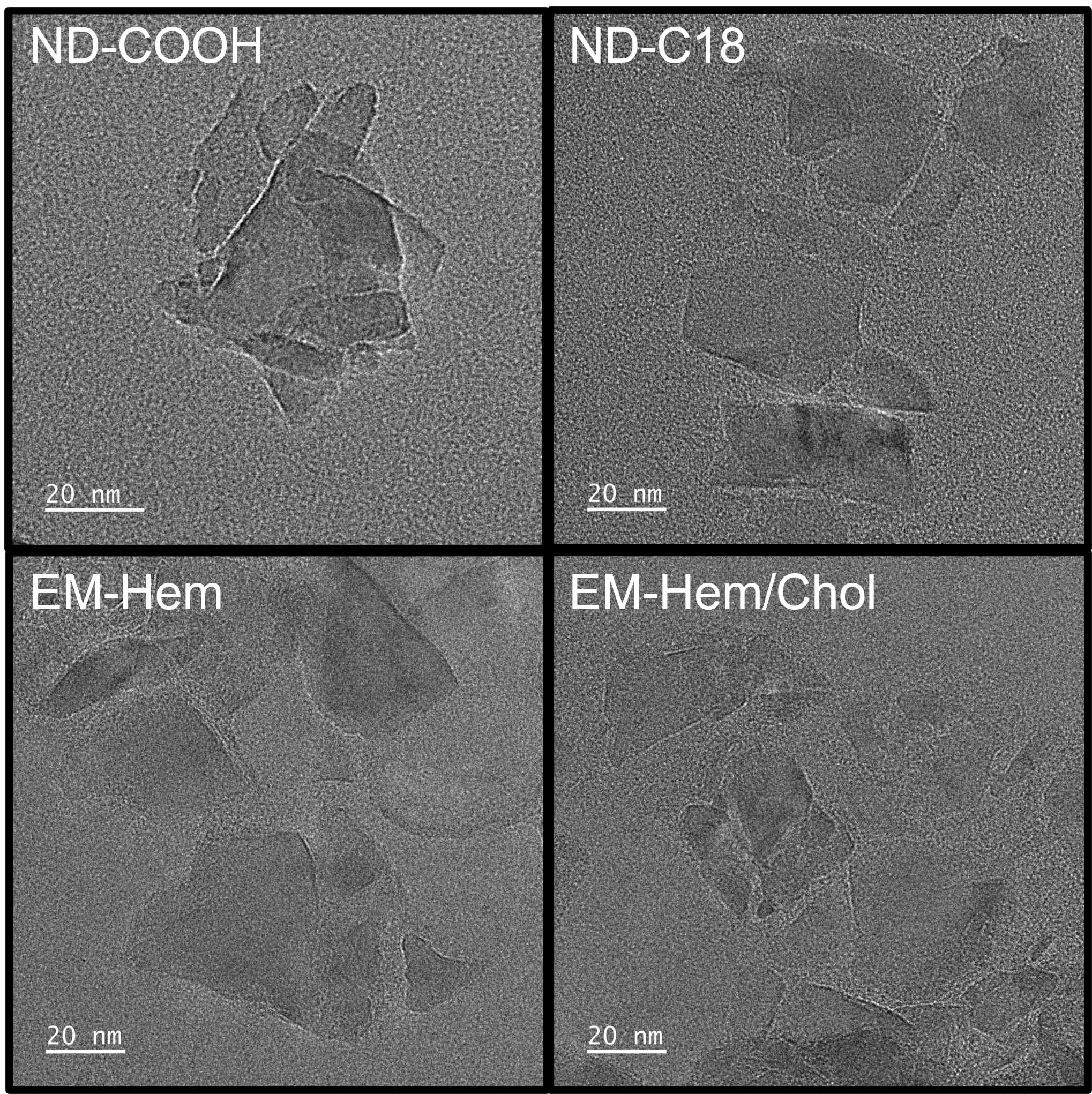}
%\justifying
\caption[Nanodiamond Emulsions TEM Small-Area]{ 
High-resolution TEM images of ND-COOH, ND-C18, EM-Hem, and EM-Hem/Chol.}
\label{fig:TEM20}
  
 \end{figure}}

\def\SuppDLSBar{\begin{figure}[t]
\renewcommand\figurename{Figure}
\centering
\includegraphics[width = \textwidth ]{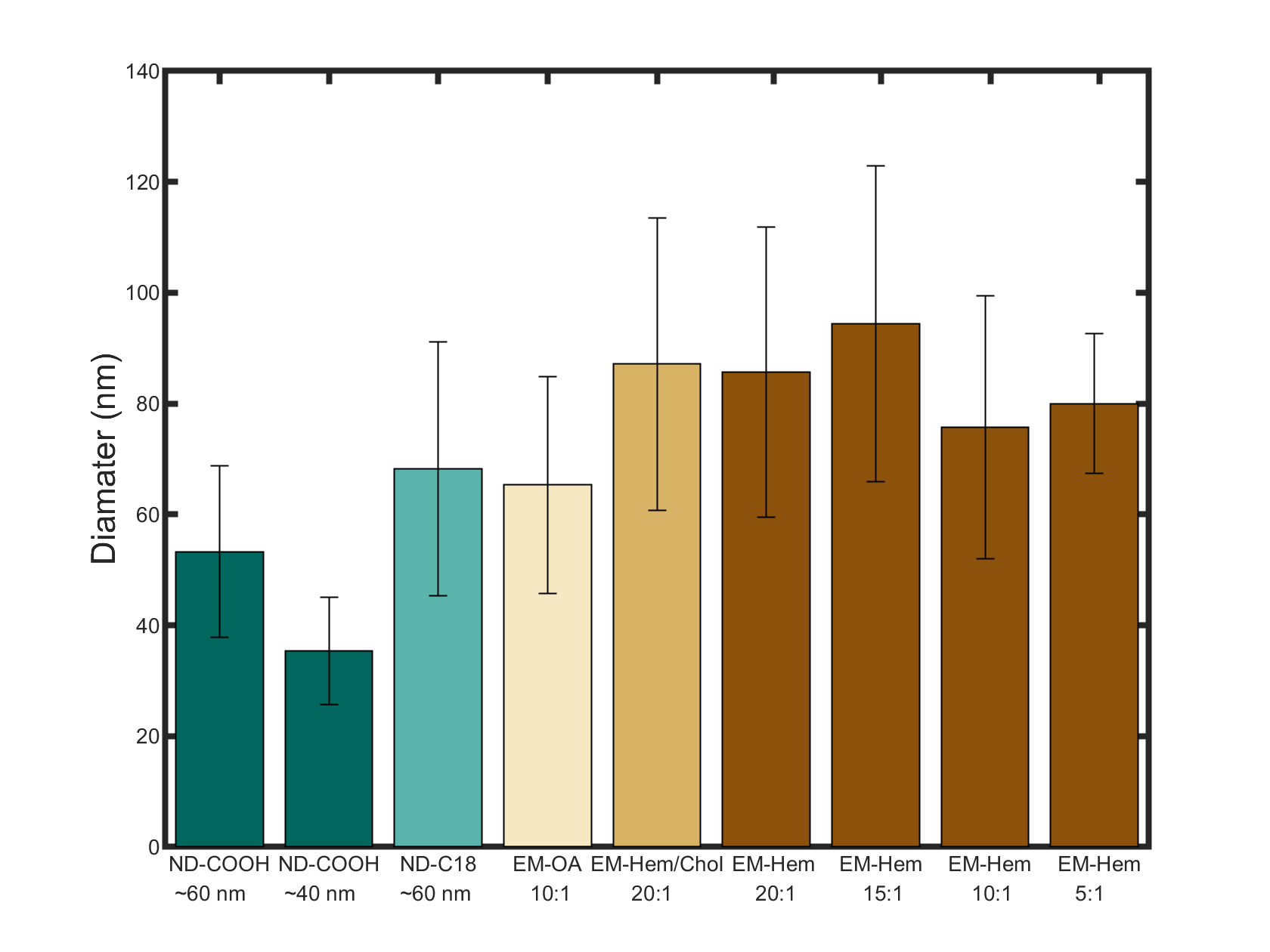}
%\centering
\caption[DLS Bar Plot For Emulsions]{Dynamic light scattering measurements (DLS) of the mean diameter for nanodiamond and emulsion samples. 
The top row of the X-axis label indicates the material type and the bottom row details the advertised milled nanodiamond diameter, for ND-COOH and ND-C18 commercial samples, or the weight ratio of the amphiphilic molecules hemin, oleic acid, or hemin-cholesteryl-TEG azide to the ND-C18 nanodiamonds used in synthesizing the emulsions. 
Error bars represent the standard deviation of the DLS size distribution.}
\label{fig:DLS Bar}
 \end{figure}}

 \def\NDEPLSpectraFigure{\begin{figure}[t]
\renewcommand\figurename{Figure}
\centering
\includegraphics[width = \textwidth ]{Supp_NDE_PL_V4.png}
%\centering
\caption[Nanodiamond and Emulsion PL Emission Spectra]{%Photoluminescence emission spectra for  60nm carboxylate 
Photoluminescence emission spectra for nanodiamond and emulsion samples. 
The top row of the labels indicates the material type and the bottom row details the advertised milled nanodiamond diameter, for ND-COOH and ND-C18 commercial samples, or the weight ratio of the amphiphilic molecules hemin, oleic acid or hemin-cholesteryl-TEG azide to the nanodiamonds. 
}
%nanodiamonds (COOH60) 40nm carboxylate nanodiamonds (COOH40),  60nm octadecane (C18) nanodiamonds, oleic acid nanodiamond emulsion (OA), nanodiamond-hemin-cholesteryl teg azide  emulsion at a 20 to 1 hemin and cholesteryl to nanodiamond ratio (HC20:1) by weight, nanodiamond-hemin emulsion at a 20 to 1 (H20:1), 15 to 1 (H15:1) 10 to 1 (H10:1) and 5 to 1 (H5:1) hemin  to nanodiamond ratio by weight emulsions. }
\label{fig:All PLSpectra}

 \end{figure}}

\def\NDEPLBar{\begin{figure}[t]
\renewcommand\figurename{Figure}
%\centering
\includegraphics[width = \textwidth ]{Supp_NDE_PL_Bar_V2.png}
%\centering
\caption[Nanodiamond and Emulsion  Integrated PL Intensity]{Integrated photoluminescent intensity  for  60nm carboxylate nanodiamonds (COOH60) 40nm carboxylate nanodiamonds (COOH40),  60nm octadecane (C18) nanodiamonds, oleic acid nanodiamond emulsion (OA), nanodiamond-hemin-cholesteryl-TEG azide  emulsion at a 20 to 1 hemin and cholesteryl to nanodiamond ratio by weight (HC20:1), nanodiamond-hemin emulsion at a 20 to 1 (H20:1), 15 to 1 (H15:1) 10 to 1 (H10:1) and 5 to 1 (H5:1) hemin to nanodiamond ratio by weight emulsions.}
\label{fig:Supp Intensity Bar}
 \end{figure}}

\def\NDEFracNV{\begin{figure}[t]
\renewcommand\figurename{Figure}
\centering
\includegraphics[width = 0.8\textwidth ]{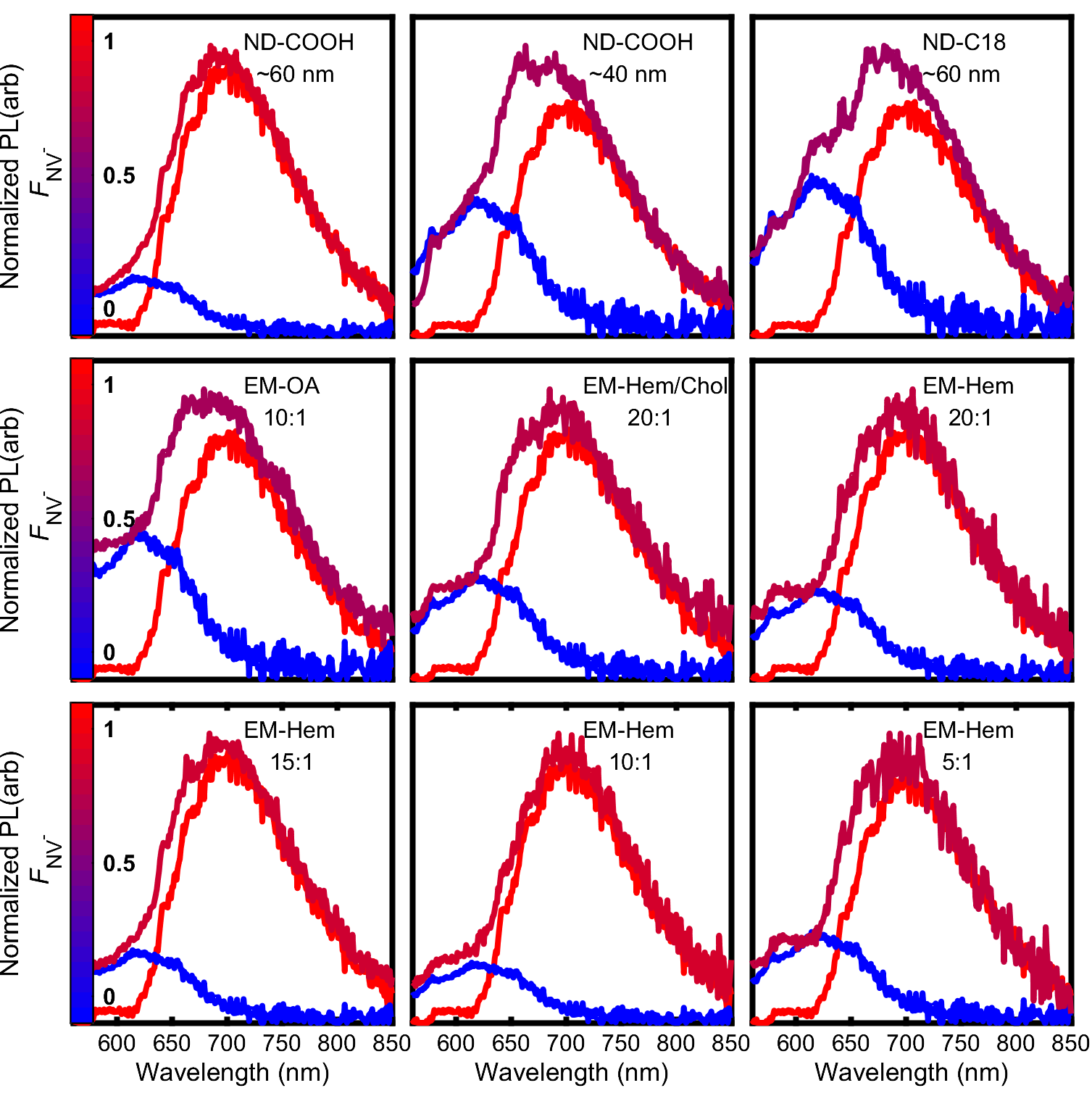}
%\centering
\caption[Nanodiamond and Emulsion NNMF Spectra]
{Photoluminescence spectra of nanodiamond samples and the nonnegative matrix factorization into $\mathrm{NV^{-}}$ (red) and $\mathrm{NV^{0}}$ (blue) charge state spectra, with the color of the measured spectra indicating the charge ratio ($F_\mathrm{NV^{-}}$) for commercial and emulsion nanodiamond samples. 
The top row of the labels indicates the material type and the bottom row details the advertised milled nanodiamond diameter, for ND-COOH and ND-C18 commercial samples, or the weight ratio of the amphiphilic molecules hemin, oleic acid, and hemin-cholesteryl-TEG azide to the nanodiamonds used in synthesizing the emulsions.}
%The top row of labels indicates the material type and the bottom row either the advertised milled nanodiamond diameter or the weight ratio of amphiphilic molecules to nanodiamond.  

%60nm carboxylate nanodiamonds (COOH60) 40nm carboxylate nanodiamonds (COOH40),  60nm octadecane (C18) nanodiamonds, oleic acid nanodiamond emulsion (OA), nanodiamond-hemin-cholesteryl teg azide  emulsion at a 20 to 1 Hemin and cholesteryl to nanodiamond ratio by weight (HC20:1), nanodiamond-hemin emulsion at a 20 to 1 (H20:1), 15 to 1 (H15:1) 10 to 1 (H10:1) and 5 to 1 (H5:1) hemin to nanodiamond ratio by weight emulsions.  
\label{fig:NDE FracNV}
 \end{figure}}

\def\NDEFracNVBar{\begin{figure}[t]
\renewcommand\figurename{Figure}
\centering
\includegraphics[width = \textwidth ]{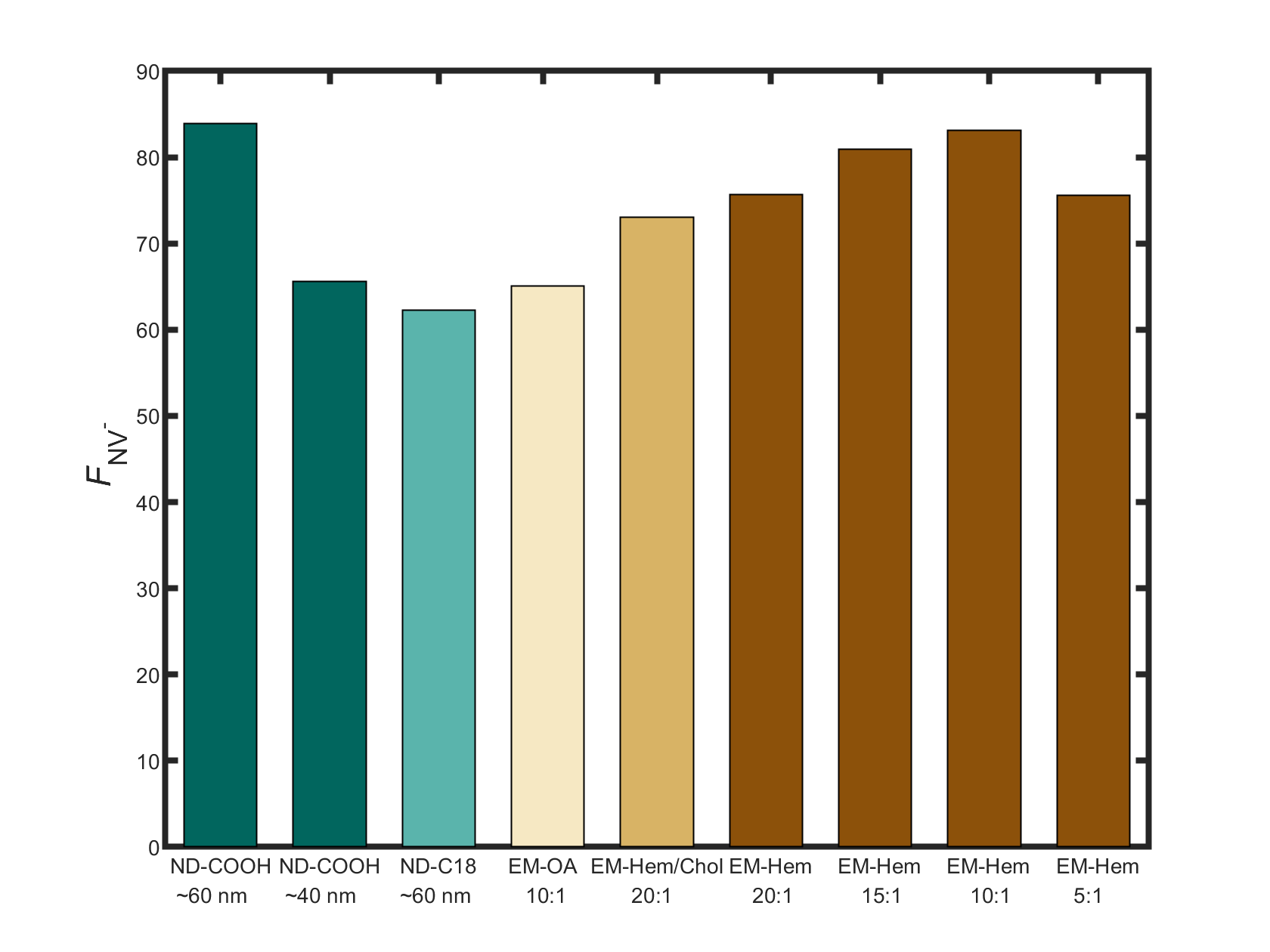}
%\centering
\caption[Nanodiamond and Emulsion $F_\mathrm{NV^{-}}$ ]{Charge ratio ($F_\mathrm{NV^{-}}$)  for commercial and emulsion nanodiamond samples. The top row of the X-axis label indicates the material type and the bottom row details the advertised milled nanodiamond diameter, for ND-COOH and ND-C18 commercial samples, or the weight ratio of the amphiphilic molecules hemin, oleic acid, or hemin-cholesteryl-TEG azide to the ND-C18 nanodiamonds used in synthesizing the emulsions.
%The top row of X-axis label indicates the material type and the bottom row either the advertised milled nanodiamond diameter or the weight ratio of amphiphilic molecules to nanodiamond.  
%60nm carboxylate nanodiamonds (COOH60) 40nm carboxylate nanodiamonds (COOH40),  60nm octadecane (C18) nanodiamonds, oleic acid nanodiamond emulsion (OA), nanodiamond-hemin-cholesteryl teg azide  emulsion at a 20 to 1 hemin and cholesteryl to nanodiamond ratio by weight (HC20:1), nanodiamond-hemin emulsion at a 20 to 1 (H20:1), 15 to 1 (H15:1) 10 to 1 (H10:1) and 5 to 1 hemin to nanodiamond ratio by weight emulsions. 
}
\label{fig:NDEFracNVBar}
 \end{figure}}

\def\NDETSBar{\begin{figure}[t]
\renewcommand\figurename{Figure}
\centering
\includegraphics[width = \textwidth ]{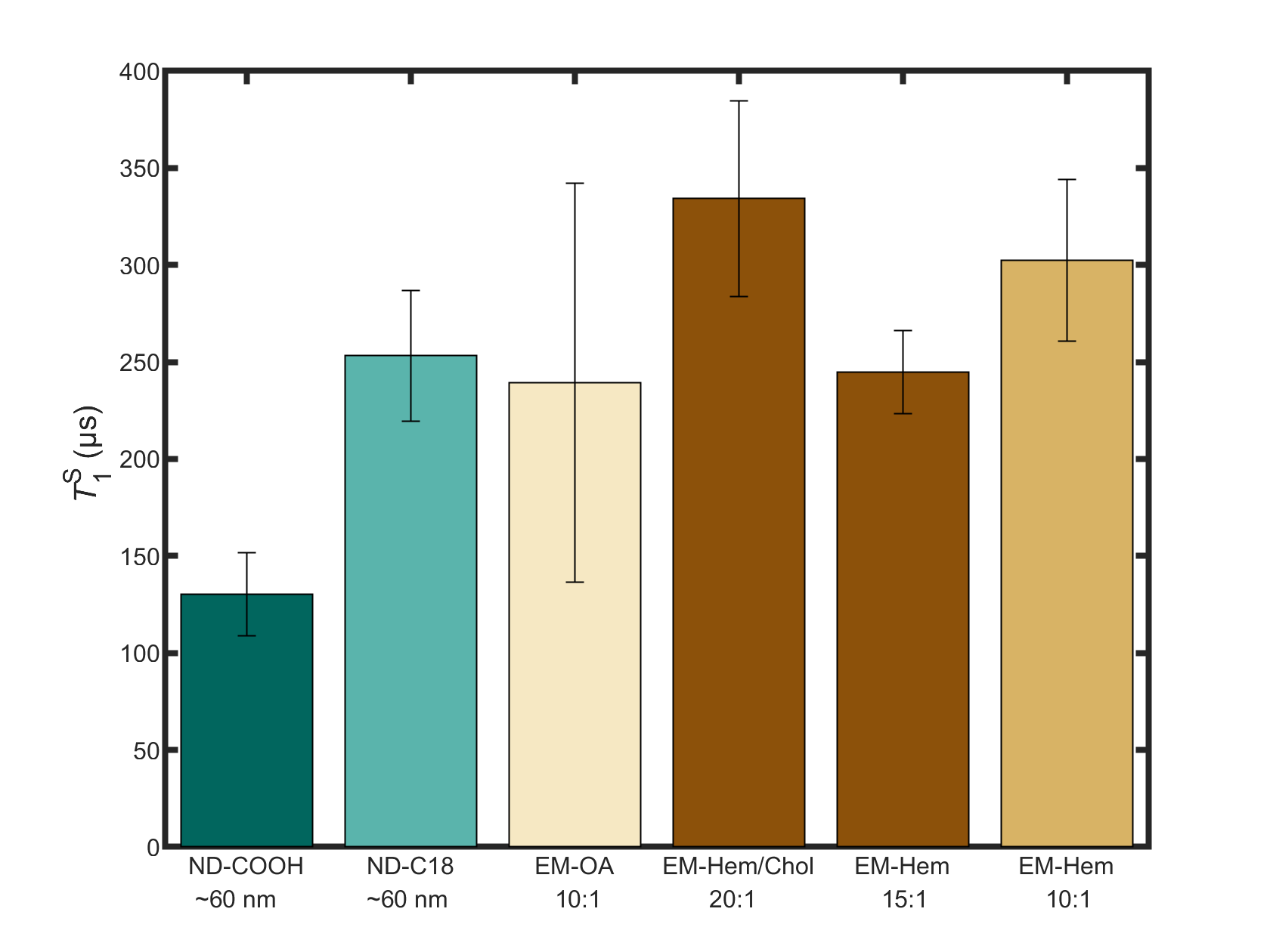}
%\centering
\caption[Nanodiamond and Emulsion ${T}_1^S$ ]{Stretched-exponential spin lifetime (${T}_1^S$) for commercial and emulsion nanodiamond samples. 
The top row of the X-axis label indicates the material type and the bottom row details the advertised milled nanodiamond diameter, for ND-COOH and ND-C18 commercial samples, or the weight ratio of the amphiphilic molecules hemin, oleic acid, or hemin-cholesteryl-TEG azide to the nanodiamonds used in synthesizing the emulsions. 
%The top row of X-axis labels indicates the material type and the bottom row either the advertised milled nanodiamond diameter or the weight ratio of amphiphilic molecules to nanodiamond.  
%60nm carboxylate nanodiamonds (COOH60), 60nm octadecane (C18) nanodiamonds, oleic acid nanodiamond emulsion (OA), Nanodiamond-hemin emulsion at a 15 to 1 (H15:1)  and 10 to 1 (H10:1) hemin to nanodiamond ratio by weight, nanodiamond-hemin-cholesteryl teg azide  emulsion at a 20 to 1 hemin (HC20:1) and cholesteryl to nanodiamond ratio by weight.
 Error bars represent the standard deviation of ${T}_1^S$ from measurements of multiple samples (for ND-COOG, ND-C18, EM-Hem 15:1, and EM-Hem 10:1) or the fit uncertainty from a single measurement (for EM-OA and EM-Hem/Chol).}
\label{fig:NDET1S}
 \end{figure}}

\def\NDEConBar{\begin{figure}[t]
\renewcommand\figurename{Figure}
\centering
\includegraphics[width = \textwidth ]{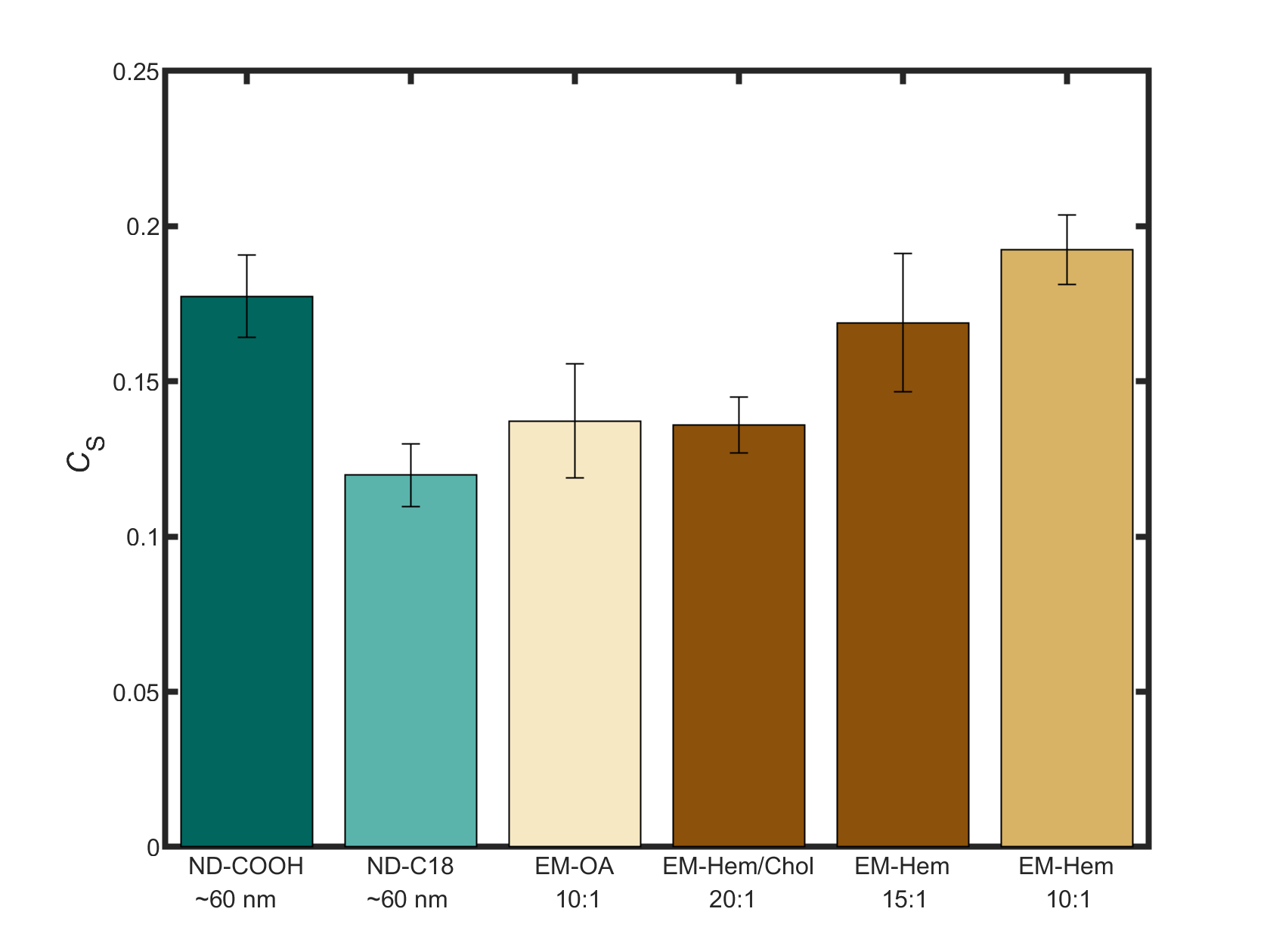}
%\centering
\caption[Nanodiamond and Emulsion T$_1$ PL Contrast]{Photoluminescence contrast (${C}_S$) from stretched-exponential spin lifetime fits for commercial and emulsion nanodiamond samples. 
The top row of the X-axis label indicates the material type and the bottom row details the advertised milled nanodiamond diameter, for ND-COOH and ND-C18 commercial samples, or the weight ratio of the amphiphilic molecules hemin, oleic acid, or hemin-cholesteryl-TEG azide to the nanodiamonds used in synthesizing the emulsions. 
%60nm carboxylate nanodiamonds (COOH60), 60nm octadecane (C18) nanodiamonds, oleic acid nanodiamond emulsion (OA), Nanodiamond-hemin emulsion at a 15 to 1 (H15:1)  and 10 to 1 (H10:1) hemin to nanodiamond ratio by weight, nanodiamond-hemin-cholesteryl teg azide  emulsion at a 20 to 1 hemin (HC20:1) and cholesteryl to nanodiamond ratio by weight.
 Error bars represent the standard deviation of ${C}_S$ from measurements of multiple samples (for ND-COOH, ND-C18, EM-Hem 15:1, and EM-Hem 10:1) or the fit uncertainty from a single measurement (for EM-OA and EM-Hem/Chol).}
\label{fig:NDECon Bar}
 \end{figure}}

\def\NDEBetaBar{\begin{figure}[t]
\renewcommand\figurename{Figure}
%\centering
\includegraphics[width = \textwidth ]{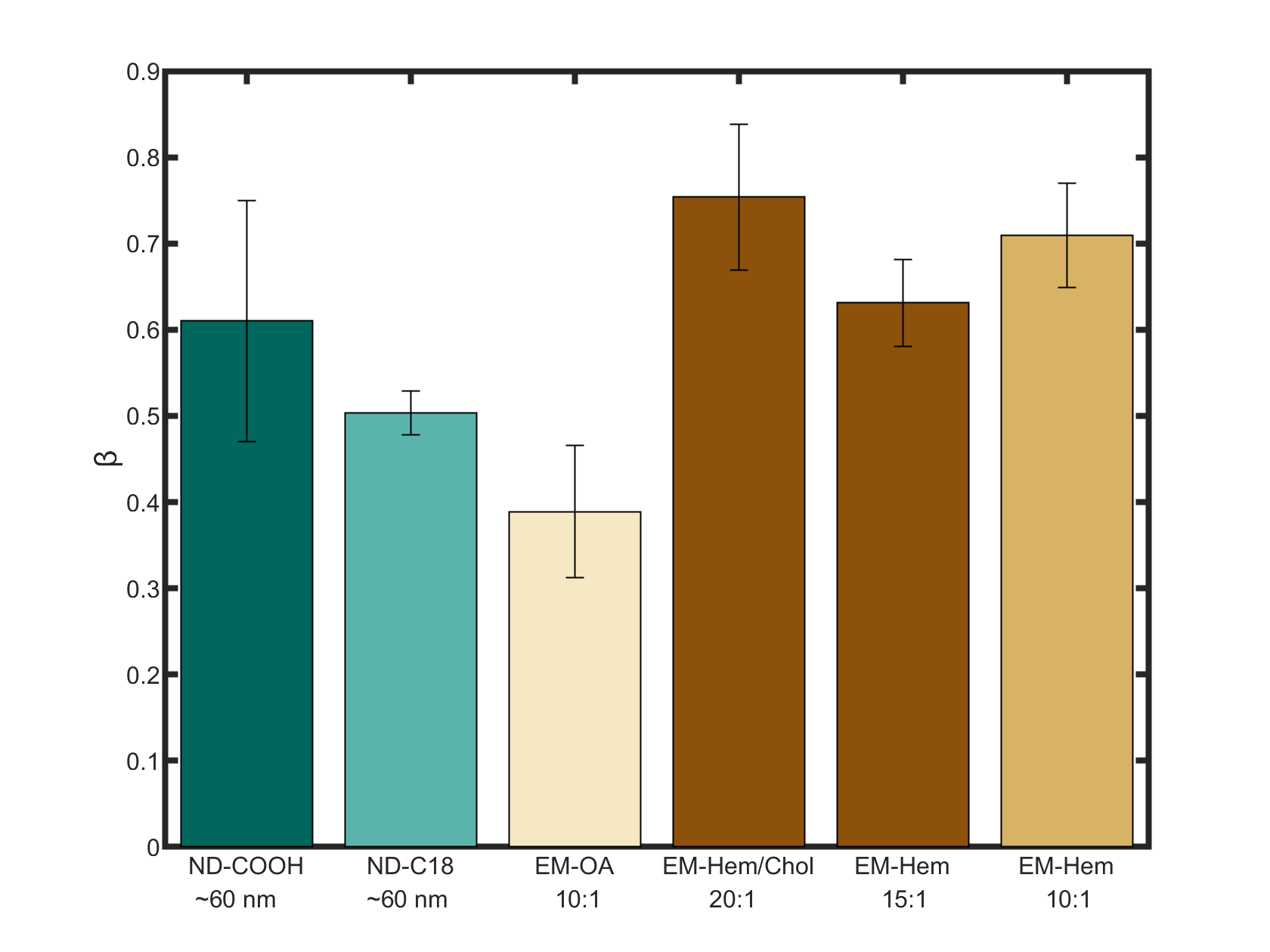}
%\centering
\caption[Nanodiamond and Emulsion T$_1$ $\beta$ Factor]{$\beta$ factor from stretched-exponential spin lifetime fits for  commercial and emulsion nanodiamond samples. 
The top row of the X-axis label indicates the material type and the bottom row details the advertised milled nanodiamond diameter, for ND-COOH and ND-C18 commercial samples, or the weight ratio of the amphiphilic molecules hemin, oleic acid, or hemin-cholesteryl-TEG azide to the nanodiamonds used in synthesizing the emulsions. 
%60nm carboxylate nanodiamonds (COOH60), 60nm octadecane (C18) nanodiamonds, oleic acid nanodiamond emulsion (OA), Nanodiamond-hemin emulsion at a 15 to 1 (H15:1)  and 10 to 1 (H10:1) hemin to nanodiamond ratio by weight, nanodiamond-hemin-cholesteryl teg azide  emulsion at a 20 to 1 hemin (HC20:1) and cholesteryl to nanodiamond ratio by weight. 
 Error bars represent the standard deviation of $\beta$ from measurements of multiple samples (for ND-COOH, ND-C18, EM-Hem 15:1, and EM-Hem 10:1) or the fit uncertainty from a single measurement (for EM-OA and EM-Hem/Chol).}
\label{fig:NDEBeta Bar}

 \end{figure}}

\def\NDETABar{\begin{figure}[t]
\renewcommand\figurename{Figure}
\centering
\includegraphics[width = \textwidth ]{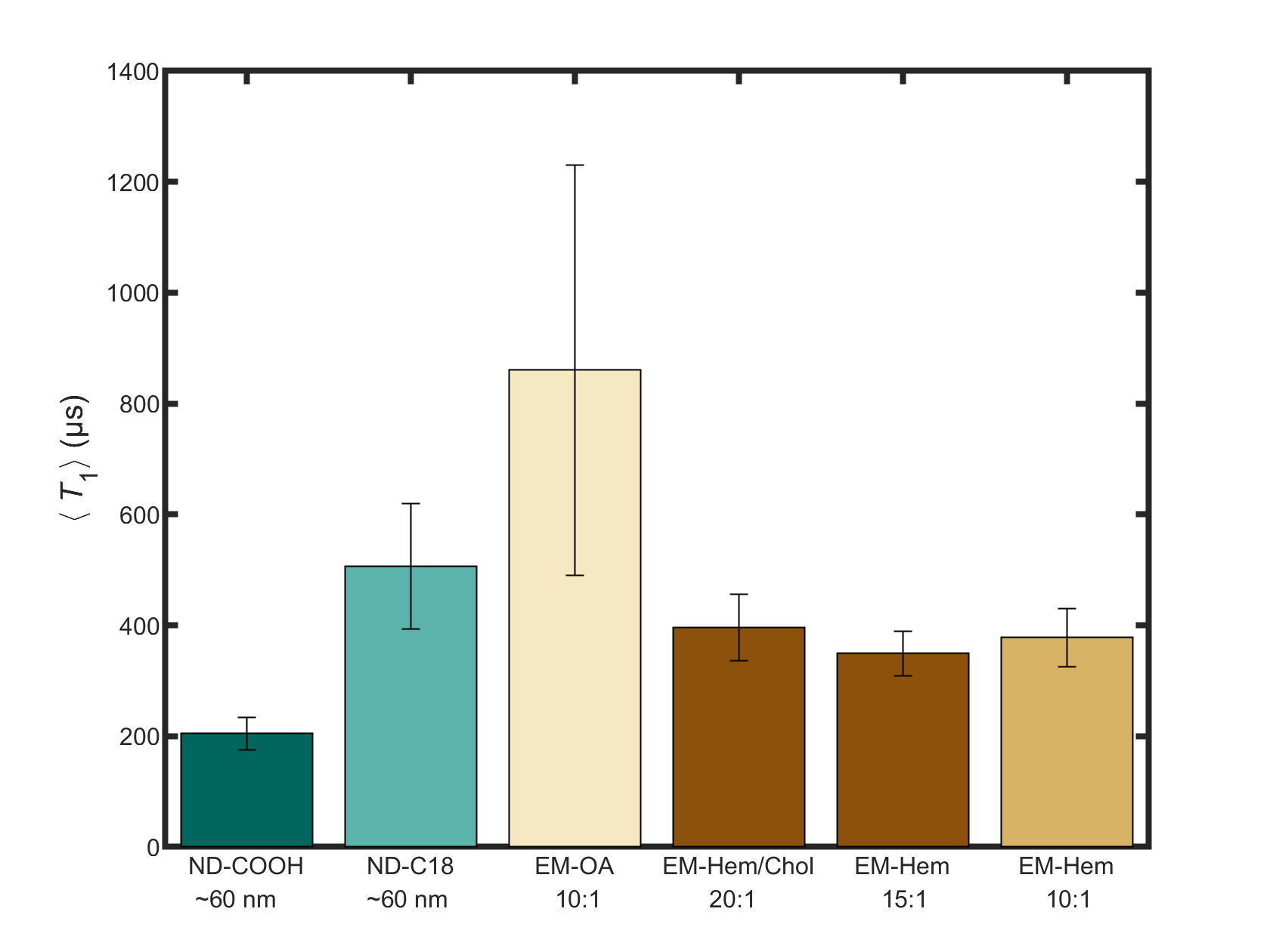}
%\centering
\caption[Nanodiamond and Emulsion ($\langle{T}_1\rangle$]{Average spin lifetime ($\langle{T}_1\rangle$) for 
commercial and emulsion nanodiamond samples. 
The top row of the X-axis label indicates the material type and the bottom row details the advertised milled nanodiamond diameter, for ND-COOH and ND-C18 commercial samples, or the weight ratio of the amphiphilic molecules hemin, oleic acid, or hemin-cholesteryl-TEG azide to the nanodiamonds used in synthesizing the emulsions. 
 % 60nm carboxylate nanodiamonds (COOH60), 60nm octadecane (C18) nanodiamonds, oleic acid nanodiamond emulusion (OA), Nanodiamond-hemin emululsion at a 15 to 1 (H15:1)  and 10 to 1 (H10:1) hemin to nanodiamond ratio by weight, nanodiamond-hemin-cholesteryl teg azide  emulsion at a 20 to 1 Hemin (HC20:1) and cholesteryl to nanodiamond ratio by weight. 
  Error bars represent the standard deviation of $\langle{T}_1\rangle$ from measurements of multiple samples (for ND-COOH, ND-C18, EM-Hem 15:1, and EM-Hem 10:1) or the fit uncertainty from a single measurement (for EM-OA and EM-Hem/Chol).}
\label{fig:NDET1A}
 \end{figure}}

\def\GDVCon{\begin{figure}[t]
\renewcommand\figurename{Figure}
\centering
\includegraphics[width = \textwidth ]{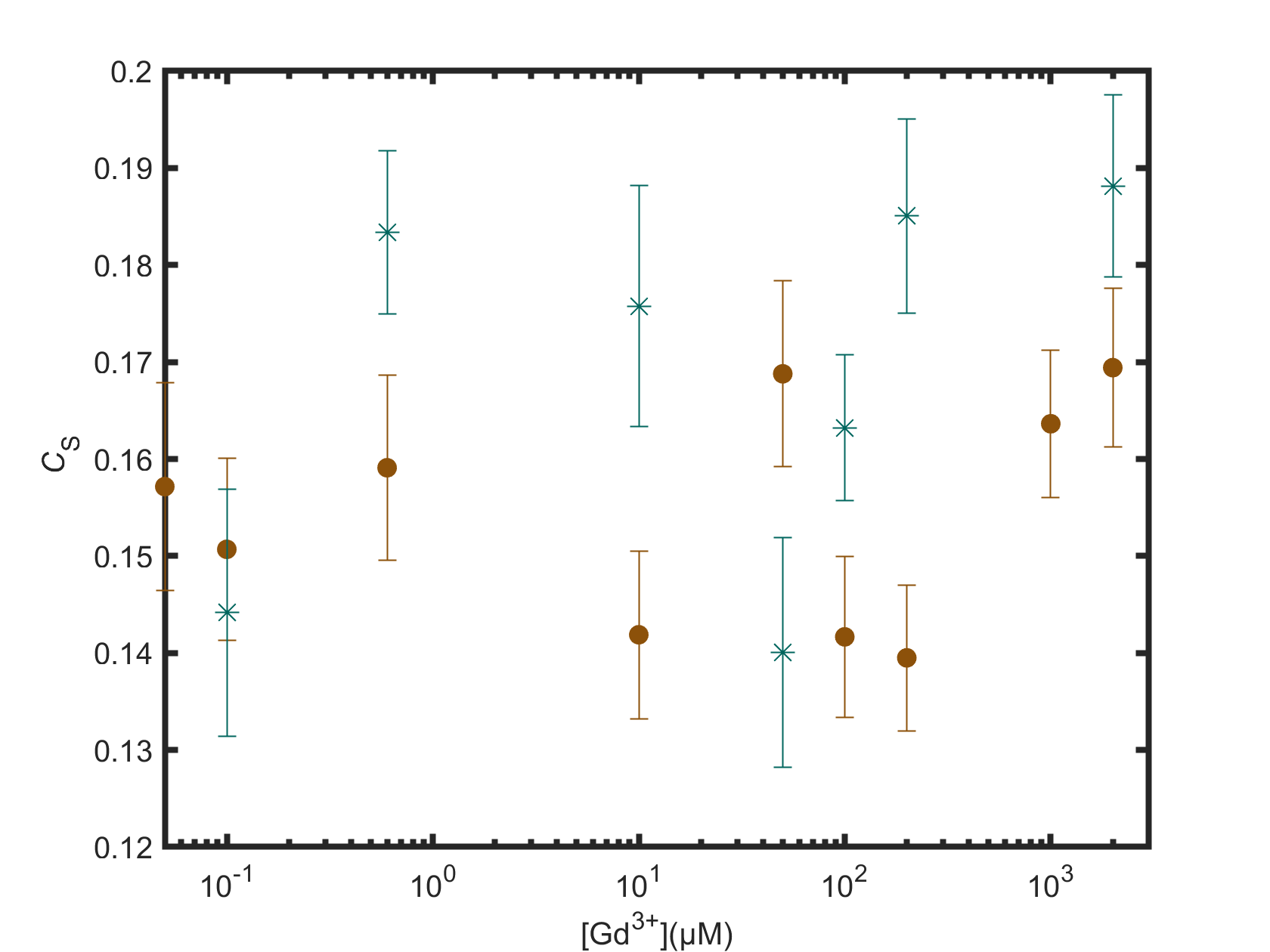}
%\centering
\caption[PL Contrast \textit{vs} Gd$^{3+}$]{Photoluminescence contrast ($C_S$) \textit{vs} [Gd$^{3+}$] for different concentrations of the Gd$^{3+}$ chelate added to aqueous dispersions of ND-COOH (green stars) and EM-Hem (brown circles) nanodiamonds. 
Error bars represent 68\% confidence intervals from fit.}
\label{fig:GDVCon}
 \end{figure}}

 \def\GDVBeta{\begin{figure}[t]
\renewcommand\figurename{Figure}
\centering
\includegraphics[width = \textwidth ]{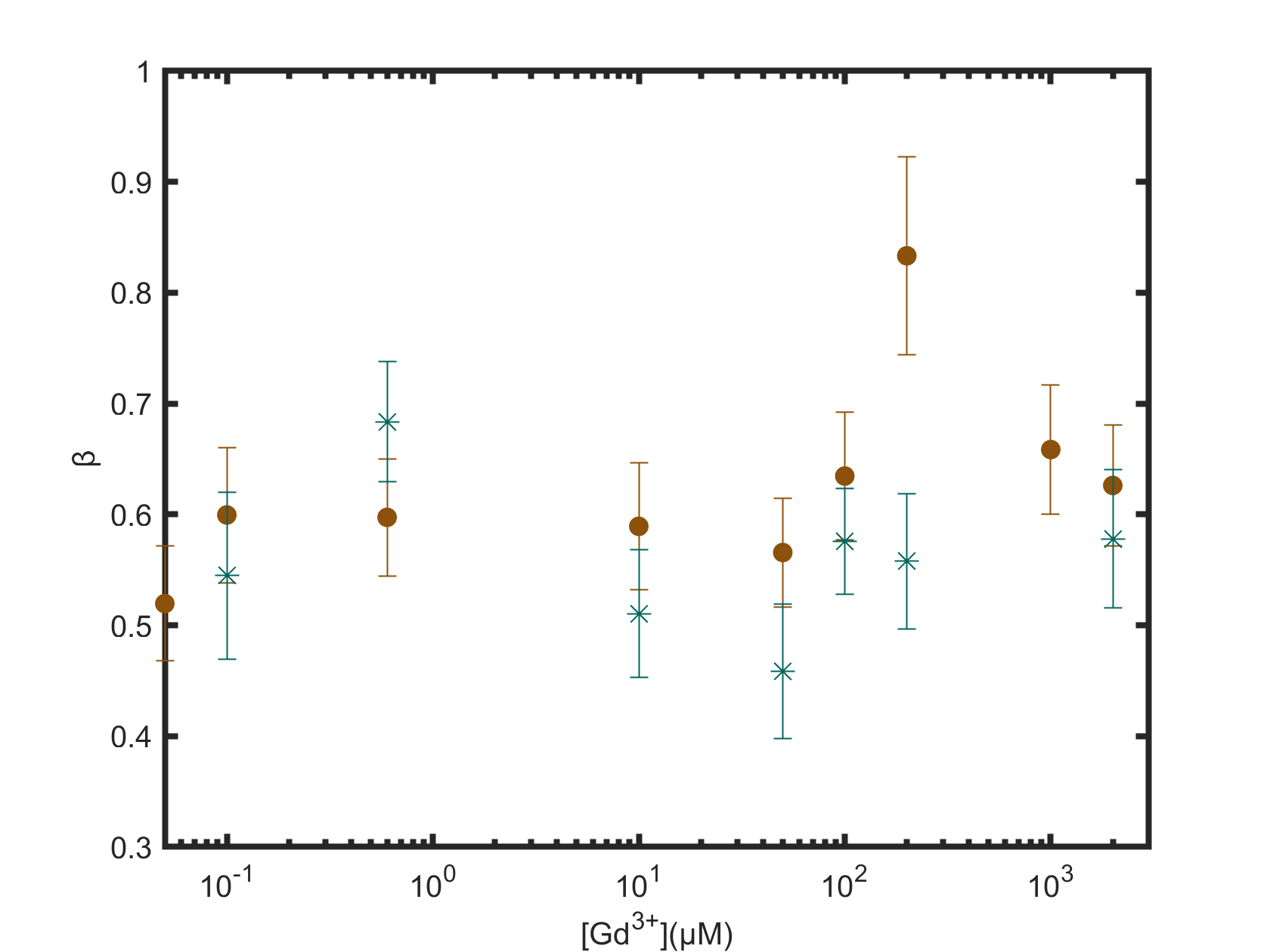}
%\centering
\caption[ $\beta$  \textit{vs} Gd$^{+3}$]{$\beta$  \textit{vs} [Gd$^{+3}$] for different concentrations of the Gd$^{3+}$ chelate added to aqueous dispersions of ND-COOH (green stars) and EM-Hem (brown circles) nanodiamonds. 
 Error bars represent 68\% confidence intervals from fit.}
\label{fig:GDVBeta}
 \end{figure}}

\def\GDVTA{\begin{figure}[t]
\renewcommand\figurename{Figure}
\centering
\includegraphics[width = \textwidth ]{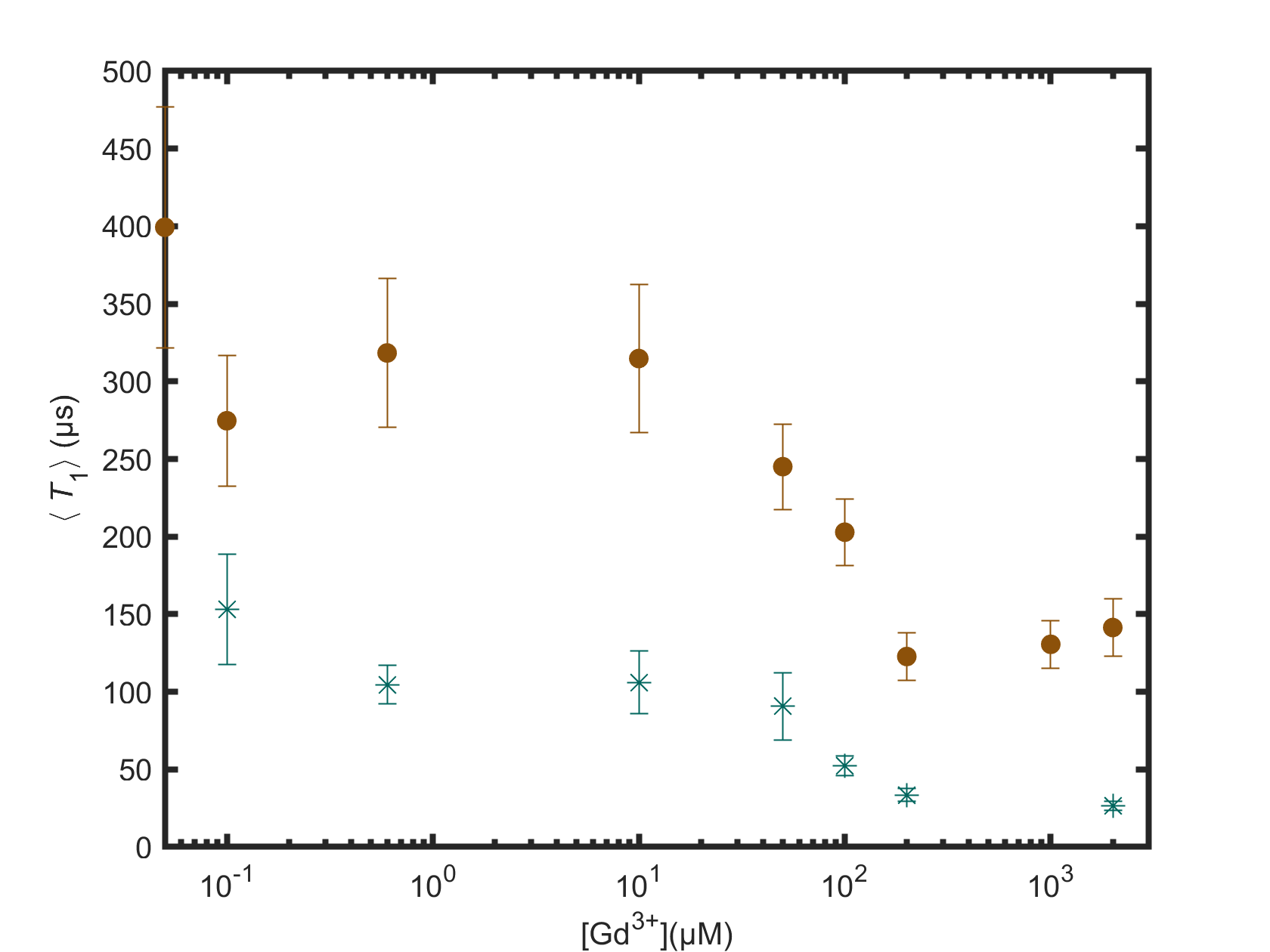}
%\centering
\caption[$\langle{T}_1\rangle$ \textit{vs}  Gd$^{+3}$]{ 
$\langle{T}_1\rangle$ \textit{vs}  [Gd$^{+3}$] for different concentrations of the Gd$^{3+}$ chelate added to aqueous dispersions of ND-COOH (green stars) and EM-Hem (brown circles) nanodiamonds. Error bars represent 68\% confidence intervals from fit.}
\label{fig:GDVT1A}
 \end{figure}}

 \def\EDCHEPES{\begin{figure}[t]
\renewcommand\figurename{Figure}
\centering
\includegraphics[width = \textwidth ]{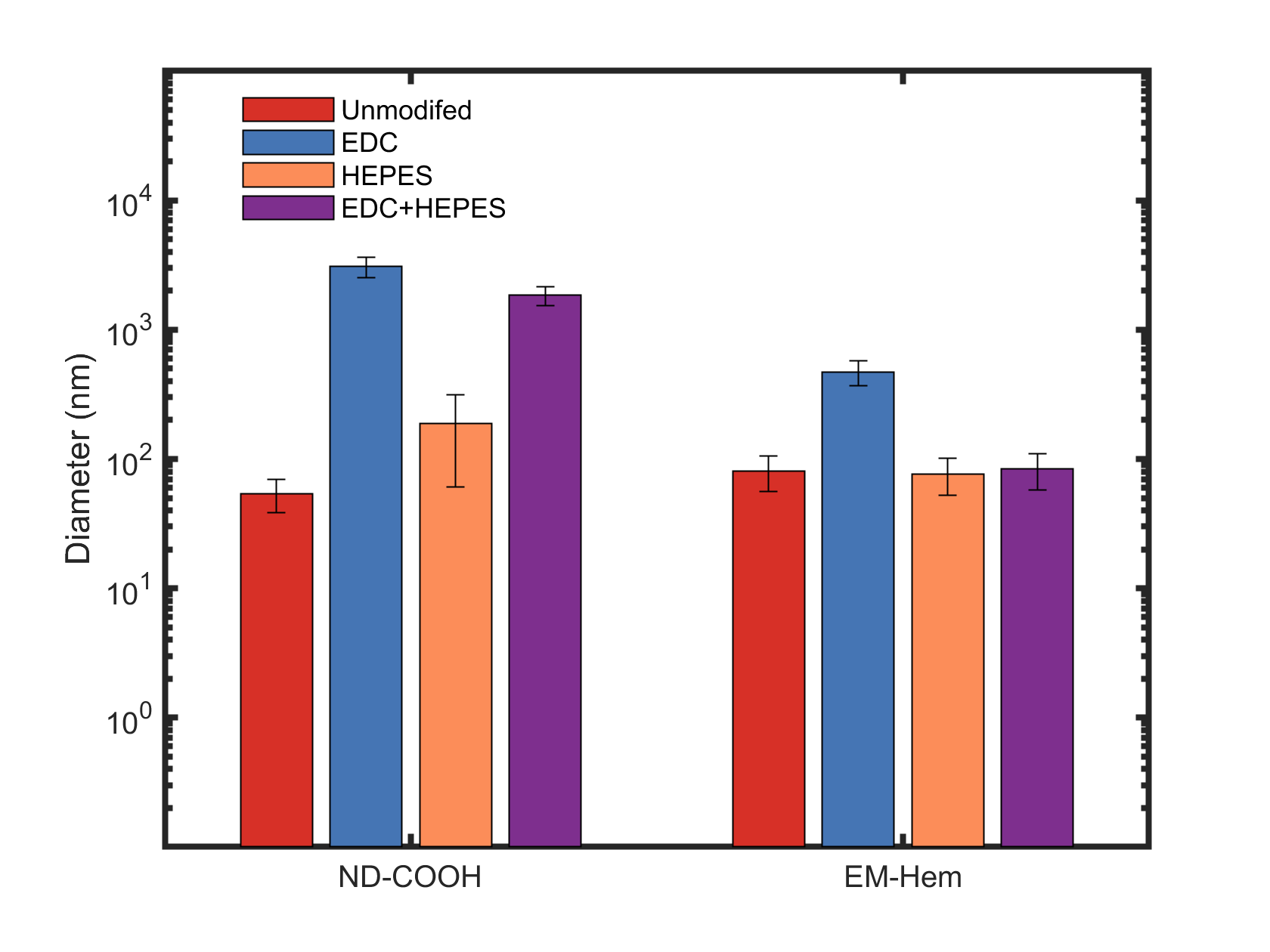}
%\centering
\caption[Agglomeration due to Crosslinking Chemicals ]{Average particle diameters from DLS measurements of samples of ND-COOH and EM-Hem unmodified (red), and upon addition of the various carbodiimide crosslinking chemicals: with EDC (blue), with HEPES buffer (orange), and with EDC and HEPES buffer (purple). Error bars represent one standard deviation of the size distribution.}
\label{fig:EDCHEPESDLS}
 \end{figure}}

\def\EDCDyeUVS{\begin{figure}[t]
\renewcommand\figurename{Figure}
\centering
\includegraphics[width = \textwidth ]{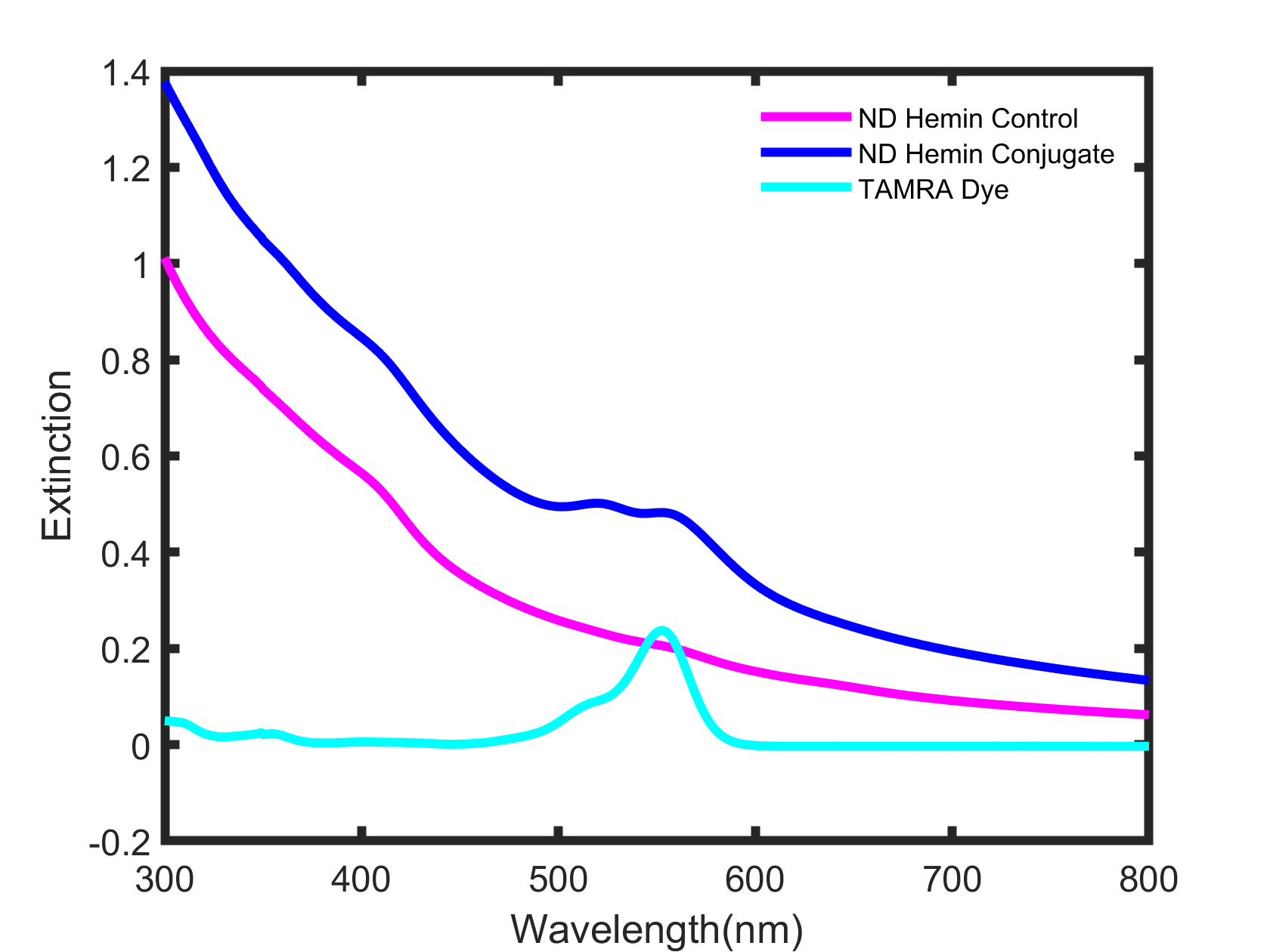}
%\centering
\caption[Dye Crosslinker Conjugation Absorption 1]{UV-Vis extinction spectra of ND-Hemin emulsions conjugated to the TAMRA dye (cyan, TAMRA dye), with EDC (blue, ND Hemin Conjugate) without EDC (magenta, ND Hemin Control) after dialysis.  }
\label{fig:EDC UVVIS S}
 \end{figure}}

 \def\EDCDyeUVSMod{\begin{figure}[t]
\renewcommand\figurename{Figure}
\centering
\includegraphics[width = \textwidth ]{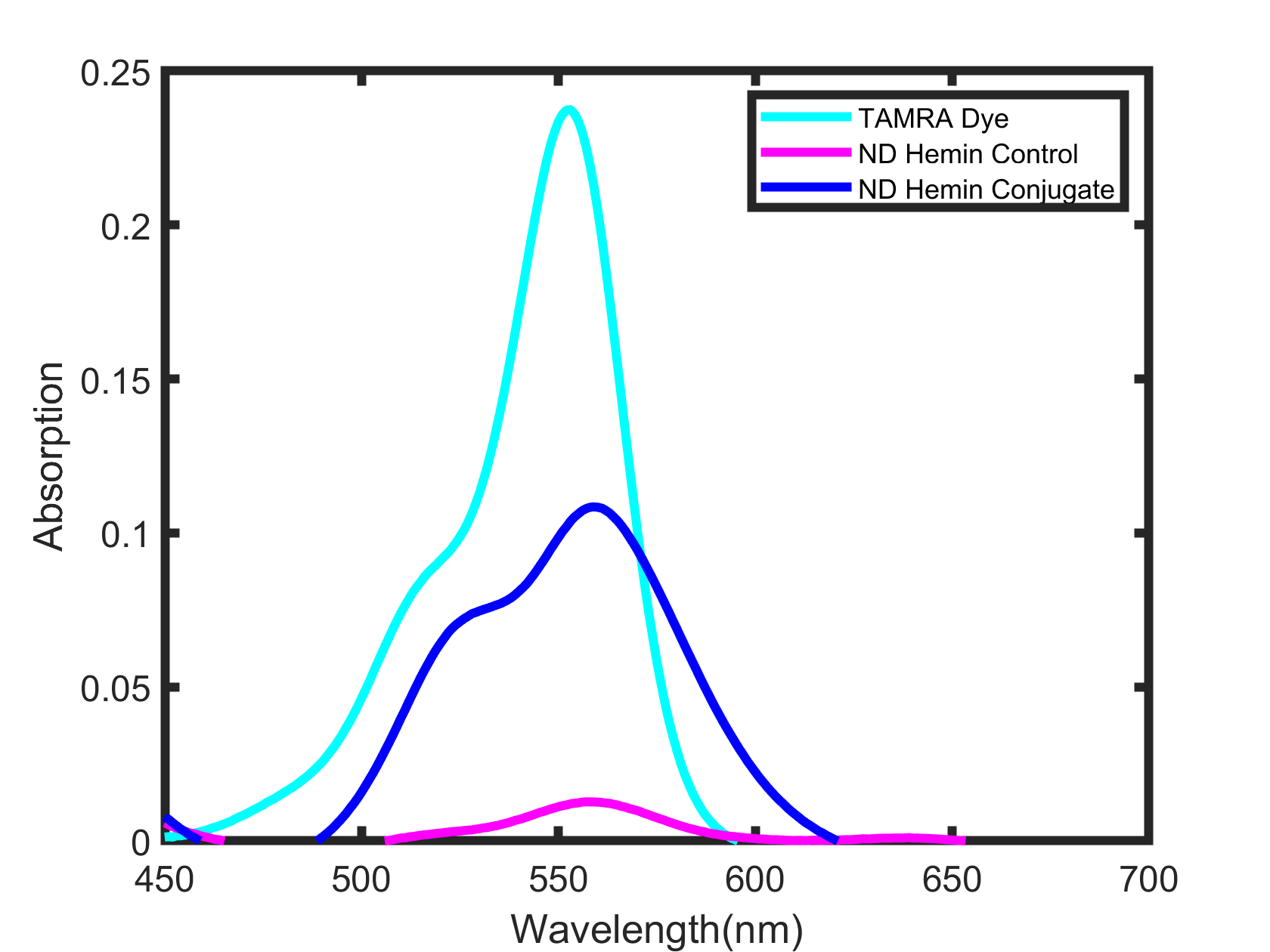
}
%\centering
\caption[Dye Crosslinker Conjugation Absorption 2]{Absorption spectra for the TAMRA dye (teal) in comparison to spectra calculated from the difference between the extinction spectra for the ND-Hemin emulsions conjugated to TAMRA, with EDC (blue, EM-Hem Conjugate) and without EDC (magenta ND Hemin Control) after dialysis.}
\label{fig:EDC UVVIS Mod}
 \end{figure}}

\def\EDCGdDLS{\begin{figure}[t]
\renewcommand\figurename{Figure}
\centering
\includegraphics[width = \textwidth ]{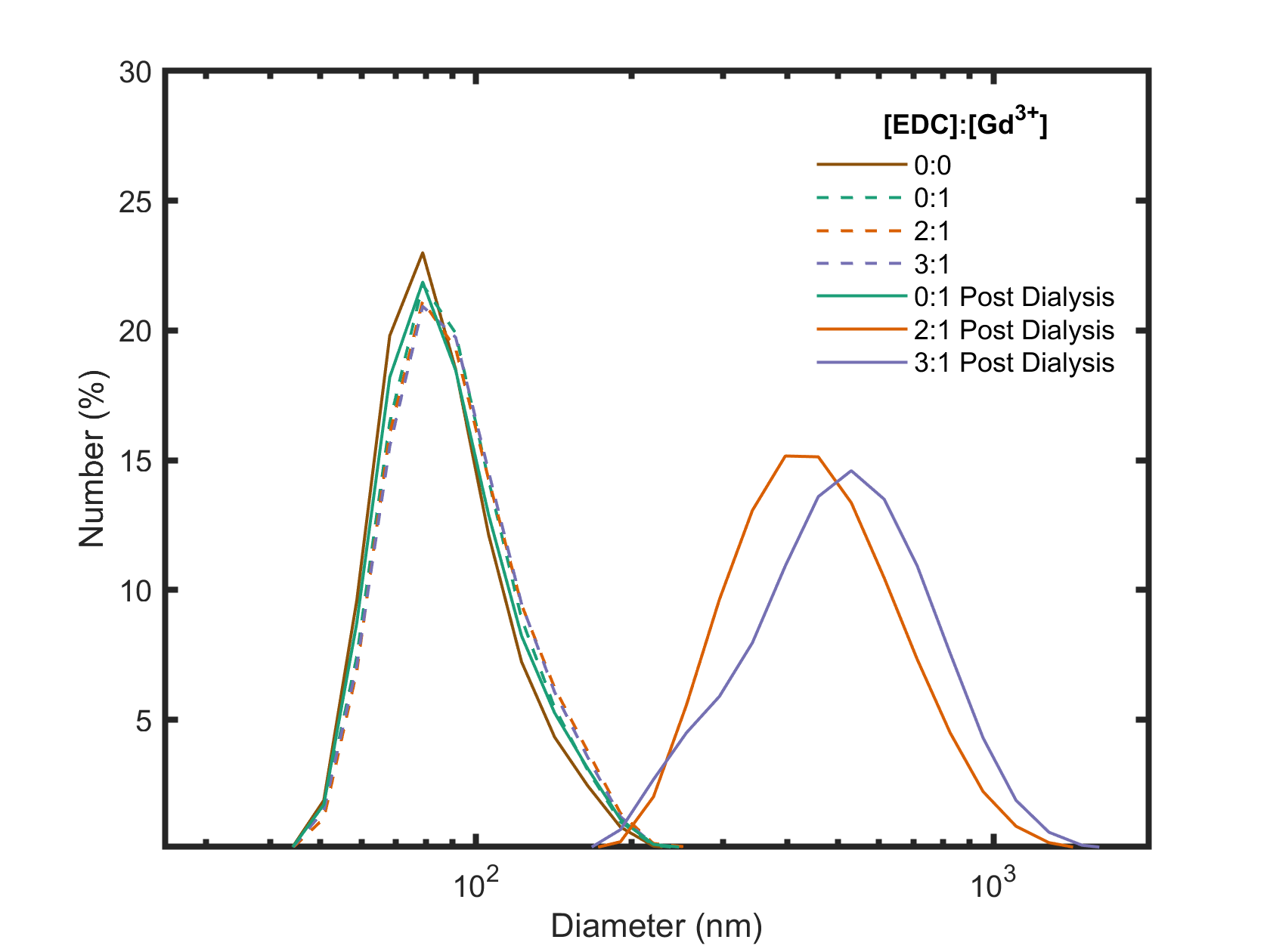}
%\centering
\caption[EDC-Gd$^{3+}$ Conjugation DLS]{DLS measurements of EM-Hem conjugation to the Gd$^{3+}$ chelate using EDC, pre- (dotted) and post- (solid) dialysis for the [EDC]:[Gd$^{3+}$] molar ratios discussed in the main text.}
\label{fig:EDCGDConDLS}
 \end{figure}}

\def\DyeClickUVVis{\begin{figure}[t]
\renewcommand\figurename{Figure}
\centering
\includegraphics[width = \textwidth ]{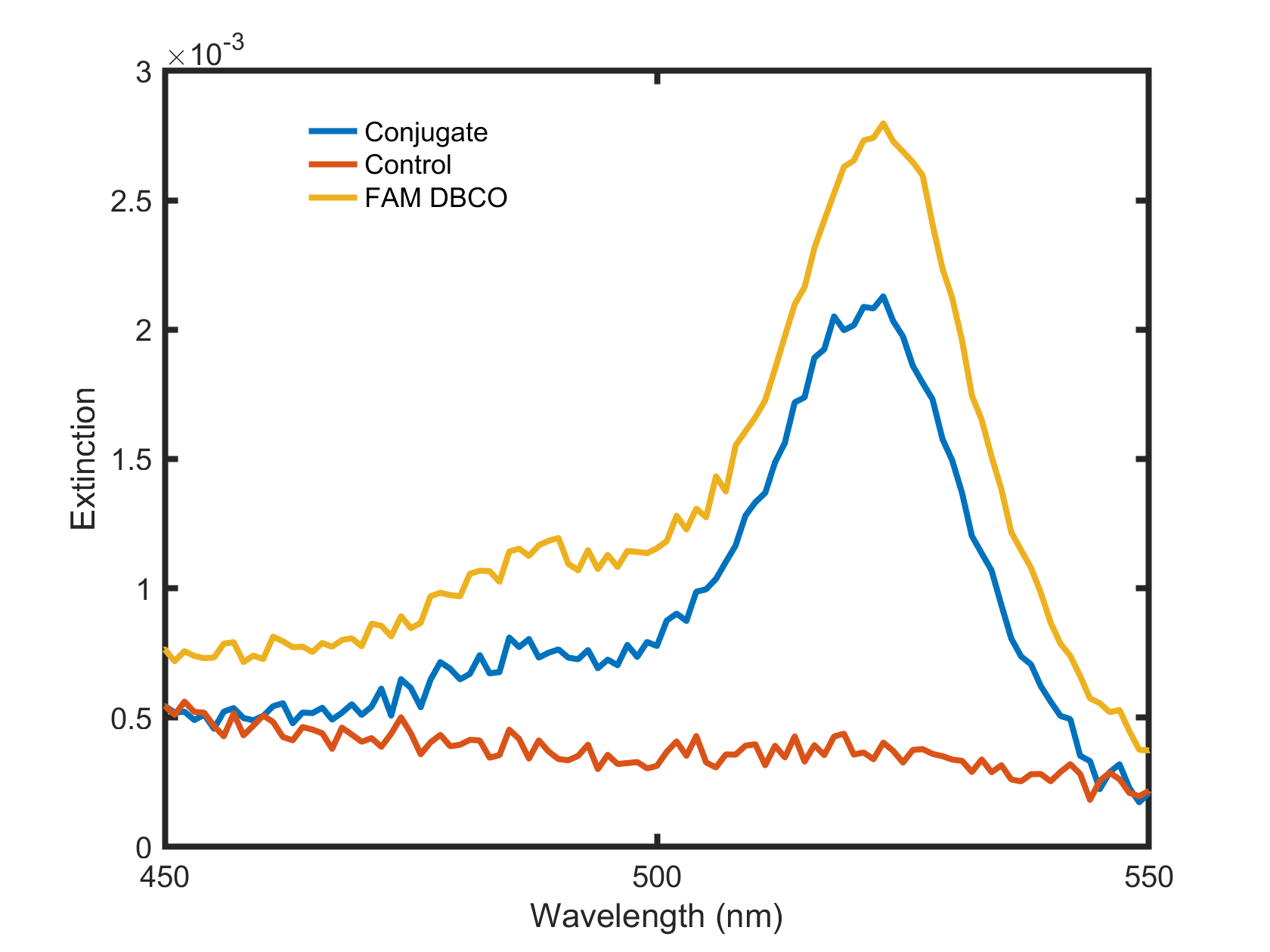}
%\centering
\caption[Dye Click Conjugation Absorption]{UV-Vis absorption data for FAM-DBCO click conjugation:  EM-Hem/Chol  conjugated to FAM DBCO after washing \textit{via} centrifugation (blue, Conjugate), EM-Hem mixed with dye after washing (red, control), and  FAM-DBCO at a1X  EM-Hem/Chol  azide molarity  (yellow, FAM DBCO). }
\label{fig:DyeClick UVVis}
 \end{figure}}

\def\DyeClickEmission{\begin{figure}[t]
\renewcommand\figurename{Figure}
\centering
\includegraphics[width =\textwidth ]{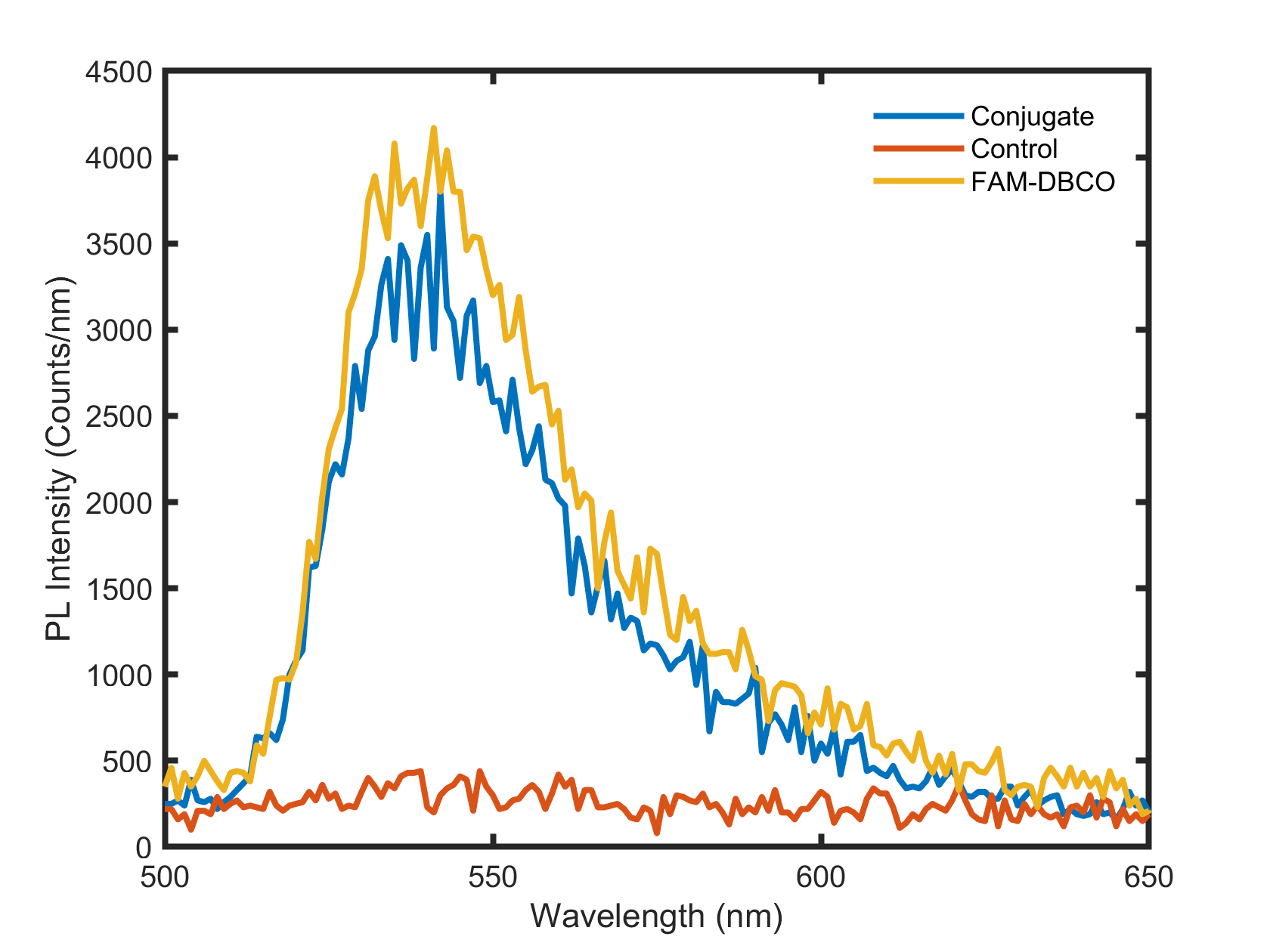}
%\centering
\caption[Dye Click Conjugation Emission]{Photoluminescence (PL) spectra for FAM-DBCO click conjugation: EM-Hem/Chol  conjugated to FAM-DBCO after washing \textit{via } centrifugation  (blue), EM-Hem mixed with FAM-DBCO after washing carried out as a control (red), and FAM-DBCO at a concentration 1x EM-Hem/Chol molarity  (yellow).}
\label{fig:DyeClick Emission}
 \end{figure}}

\def\FullFTIR{\begin{figure}[t]
\renewcommand\figurename{Figure}
\centering
\includegraphics[width = \textwidth ]{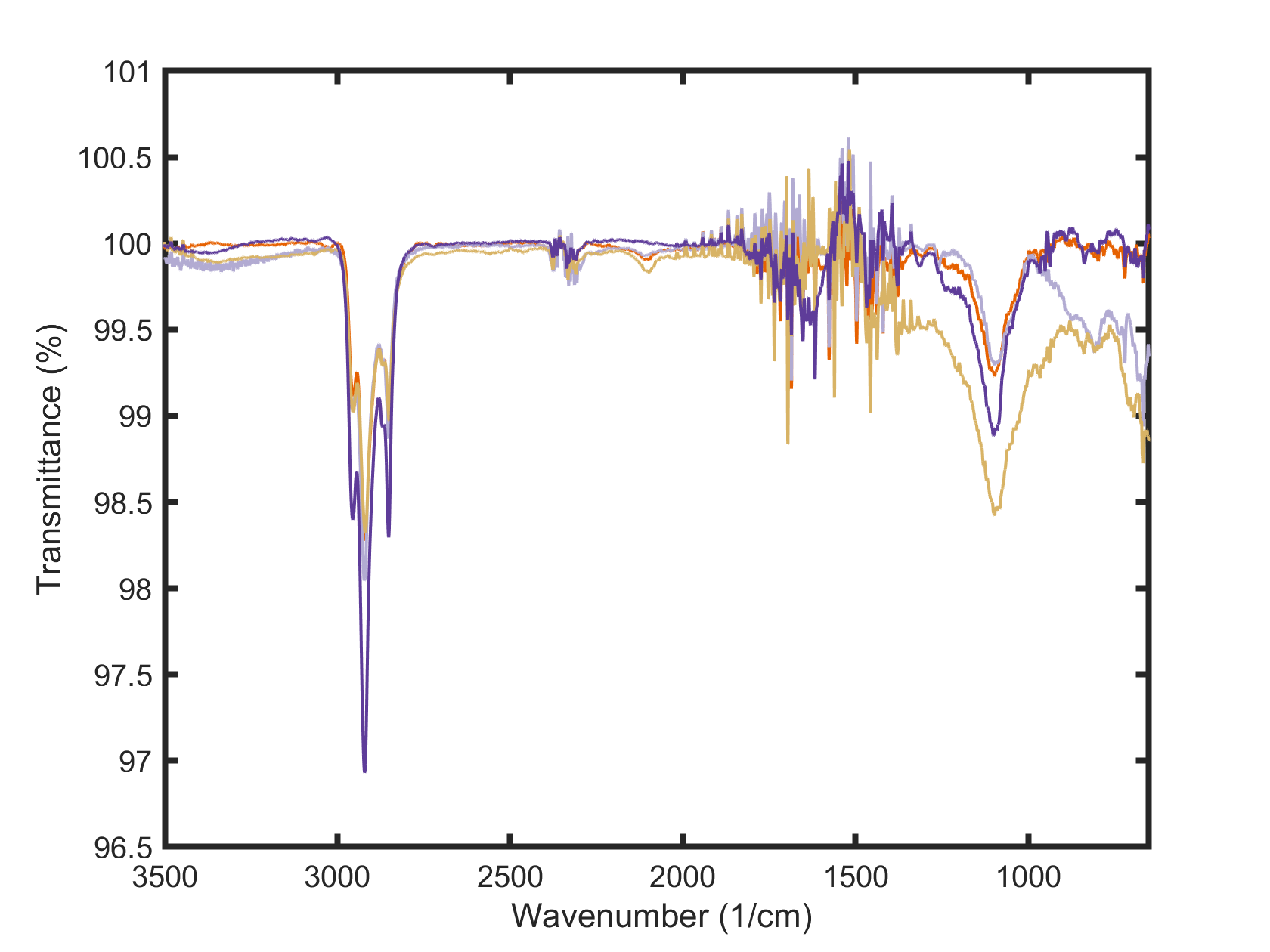}
%\centering
\caption[Full FTIR Curve of  Gd$^{3+}$ Click Conjugation]{Full FTIR spectra for EM-Hem/Chol  dispersions conjuagted to Gd$^{3+}$ 
 dibenzocyclooctyne (DBCO) at 0:1 (yellow), 1:4 (orange), 2:1 (lavender), and 10:1 (purple) Gd$^{3+}$ 
 dibenzocyclooctyne (DBCO) to  cholesteryl-TEG azide ($-N_{3}$) ratios after dialysis.}
\label{fig:Full FTIR}
 \end{figure}}

\def\DBCOCtrlTS{\begin{figure}[t]
\renewcommand\figurename{Figure}
\centering
\includegraphics[width = \textwidth ]{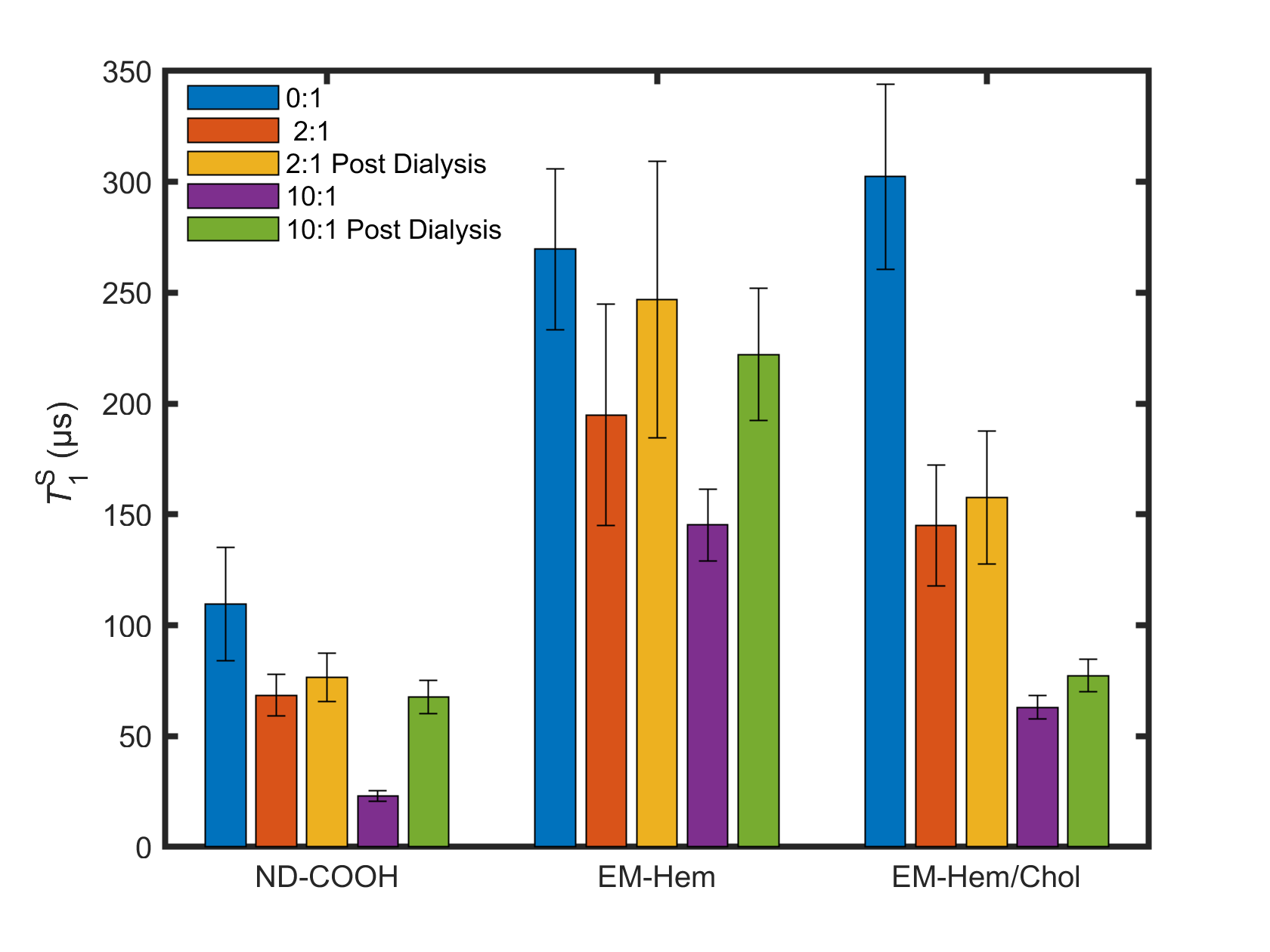}
%\centering
\caption[Gd$^{3+}$ DBCO Click Conjugation Control: ${T}_1^S$]{ ${T}_1^S$ for click chemistry conjugation experiments with NC-COOH, EM-Hem, and EM-Hem/Chol dispersions and the Gd$^{3+}$ 
 dibenzocyclooctyne chelate. For EM-Hem/Chol dispersions, experiments are carried out for dibenzocyclooctyne (DBCO) to azide ($-N_{3}$) molar ratios of 0:1 (blue),  2:1 (before dialysis: orange; after dialysis: yellow), and 10:1 (before dialysis: purple; after dialysis: green). For  ND-COOH and EM-Hem, which do not have surface functional groups with azide termination, the ratios refer to the addition of identical Gd$^{3+}$ concentrations as used for EM-Hem/Chol conjugation reactions. 
Error bars represent the best-fit confidence intervals.}
\label{fig:DBCOCtrlTS}
 \end{figure}}

\def\DBCOCtrlBeta{\begin{figure}[t]
\renewcommand\figurename{Figure}
\centering
\includegraphics[width = \textwidth ]{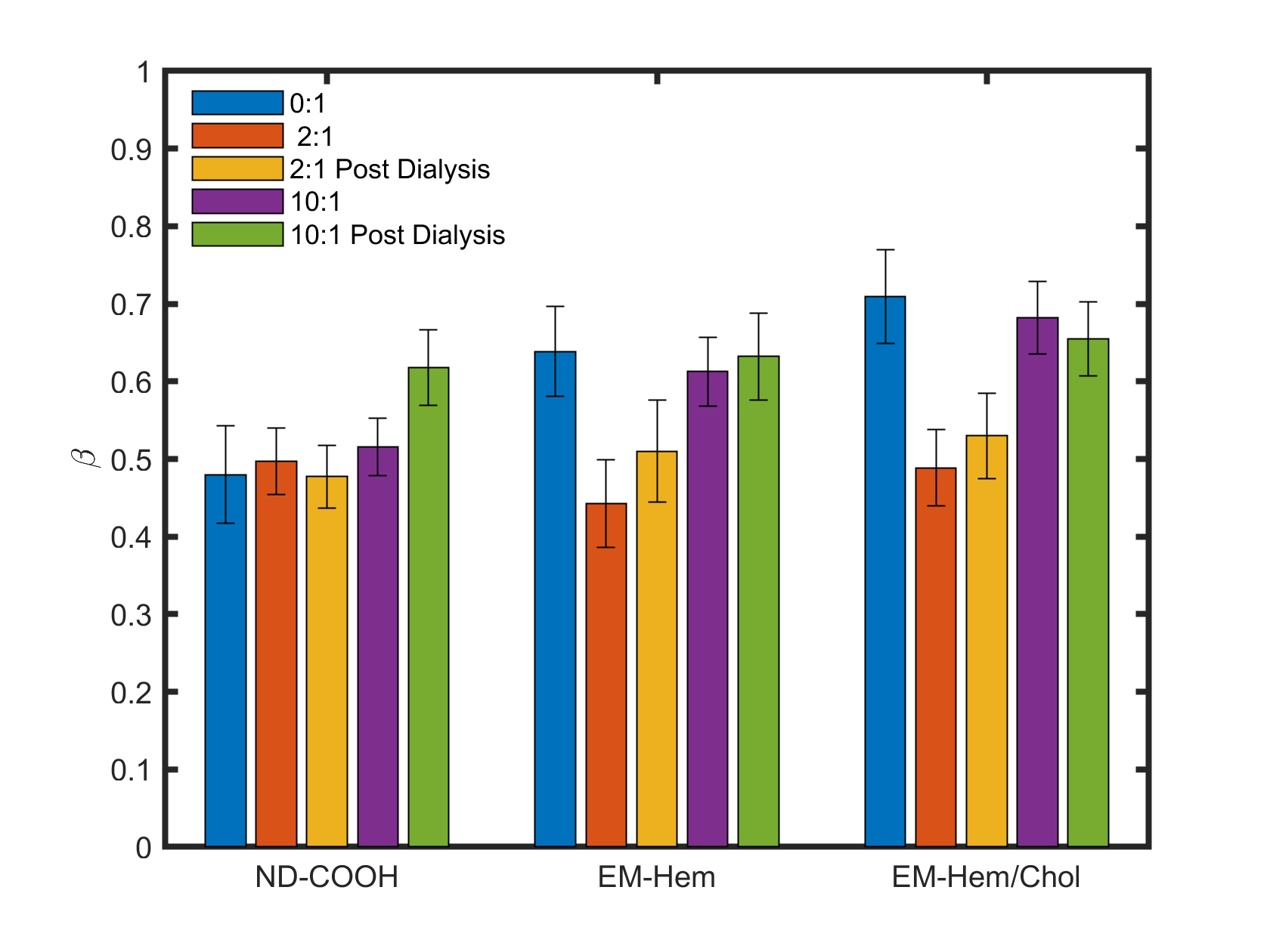}
%\centering
\caption[Gd$^{3+}$ DBCO Click Conjugation Control: $\beta$]{$\beta$ for  click chemistry conjugation experiments with NC-COOH, EM-Hem, and EM-Hem/Chol dispersions and the Gd$^{3+}$ 
 dibenzocyclooctyne chelate. For EM-Hem/Chol dispersions, experiments are carried out for dibenzocyclooctyne (DBCO) to azide ($-N_{3}$) molar ratios of 0:1 (blue),  2:1 (before dialysis: orange; after dialysis: yellow) and 10:1 (before dialysis: purple; after dialysis: green) Dibenzocyclooctyne (DBCO) to azide ($-N_{3}$) molar ratios samples, for  ND-COOH, EM-Hem, and EM-Hem/Chol. ND-COOH and EM-Hem do not have azide termination. Therefore ratios refer to  the addition of identical Gd$^{3+}$ concentrations as EM-Hem/Chol. Error bars represent the best-fit confidence intervals.}
\label{fig:DBCOCtrlBeta}
 \end{figure}}

\def\DBCOCtrlCon{\begin{figure}[t]
\renewcommand\figurename{Figure}
\centering
\includegraphics[width = \textwidth ]{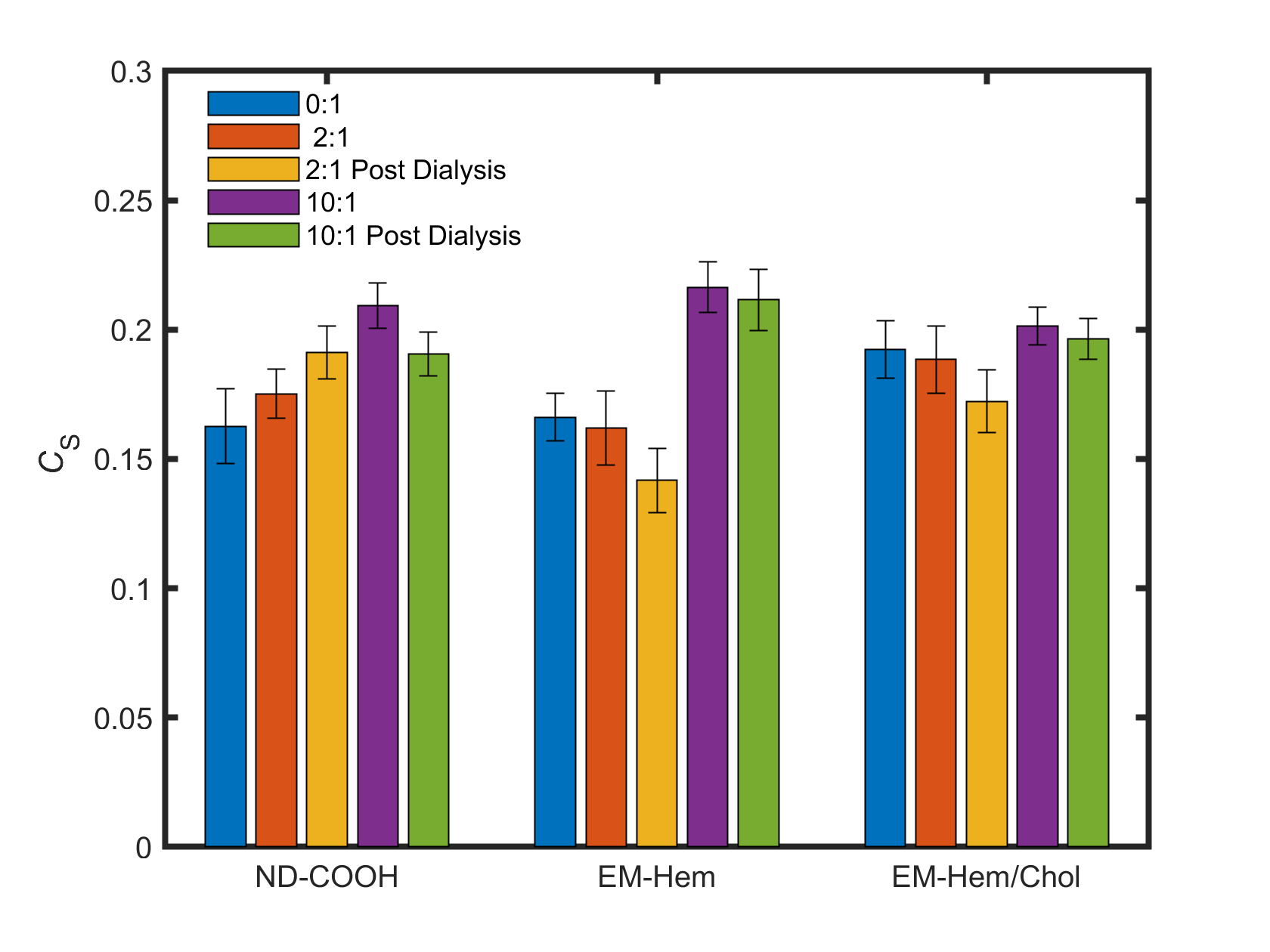}
%\centering
\caption[Gd$^{3+}$ DBCO Click Conjugation Control:PL Contrast]{Photoluminescence contrast ($C_S$) for click chemistry conjugation experiments with NC-COOH, EM-Hem, and EM-Hem/Chol dispersions and the Gd$^{3+}$ 
 dibenzocyclooctyne chelate. For EM-Hem/Chol dispersions, experiments are carried out for dibenzocyclooctyne (DBCO) to azide ($-N_{3}$) molar ratios of 0:1 (blue),  2:1 (before dialysis: orange; after dialysis: yellow), and 10:1 (before dialysis: purple; after dialysis: green). For  ND-COOH and EM-Hem, which do not have surface functional groups with azide termination, the ratios refer to the addition of identical Gd$^{3+}$ concentrations as used for EM-Hem/Chol conjugation reactions. Error bars represent the best-fit confidence intervals. }
\label{fig:DBCOCtrlCon}
 \end{figure}}

\def\DBCOCtrlTA{\begin{figure}[t]
\renewcommand\figurename{Figure}
\centering
\includegraphics[width = \textwidth ]{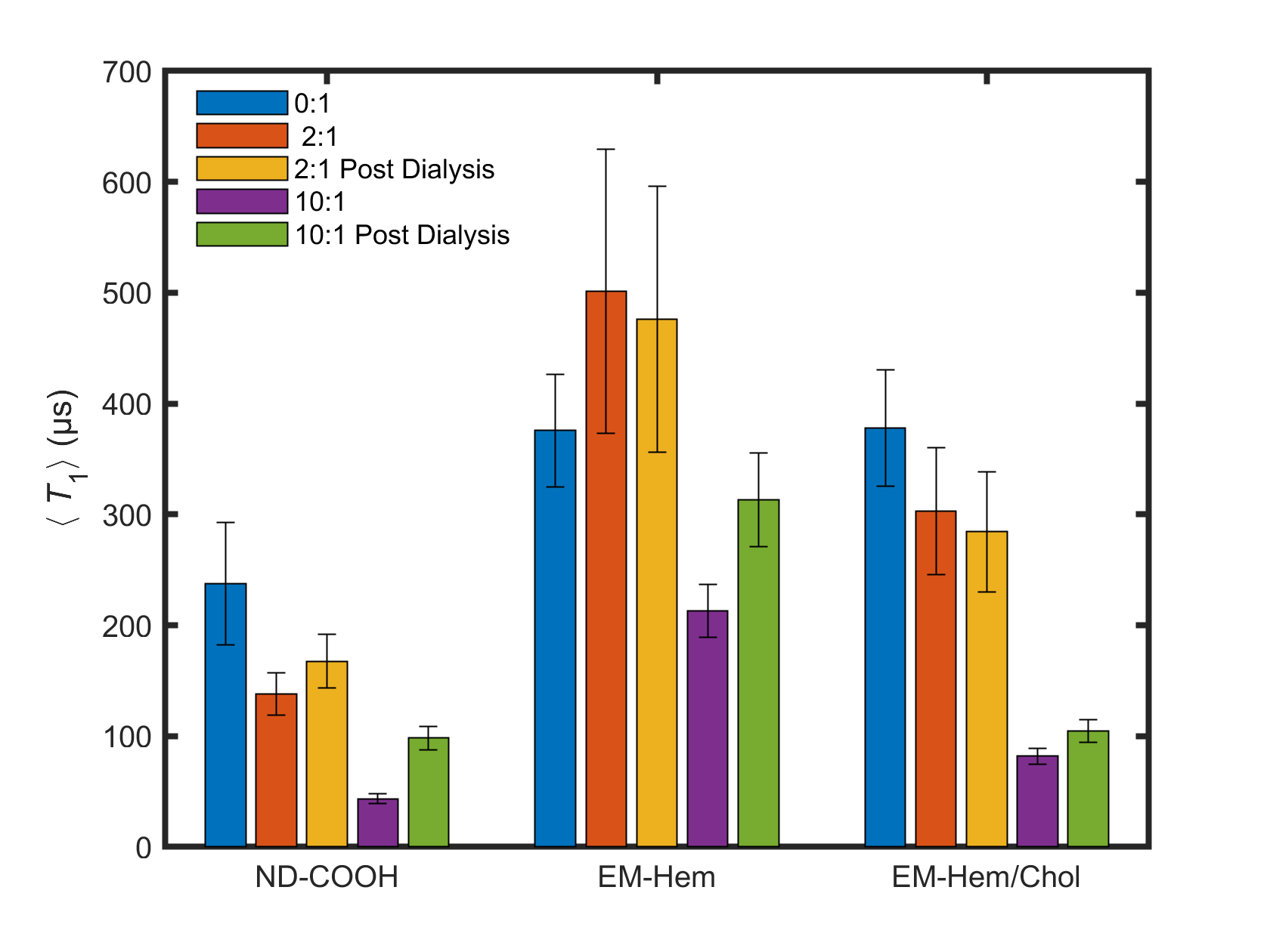}
%\centering
\caption[Gd$^{3+}$ DBCO Click Conjugation Control: $\langle{T}_1\rangle$]{$\langle{T}_1\rangle$ for click chemistry conjugation experiments with NC-COOH, EM-Hem, and EM-Hem/Chol dispersions and the Gd$^{3+}$ 
 dibenzocyclooctyne chelate. For EM-Hem/Chol dispersions, experiments are carried out for dibenzocyclooctyne (DBCO) to azide ($-N_{3}$) molar ratios of 0:1 (blue),  2:1 (before dialysis: orange; after dialysis: yellow), and 10:1 (before dialysis: purple; after dialysis: green). For  ND-COOH and EM-Hem, which do not have surface functional groups with azide termination, the ratios refer to the addition of identical Gd$^{3+}$ concentrations as used for EM-Hem/Chol conjugation reactions.  Error bars represent the best-fit confidence intervals. }
\label{fig:T1A GD Con}
 \end{figure}}

\def\DBCOCtrlICP{\begin{figure}[t]
\renewcommand\figurename{Figure}
\centering
\includegraphics[width = \textwidth ]{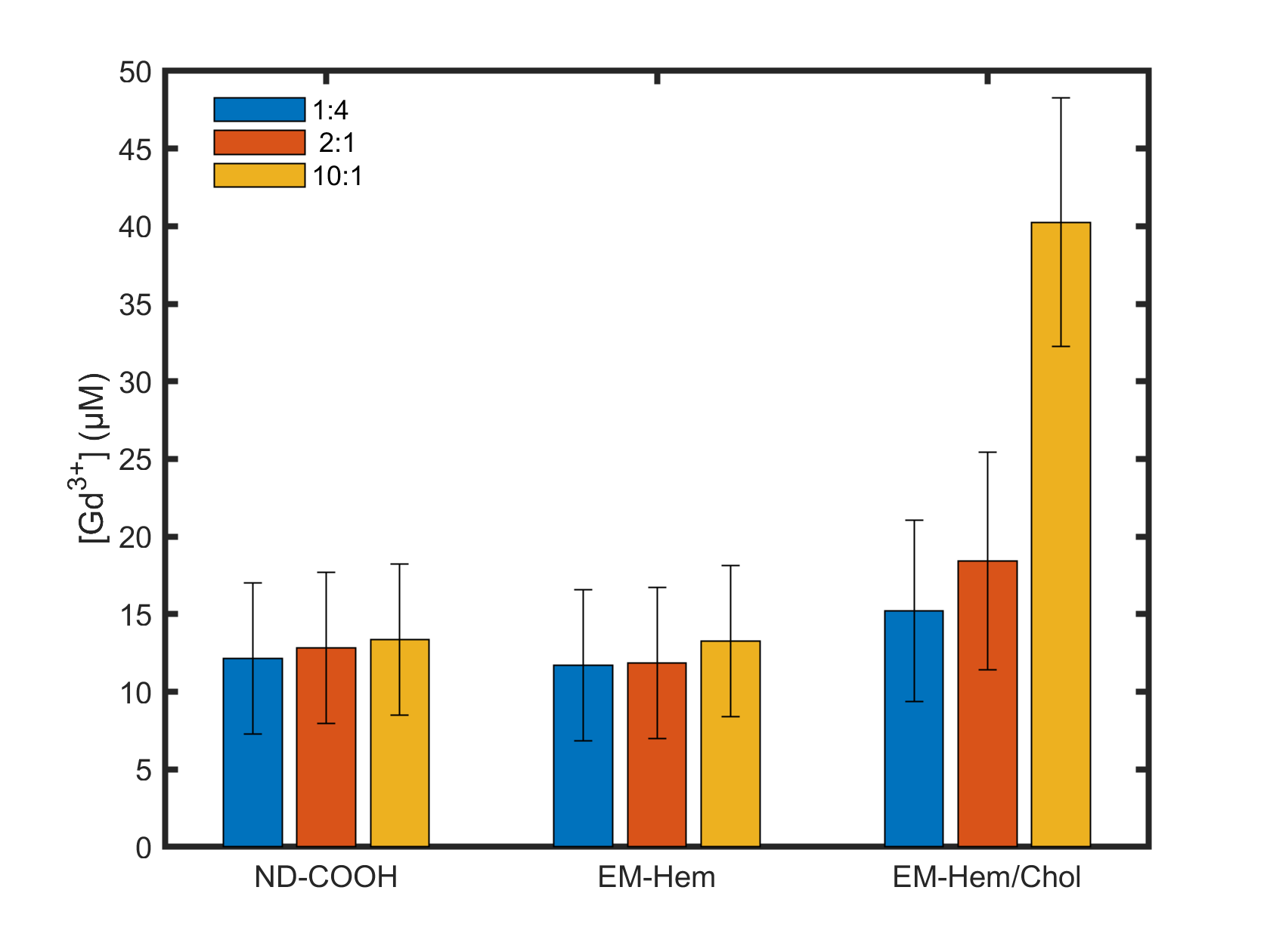}
%\centering
\caption[Gd$^{3+}$ DBCO Click Conjugation Control: ICP-OES]{ICP-OES measurements of remaining [Gd$^{3+}$] for click chemistry conjugation experiments with NC-COOH, EM-Hem, and EM-Hem/Chol dispersions and the Gd$^{3+}$ 
 dibenzocyclooctyne (DBCO) chelate. For EM-Hem/Chol dispersions, experiments are carried out for DBCO to azide ($-N_{3}$) molar ratios of 0:1 (blue),  2:1 (orange), and 10:1 (yellow). For  ND-COOH and EM-Hem, which do not have surface functional groups with azide termination, the ratios refer to the addition of identical Gd$^{3+}$ concentrations as used for EM-Hem/Chol conjugation reactions.  Error bars represent propagated uncertainty from the best-fit confidence intervals for the OES-ICP gadolinium calibration curve.}
\label{fig:Click ICP Full}
 \end{figure}}

\def\TOneSurfModel{\begin{figure}[t]
\renewcommand\figurename{Figure}
\centering
\includegraphics[scale= 0.7]{surf_all.pdf}
\caption[$T_1$ Fit to Surface Model]{$\left\langle T_1 \right\rangle$ \textit{vs} $\rho_{Gd^{+3}}$ for EM-Hem (brown) along   (Left) fit for different value of $ \sigma_{\infty}$ for $K=75 \mu\text{M}^{-1}$ and (Right) fit for different value of $ \sigma_{\infty}$ for $K=250 \mu\text{M}^{-1}$.}
\label{fig:all_surfl}
 \end{figure}}

 \def\TOneVolModel{\begin{figure}[t]
\renewcommand\figurename{Figure}
\centering
\includegraphics[scale= .3]{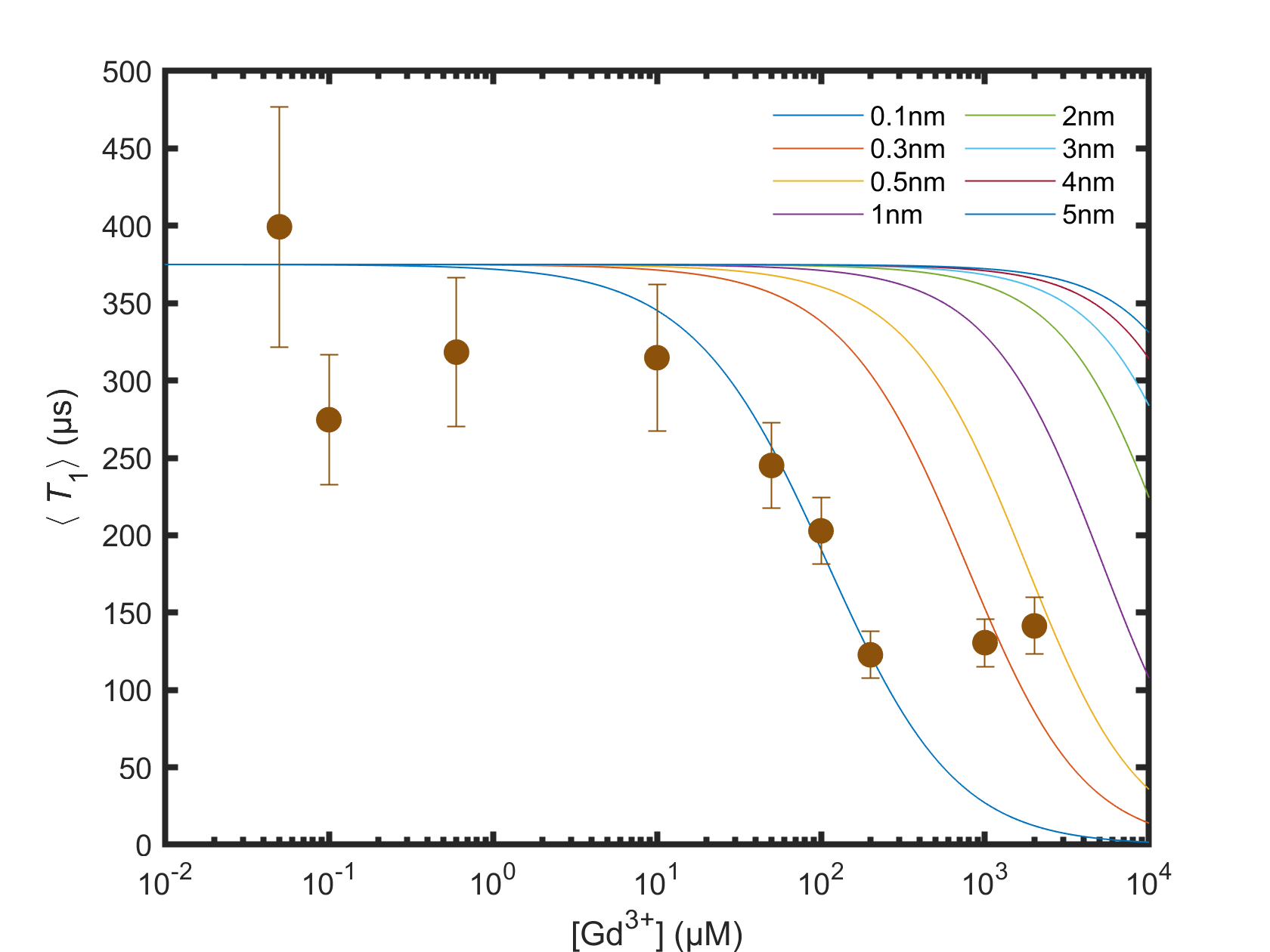}
\caption[$T_1$ free volume model]{$\left\langle T_1 \right\rangle$ \textit{vs} [$Gd^{+3}$] for EM-Hem (brown),  along with $T_1$ curves calculated using Equations~\ref{eqn:nvOneoverT1},~\ref{eqn:Bvolume}, and~\ref{eqn:Rdip}, with values of $r_\text{excl}$ ranging from 0.1 nm to 5.0 nm.}
\label{fig:all_vol}
 \end{figure}}

\def\TOneCleanExamples{\begin{figure}[t]
\renewcommand\figurename{Figure}
\centering
\includegraphics[width = \textwidth ]{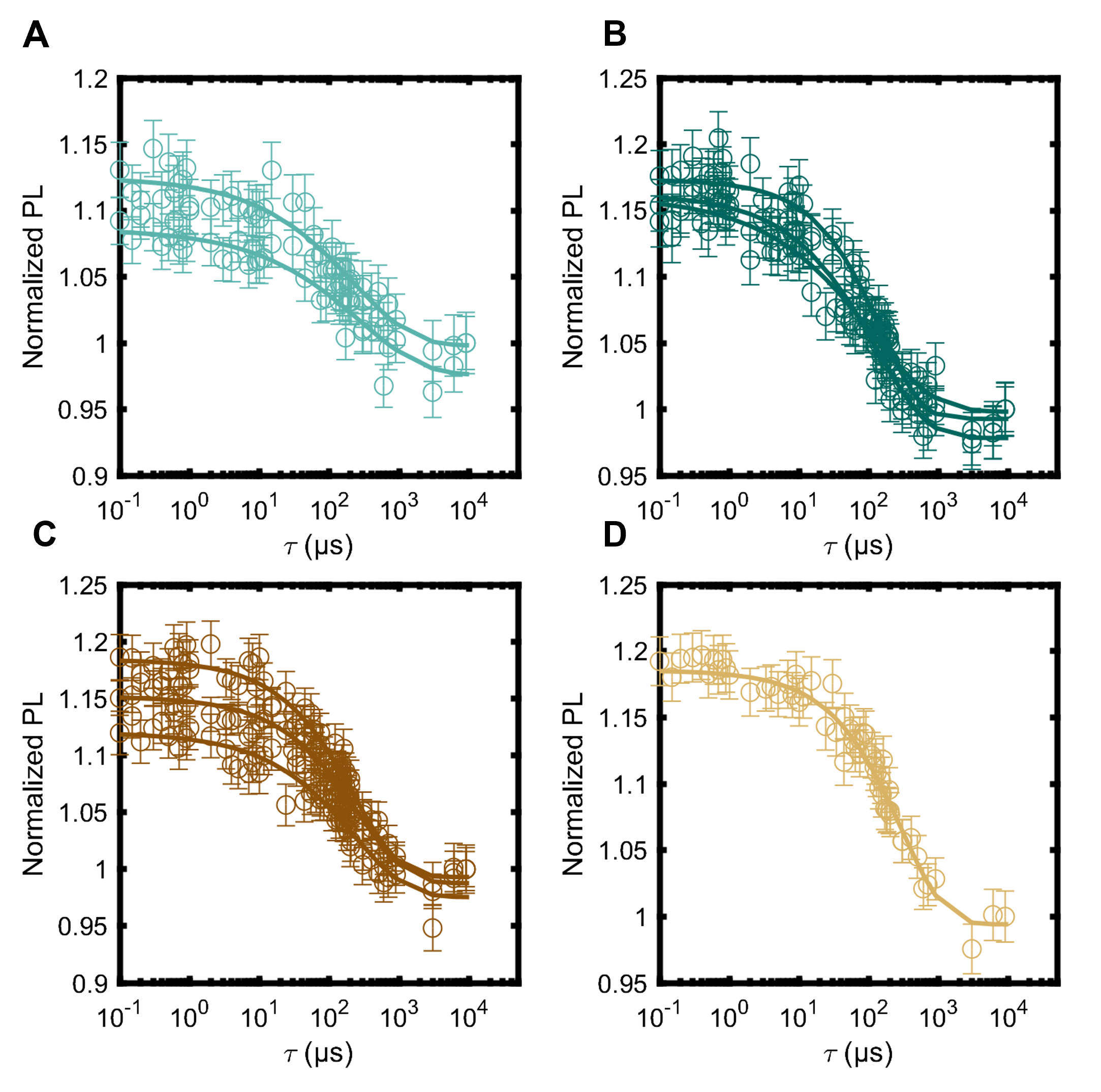}
%\centering
\caption[Spin Lifetime Curves for Nanodiamonds and Emulsion Samples]{Photoluminescence (PL) intensity as a function of delay time, $\tau$, between an optical initialization pulse and subsequent readout pulse, for (A) ND-C18, (B) ND-COOH  (C) EM-Hem and (D) EM-Hem/Chol dispersions. 
Solid curves are fits using Equation~2 of the Main Text.
The PL signal is normalized to the intensity at $\tau = 9$~ms, and error bars represent the uncertainty from photon shot noise. 
The curves within panel A-C represent multiple measurements of the same sample.
}
\label{fig:Clean T1}
 \end{figure}}

 \def\AqNDPlusGD{\begin{figure}[t]
\renewcommand\figurename{Figure}
\centering
\includegraphics[width = \textwidth ]{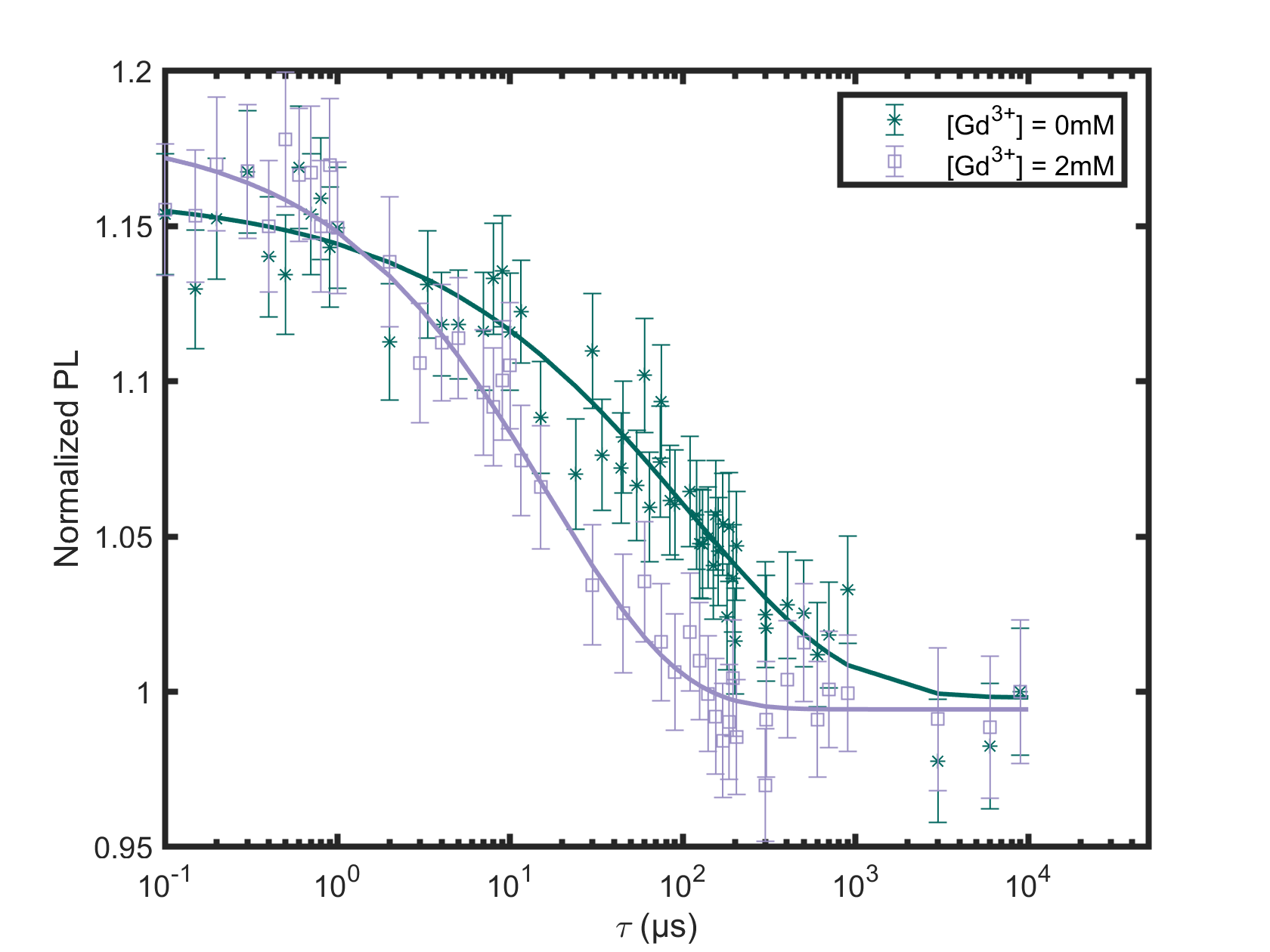}
%\centering
\caption[Spin Lifetime Curves for Nanodiamonds and Emlusion Samples]{Photoluminescence (PL) intensity as a function of delay time, $\tau$, between an optical initialization pulse and subsequent readout pulse, for a  ND-C18 dispersion before (green, stars) and after (purple boxes) the addition of 2 mM of Gd$^{3+}$ chelate.
Solid curves are fits using Equation~2 of the Main Text.
The PL signal is normalized to the intensity at $\tau = 9$~ms, and error bars represent the uncertainty from photon shot noise. 
}
\label{fig:ND-C18 + Gd}
 \end{figure}}

 \def\EMHemPlusGD{\begin{figure}[t]
\renewcommand\figurename{Figure}
\centering
\includegraphics[width = \textwidth ]{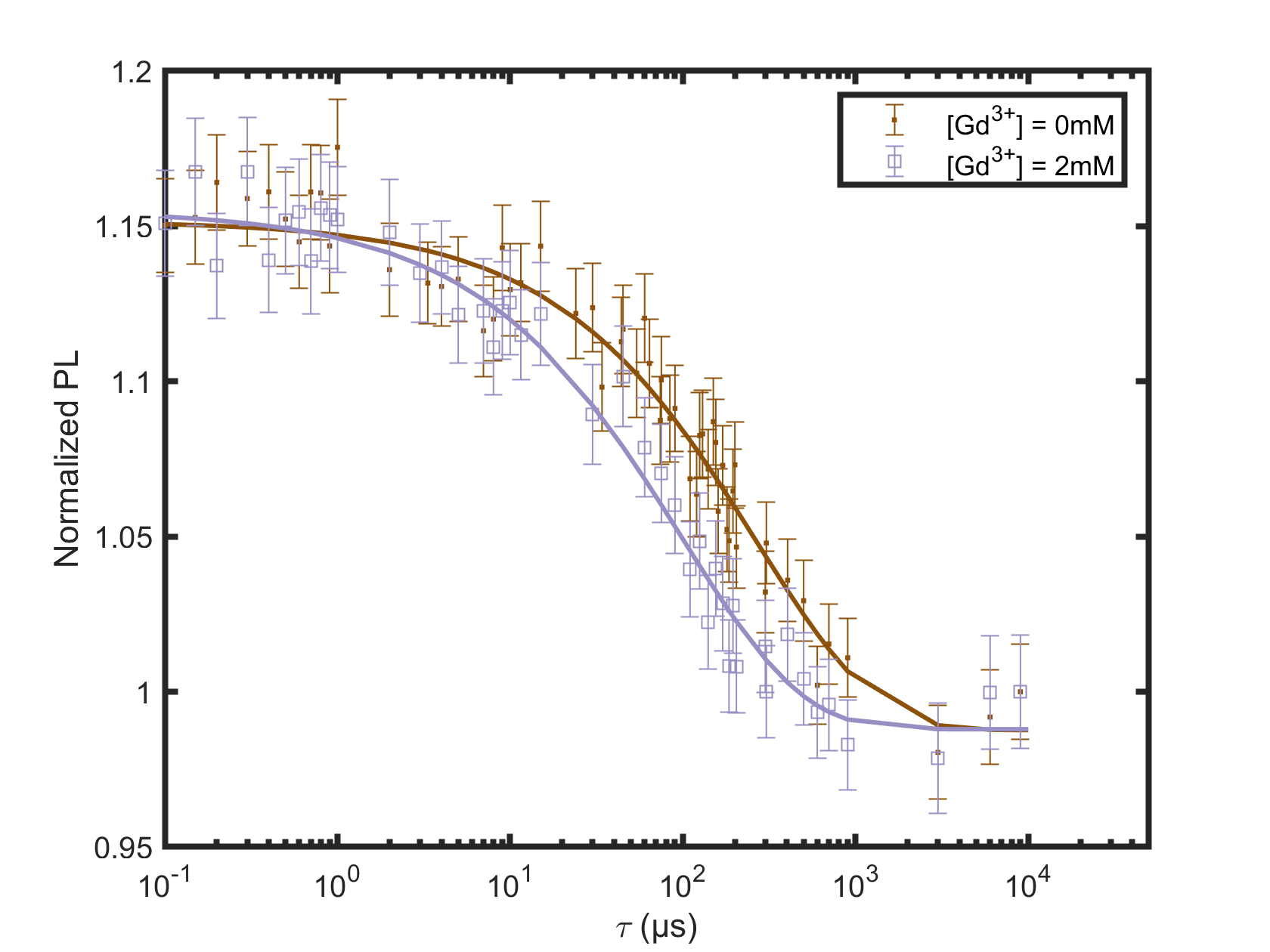}
%\centering
\caption[Spin Lifetime Curves for Nanodiamonds and Emulsion Samples]{Photoluminescence (PL) intensity as a function of delay time, $\tau$, between an optical initialization pulse and subsequent readout pulse, for an EM-Hem dispersion before (brown, dots) and after (purple boxes) the addition of  2 mM of Gd$^{3+}$ chelate.
Solid curves are fits using Equation~2 of the Main Text.
The PL signal is normalized to the intensity at $\tau = 9$~ms, and error bars represent the uncertainty from photon shot noise. 
}
\label{fig:EM-Hem + Gd}
 \end{figure}}

  \def\HeminSpinModel{\begin{figure}[t]
\renewcommand\figurename{Figure}
\centering
\includegraphics[width = 0.8\textwidth ]{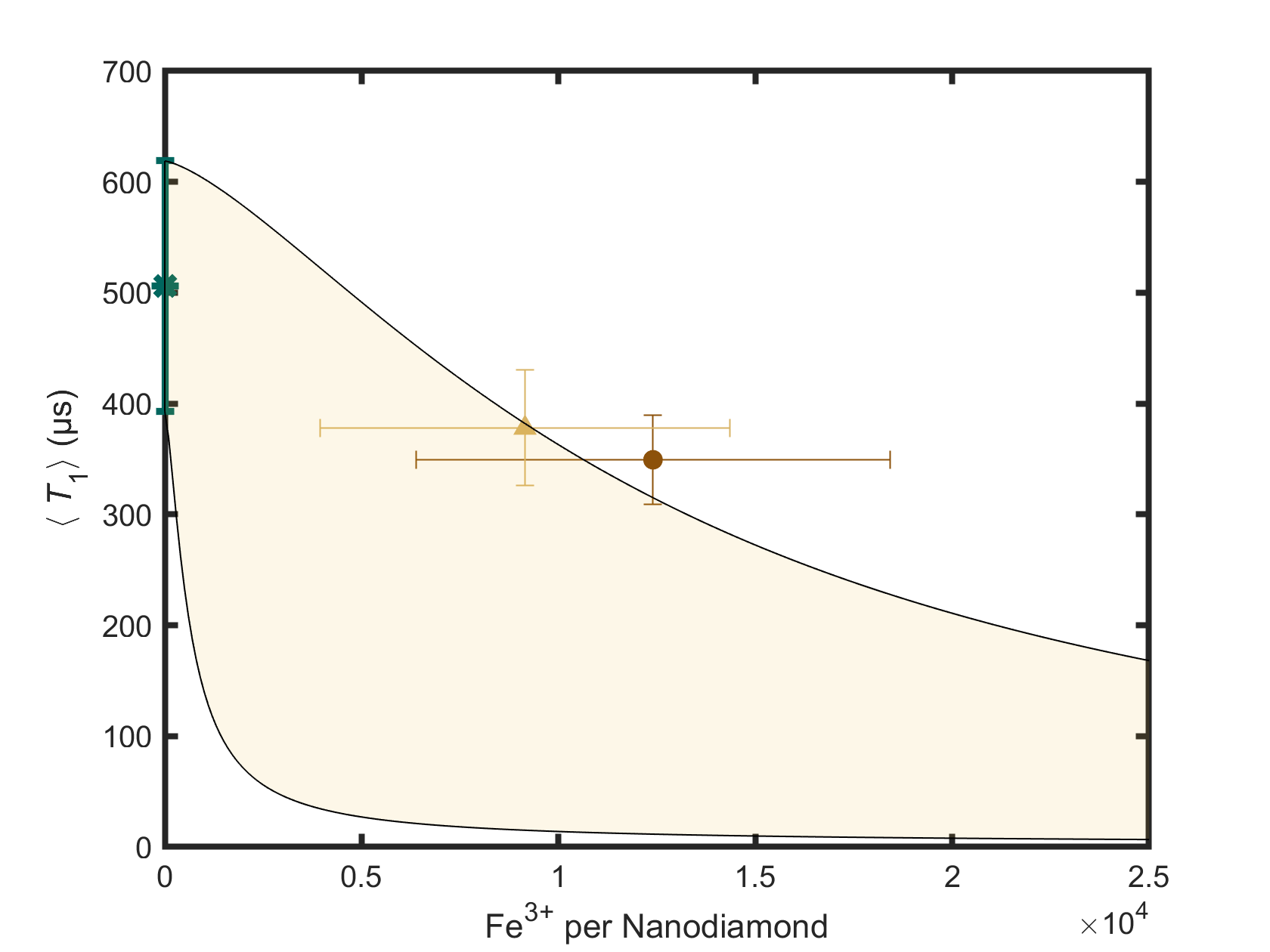}
%\centering
\caption[Hemin Coating Impact on Spin Lifetime Model]{Modeled impact of Fe$^{3+}$ on $\left\langle {T}_1 \right\rangle$ for a 68~nm diameter nanodiamond. The upper bound of the shaded region describes a model where r$_{sl}$ = 4.0~nm, $D_{min}$ = 1.57~nm, and $T_1^{\text{0}}$ = 618~$\mu$s and the lower bound is a model for r$_{sl}$ = 0.5~nm, $D_{min}$ = 0.34~nm,  and $T_1^{\text{0}}$ =392~$\mu$s. $\left\langle {T}_1 \right\rangle$ are shown for ND-C18 (green star), EM-Hem (brown dot), and EM-Hem/Chol (yellow triangle) dispersions.
Vertical error bars represent the standard deviation of $\langle{T}_1\rangle$ from measurements of multiple samples for ND-C18 and EM-Hem or the fit uncertainty from a single measurement of EM-Hem/Chol.
Horizontal error bars represent the propagated uncertainty of the hemin concentration estimate. 
}
\label{fig:HeminSpinModel}
 \end{figure}}

%% Adds the main heading for the SI text. Comment out this line if you do not have any supporting information text.
%\SItext

\section{Varying Nanodimaond Size and Surface Coating}\label{subsec:level1}
In this subsection, in addition to the materials described in the Main Text, we also consider smaller milled nanodiamonds (Ad\'amas Nanotechnologies), emulsions (EM-OA) formed with oleic acid (Thermo Fisher), and emulsions formed with varying amounts of hemin and hemin-cholesteryl-TEG azide. 
Accordingly, the samples are labeled by the material type and either the advertised size for ND-COOH and ND-C18 or the weight ratio of amphiphilic molecules to nanodiamonds utilized in preparing the emulsions. 
Due to the variability of the milling process the advertised size of the nanodiamonds is often different than the measured size as seen in Figs.~\ref{fig:DLS Raw} and \ref{fig:DLS Bar} and Main Text Figure~1. 

\SuppDLSRaw
\SuppDLSBar

\clearpage

%data from emulsions using different amounts of hemin than those mentioned in the Main Text, as well as emulsions formed with oleic acid (Thermo Fisher) as the amphiphilic material.
 %Accordingly, we adopt a modified naming convention in which the 60nm carboxylate nanodiamonds, previously referred to as ND-COOH in the Main Text, are now designated as COOH60. Similarly, the 40nm carboxylate nanodiamonds are labeled as COOH40.
%The 60nm hydrophobic octadecane-terminated nanodiamonds mentioned as ND-C18 in the Main Text will be referred to as C18 here.
%The nanodiamond-hemin emulsion formed at weight ratios of 20 to 1, 15 to 1, 10 to 1 (referred to as EM-Hem in the Main Text), and 5 to 1  are denoted here as H20:1, H15:1, H10:1, and H5:1, respectively.
%The nanodiamond-hemin-cholesteryl teg azide emulsion, previously referred to as EM-Hem/Chol in the Main Text, will be labeled as HC20:1.

%\NDEPLSpectraFigure
%\NDEPLBar

\subsection{Non-negative Matrix Factorization}
We employ the MATLAB function \texttt{nnmf} to perform a rank two non-negative matrix factorization (NNMF) of  the measured nanodiamond and emulsion spectra \cite{berry2007algorithms}. % in Fig. S~\ref{fig:All PLSpectra} \cite{berry2007algorithms}. 
NNMF allows us to  extract the NV$^{0}$ and NV$^{-}$ emission spectra ($S_\mathrm{NV^{-}}$ and $S_\mathrm{NV^{0}}$) and the PL intensities of those states relative to the total emission spectra such that:
\begin{equation}
    \mathrm{Measured\,Spectrum} \approx S_\mathrm{NV^{0}} * \mathrm{NV}^{0} + S_\mathrm{NV^{-}} * \mathrm{NV}^{-}
\end{equation}
%The upper and lower uncertainty limits of this factorization for each nanodiamond solution, is calculated by scaling $S_\mathrm{NV^{-}}$ and $S_\mathrm{NV^{0}}$ until the reduced chi-squared between the measured spectrum and  $S_\mathrm{NV^{0}} * NV^{0} + S_\mathrm{NV^{-}} * NV^{-}$  exceeds 2. 
The results of this factorization can be seen in Figure 3 of the Main Text and Fig.~\ref{fig:NDE FracNV}.
Here our NNMF analysis resulted in a  root mean square residual of 4.1 counts/nm and a scatter index of 7\%.

\NDEFracNV
\NDEFracNVBar

\NDETSBar
\NDEConBar
\NDEBetaBar
\NDETABar

\clearpage

 %Error bars represent the limits of acceptable factorization from the NNMF anaylsis.  
\section{Transmission Electron Microscopy}
Transmission electron microscopy (TEM) images are taken on a JEOL F200 scanning /transmission electron microscope operating at 200 kV.
The nanodiamond samples are drop-cast onto 300 mesh copper TEM grids and dried overnight in a vacuum chamber.
As expected, the milled nanodiamonds have an irregular shape and size. The milled nanodiamonds exhibit varying orientations and thicknesses, which are evident from the varying contrast and focus observed in different areas of the TEM images.
We did not observe any significant difference between the images for the four nanodiamond samples.
\TEMFiveHundred
\TEMOneHundred
\TEMTwent

\clearpage

\section{Estimating Hemin Concentration}\label{Hemin per ND emistmates}

The hemin concentrations for the emulsions are calculated from the extinction curves of the samples (Main Text Figure~2) and a molar extinction coefficient of $12.2 \pm 3*10^{4}$ M$^{-1}$cm$^{-1}$ \cite{inada1962soret}.
Here we fit the extinction curves for EM-Hem and EM-Hem/Chol to a linear combination of ND-C18 and hemin extinction curves to isolate the hemin contribution for each emulsion. 
If we assume monodisperse, 68 nm diameter spherical nanodiamond samples and propagate the uncertainty from the hemin concentration analysis, for EM-Hem we estimate $12400 \pm 6000$ hemin molecules per nanodiamond and for EM-Hem/Chol we estimate $9200 \pm 5200$ hemin molecules per nanodiamond. 
Given the large size (Fig.~\ref{fig:DLS Raw}) and shape (Fig.~\ref{fig:TEM500})  dispersity of milled nanodiamond samples and the unknown density of -C18 and -COOH surface groups inherited from the parent C-18 NC samples\cite{ito2016hydrophobic}, these estimates are consistent with a sparse or patchy, multilayer hemin coating.
%These estimates imply a sparse or patchy hemin coating on the nandiamonds.
%However, given the large size (Fig.~\ref{fig:DLS Raw}) and shape (Fig.~\ref{fig:TEM500}) dispersity of milled nanodiamond samples and the unknown density  of -C18 and -COOH inherited from the parent C18-ND sample\cite{ito2016hydrophobic}, which would inhibit perfect coverage, it is difficult to characterize the exact nature of the coating layer.

\TOneCleanExamples

\clearpage

\section{Additional Information for NV-Center Spin Lifetime}\label{Sub: T1Modeling}

Our modeling of $T_1$ for a NV center in a nanodiamond builds on the analysis reported by Tetienne \textit{et al}.\cite{tetienne2013spin}. 
The relaxation rate $1/T_1$ for an NV center placed in a zero mean magnetic field $B(t)$ with a variance of $B_\perp^2$ can be expressed as follows:
\begin{equation}
\frac{1}{T_1} = \frac{1}{T_1^{\text{0}}} + 3 \gamma_e^3 B_\perp^2 \frac{\tau_c}{1+ \omega_0^2 \tau_c^2}
\label{eqn:nvOneoverT1}
\end{equation}
where $T_1^{\text{0}}$ is the spin lifetime in the absence of a magnetic field, $\tau_c$ is the correlation time of the bath, $\gamma_e$ is the electron gyromagnetic ratio, and $\omega_0$ is the electron spin resonance frequency ($\omega_0 = 2\pi \times2.87$ GHz). 
The bath's correlation time is given by: 
\begin{align}
\tau_c = \frac{1}{R_{\text{dip}}+ R_{\text{vib}}}
\label{eqn:tcgen}
\end{align}
where the rate $R_{\text{vib}}$ arises from the intrinsic vibrational spin relaxation of the bath and the rate $R_{\text{dip}}$ originates from inter-spin dipolar interactions, which can be approximated by $\hbar R_{\text{dip}} = \sqrt{\sum_{j \neq i} \left\langle H_{ij}^2 \right\rangle}$ where $H_{ij}$ is the magnetic dipolar interaction between spin $i$ and $j$ of the bath.

\begin{figure}[t]
\renewcommand\figurename{Figure}
    \centering
    \includegraphics[scale=1]{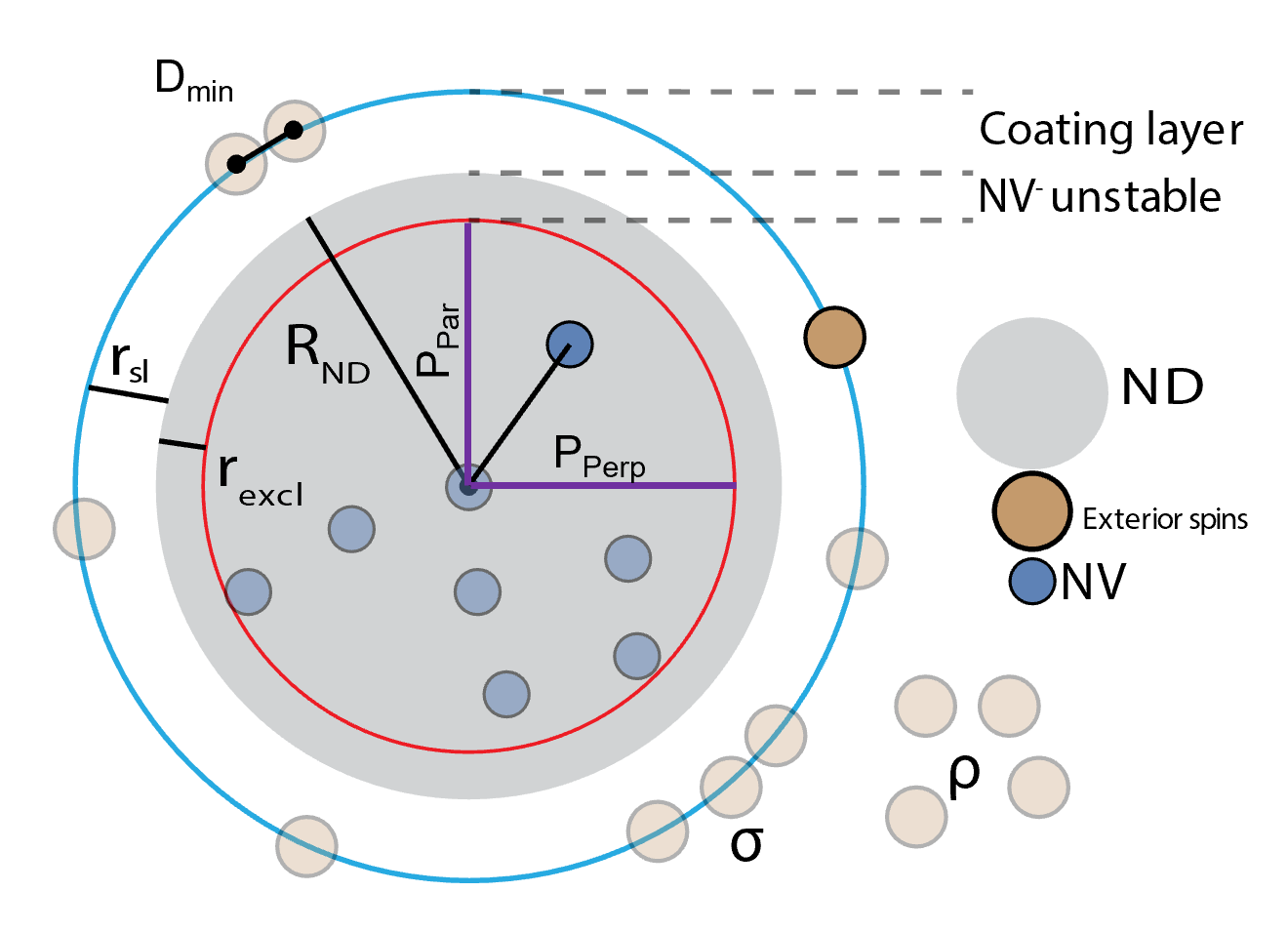}
    \caption[Schematic Representation of a nanodiamond in a spin bath]{Schematic representation of a nanodiamond (grey-filled circle) with radius $R_{ND}$ and a surface coating of thickness $r_{sl}$, denoted by the blue circle. NV centers (blue-filled circles) are randomly distributed in the nanodiamond, but consistent with literature, do not form within an exclusion zone $r_{excl}$ of the nanodiamond surface, denoted by the red circle. The exterior spin bath (gold-filled circles) can exist either as a surface distribution along the blue circle (denoted as $\sigma$) or as a volumetric distribution (denoted as $\rho$). The minimum distance between exterior spins is given by $D_{min}$.
    }
    \label{fig:nd_schem}
\end{figure}

In our analysis, we treat the nanodiamond as a sphere with a radius $R_{\text{ND}}$. 
Inside this sphere, the NV centers are distributed randomly, but they are restricted to a maximum radius that excludes a shell of thickness $r_\text{excl}$. 
The exclusion thickness $r_\text{excl}$  for NV centers in nanodiamonds is estimated to range from 2~nm to 3~nm \cite{bluvstein2019identifying}.
Exterior spins can be excluded from a shell of thickness $r_\text{sl}$ defined by the thickness of the coating layer. 
%Each of these exterior spins must maintain a minimum distance from each other, defined by the spin particle's size, denoted as $R_\text{min}$, indicating that they cannot be positioned closer to each other than this threshold.
Each of these exterior spins cannot be positioned closer to each other than a threshold ($D_\text{min}$) which is defined by the spin molecule's size and packing density. 
For a surface coating of spins, we can model the average variance $B_\perp^2$ experienced by an NV center according to:
\begin{align}
B_{\perp}^2  &= \left(\frac{\mu_0 \gamma_e \hbar}{4 \pi} \right)^2 \frac{S(S+1)}{3}\notag  \\
& \qquad \qquad \qquad \qquad \sigma \int_{0}^{2\pi} d\phi \int_{0}^\pi d\theta \sin \theta  \frac{2+ 3 \sin(\alpha(\theta,\phi)^2}{r(\theta, \phi)^{4}}
\label{eqn:BiSurf}
\end{align}
where $\sigma$ is the density of surface spins with total spin $S$, and $r(\theta, \phi)$ and $\alpha(\theta,\phi)$ are respectively the distance and relative angle between the NV center and a spin located at angular coordinates $(\theta, \phi)$. 
%For an NV center at the center of a nanodiamond we set $\alpha(\phi, \theta) = \theta$ and $r(\phi,\theta) = \frac{D_{0}}{2}$.
To capture the effect of NV orientation, we consider NV centers displaced from the center of the particle by a distance $\delta r$ along the line parallel (P$_\mathrm{Par}$) or perpendicular (P$_\mathrm{Perp}$) to the NV center's spin axis. 
These two paths provide the highest and lowest sensitivity to the external fields, respectively.
For NV centers along $P_\mathrm{Par}$, we set $r(\theta, \phi) =\sqrt{\frac{D_0}{2}^2 -\delta r^2 \sin(\theta)^2} + \delta r \cos(\theta)$ and $\alpha(\theta, \phi) = \theta$, where $D_0=2(R_{ND} + r_{sl})$, to represent the position and relative angle of the spins on the particle surface, with a coordinate system centered on the NV center.
For NV centers along P$_\mathrm{Perp}$, we set $ r(\theta, \phi) =\sqrt{\frac{D_0}{2}^2 -\delta r^2 \sin(\theta)^2} + \delta r \cos(\theta)$ and $\alpha(\theta, \phi) = \cos(\theta)^2 + \sin(\theta)^2\sin(\phi)^2$.
We then calculate a weighted average of $B_\perp^2$ for NV centers along either path from $\delta r$ = 0 to $\delta r  = R_{ND} - r_\text{excl}$, weighting according to the geometric likelihood of finding an NV center:
\begin{align}
 \left\langle B_{\perp}^2 \right\rangle &= \frac{\int_{0}^{R_{ND}-r_\text{excl}} B_{\perp}^2 (D_0, \delta r) \delta r^2  d\delta r }{\int_{0}^{R_{ND}-r_\text{excl}}\delta r^2  d\delta r}
\label{eqn:BiSurfAverage}
\end{align}
Subsequently, we average the resulting estimates from the two paths.

For spins arranged in a surface-coating layer of density $\sigma$, R$_{\text{dip}}$ can be calculated according to:
\begin{align}
R_{\text{dip}} &= \frac{1}{\hbar} \sqrt{\sum_{j\neq i} \left\langle H_{ij}^2 \right\rangle } \notag\\
&= \frac{\mu_0 \gamma_e^2 \hbar  S(S+1) }{2 \sqrt{6}  \pi}  \sqrt{\sigma\int_{D_{\text{min}}}^\infty  \frac{2\pi dr}{r^5}} 
\label{eqn:RdipSurface }
\end{align}
To model the impact of the Fe$^{3+}$ present in the hemin molecules on the $T_1$ of the emulsions, we set R$_{Vib}$ = 0.1 GHz \cite{sur1995nuclear}, S = 5/2 \cite{atak2014electronic},  $r_\text{excl}$ = 3.0~nm, and $T_1^{\text{0}}$ as 505 $\pm$ 113 $\mu s$, the $\left\langle T_1 \right\rangle$ of ND-C18.
The results of this model for 68~nm diameter nanodiamonds, $r_{sl}$ ranging from 0.5~nm to 4~nm, and $D_{min}$ ranging from  0.34~nm (the intermolecular distance in hemin assemblies\cite{maricondi1972moessbauer}) to 1.57~nm (the length of the hemin molecule), studied to account for the potential orientations of the amphiphiles on the nanodiamond surfaces, are shown in Fig.~\ref{fig:HeminSpinModel}. 
Our model is sensitive to these geometric considerations.
A more realistic model for the nanodiamond shape ($R_{\text{ND}}$), the placement of NV centers ($r_{excl}$, P$_\mathrm{Par}$ and P$_\mathrm{Perp}$), and the arrangement of surface layer distribution ($r_d$, $D_{min}$ and $\sigma$) would likely have a significant effect on the model.
%The measured $\left\langle T_1 \right\rangle$ imply that the hemin molecules are packed more tightly then their maximum diameter but do not perfectly stack on top of one another. 
%Additionally, Fig.~\ref{fig:HeminSpinModel} indicates 
%The measured $\left\langle T_1 \right\rangle$ indicates that the hemin molecules are not tightly packed onto the nanodiamond's surface but are instead offset from the diamond and each other by a nanoscale distance, which corroborates the findings from dynamic light scattering (DLS) measurements.

\HeminSpinModel

Here we see that the upper bound of our model, representing lower sensitivity of the nanodiamonds to the ions, (Fig.~\ref{fig:HeminSpinModel}) offers a more accurate description of the hemins' impact on $\left\langle T_1 \right\rangle$. 
This result implies that the hemin molecules are held slightly farther from the diamond surface than we initially expected for our model.
Alternatively, we could be observing competing effects on $T_1$ where the emulsion process quenches some surface spins on the nanodiamonds thereby counteracting the reduction from the addition of Fe$^{3+}$ spins. 
We note an alternative amphiphilic molecule could reduce or eliminate this observed reduction in $T_1$ (Fig.~\ref{fig:NDET1A}).
Simply changing hemin-chloride to hemin-bromide, which is structurally similar and thus should not impact the stability of the emulsions, is reported to have a lower Fe$^{3+}$ spin\cite{reed1996magnetochemical}, and thus is expected to decrease the hemin's effect on $\left\langle T_1 \right\rangle$. 
%from S = 5/2 to S = 3/2,%would decrease the impact of the  Fe$^{3+}$ on $\left\langle T_1 \right\rangle$ of EM-Hemin by a factor of 2. 

To model the impact of the Gd$^{3+}$ chelates, we set $R_{Vib}$ = 1 GHz \cite{caravan1999gadolinium}, $S = 7/2$, and $D_{min}$ = 0.5~nm.
%According to Equation~\ref{eqn:nvOneoverT1}, the higher S and $R_{Vib}$ values of the Gd$^{3+}$ chelates and dense is expected to reduce $T_1$ approximately twice as much as the Fe$^{3+}$ in the hemin for the same ion to nanodiamond ratio. 
According to Equation~\ref{eqn:nvOneoverT1}, the higher S and $R_{Vib}$ values and denser packing  of the Gd$^{3+}$ chelates is expected to reduce $T_1$ more significantly than the Fe$^{3+}$ in the hemin.
For chemical sensitivity experiments, where Gd$^{3+}$ chelates are in solution, and thus present beyond the surface of the nanodiamond, we must modify $B_\perp^2$ and $R_{dip}$ to account for these additional spins.
%To account for a free volume concentration of Gd$^{3+}$, we must modify $B_\perp^2$ and $R_{dip}$.
 When the Gd$^{3+}$ are not only bound to the nanodiamond surface, $B_{\perp_{vol}}^2$ takes the form:
\begin{align}
B_{\perp,\text{vol}}^2(\xi,\theta)  &= \left(\frac{\mu_0 \gamma_e \hbar}{4 \pi} \right)^2 \frac{S(S+1)}{3}\notag  \\
& \qquad \qquad \qquad \qquad \rho \int_{\frac{D_0}{2}}^{\infty} dr \int_{0}^{2\pi} d\phi \int_{0}^\pi d\theta \sin \theta  \frac{2+ 3 \sin(\alpha(\theta, \phi)^2}{r(\theta, \phi)^{4}}
\label{eqn:Bvolume}
\end{align}
Once again, we consider NV centers along P$_{Par}$ and P$_{Perp}$, as described above, and take a weighted average of the components to derive $\left\langle B_{\perp}^2 \right\rangle$.
Similarly, for unbounded spins, $R_{\text{dip}}$ takes the form:
\begin{align}
R_{\text{dip}} &= \frac{1}{\hbar} \sqrt{\sum_{j\neq i} \left\langle H_{ij}^2 \right\rangle } \notag\\
&= \frac{\mu_0 \gamma_e^2 \hbar  S(S+1) }{2 \sqrt{6}  \pi}  \sqrt{\rho\int_{D_{\text{min}}}^\infty  \frac{4\pi dr}{r^4} }
\label{eqn:Rdip}
\end{align}

In Fig.~\ref{fig:all_vol}, the results of this model are presented for a $T_1^{\text{0}}$ of 375 $\mu$s, taken as the $\left\langle T_1 \right\rangle$ measured for EM-Hemin, and a range of $r_\text{excl}$ from 0.1 nm to 5 nm. 
However, since a $r_\text{excl}$ value of 0.1 nm is not physical, we conclude and describe in the Main Text that the Gd$^{3+}$ chelates must adsorb to the nanodiamond surface, and the dependence of $T_1$ on the concentration of Gd$^{3+}$ chelates is well described by a simple Langmuir model.

\TOneVolModel

\GDVCon
\GDVBeta
\GDVTA
\AqNDPlusGD
\EMHemPlusGD

\clearpage

\section{Additional Information for Carbodiimide Crosslinker Conjugation}\label{Sup Gd Emuls EDC}

We perform carbodiimide crosslinker conjugation between EM-Hem (ND-Hemin) emulsions and the dye molecule n-succinimidlyl ester (TAMRA amine, 5-isomer), purchased from Lumiprobe, under similar conditions to those found in the Main Text.

\EDCHEPES

\EDCDyeUVS
%\EDCDyeUVSMod
\EDCGdDLS
\clearpage

\section{Additional Information for Click Chemistry Conjugation}\label{Sup Gd Emuls DBCO}

We perform click chemistry conjugation with EM-Hem/Chol emulsions and fluorescein dibenzocyclooctyne (FAM-DBCO), purchased from Lumiprobe.
Control measurements are taken with EM-Hem emulsions which do not have an choesteryl-TEG azide, and thus do not have the requisite azide termination for conjugation.

%\DyeClickUVVis
\DyeClickEmission
\FullFTIR
\clearpage

As a control, here we compare the click chemistry conjugation of EM-Hem/Chol emulsions with DBCO-terminated Gd$^{+3}$ chelate spin labels to samples of ND-COOH and EM-Hem,  which do not have surface species with azide termination, that are requisite for click chemistry conjugation.
\DBCOCtrlTS
\DBCOCtrlBeta
\DBCOCtrlCon
\DBCOCtrlTA
\DBCOCtrlICP

\clearpage

\def\GDCtrlButTS{\begin{figure}[t]
\renewcommand\figurename{Figure}
%\centering
\includegraphics[width = \textwidth ]{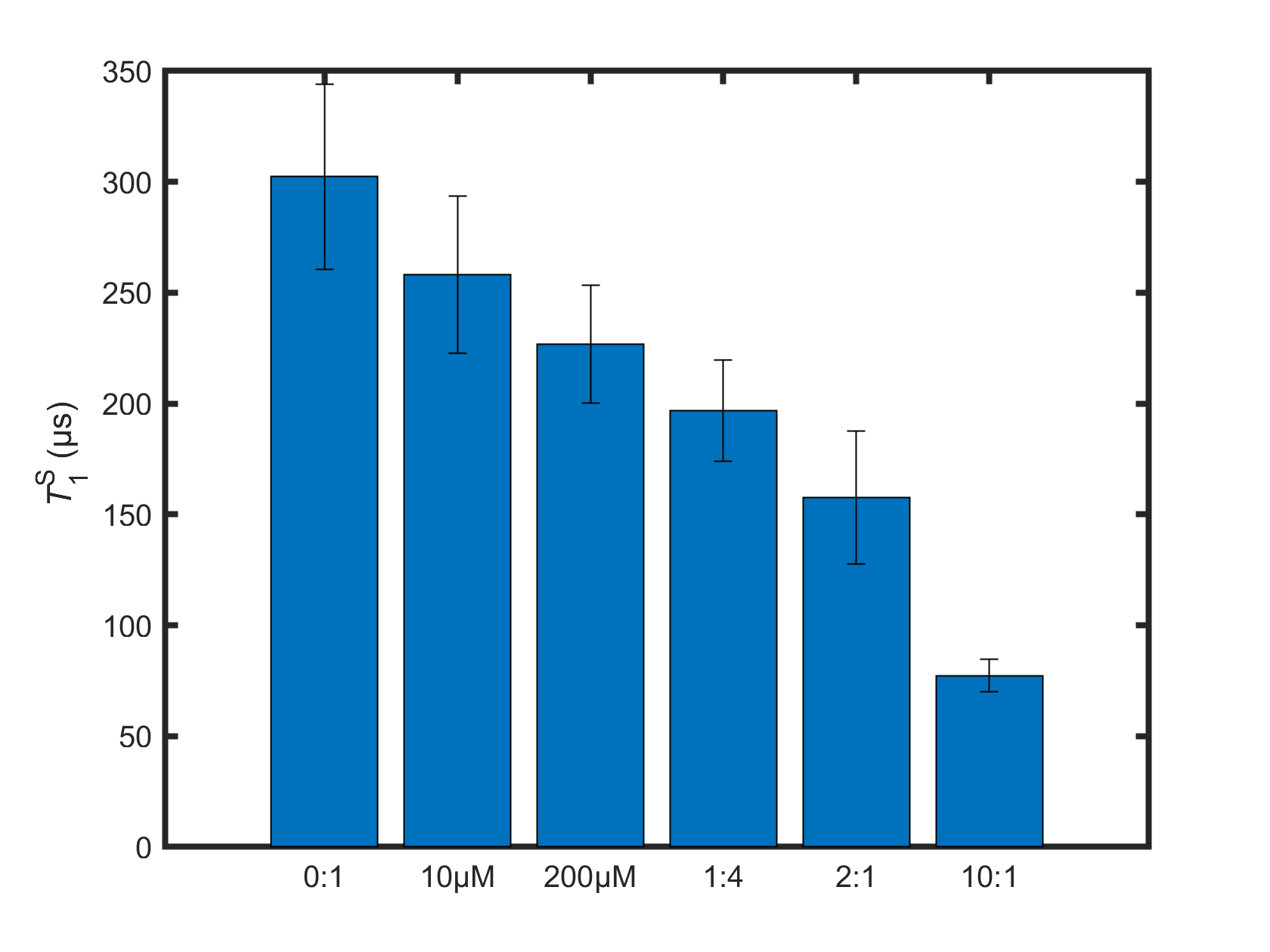}
%\centering
\caption[Free Gd$^{3+}$ Click Conjugation Control: ${T}_1^S$]{Comparison of  ${T}_1^S$ for samples of EM-Hem/Chol without any Gd$^{3+}$ (0:1), with 10 $\mu$M and 200 $\mu$M concentrations of amine-terminated Gd$^{3+}$, and post dialysis of EM-Hem/Chol conjugated to DBCO-terminated Gd$^{3+}$ at initial DBCO to azide ($-N_{3}$) molar ratios of 1:4, 2:1 and 10:1 respectively.
 These ratios translate to initially added [Gd$^{3+}$] of approximately 25 $\mu$M, 200 $\mu$M, and 1000 $\mu$M, respectively. 
See Figure 6 in the Main Text for  ICP-OES measurements of remaining  Gd$^{3+}$ in the click conjugated samples. Error bars represent the best-fit confidence intervals. }
\label{fig:GDCtrlButTS}

 \end{figure}}

\def\GDCtrlButBeta{\begin{figure}[t]
\renewcommand\figurename{Figure}
%\centering
\includegraphics[width = \textwidth ]{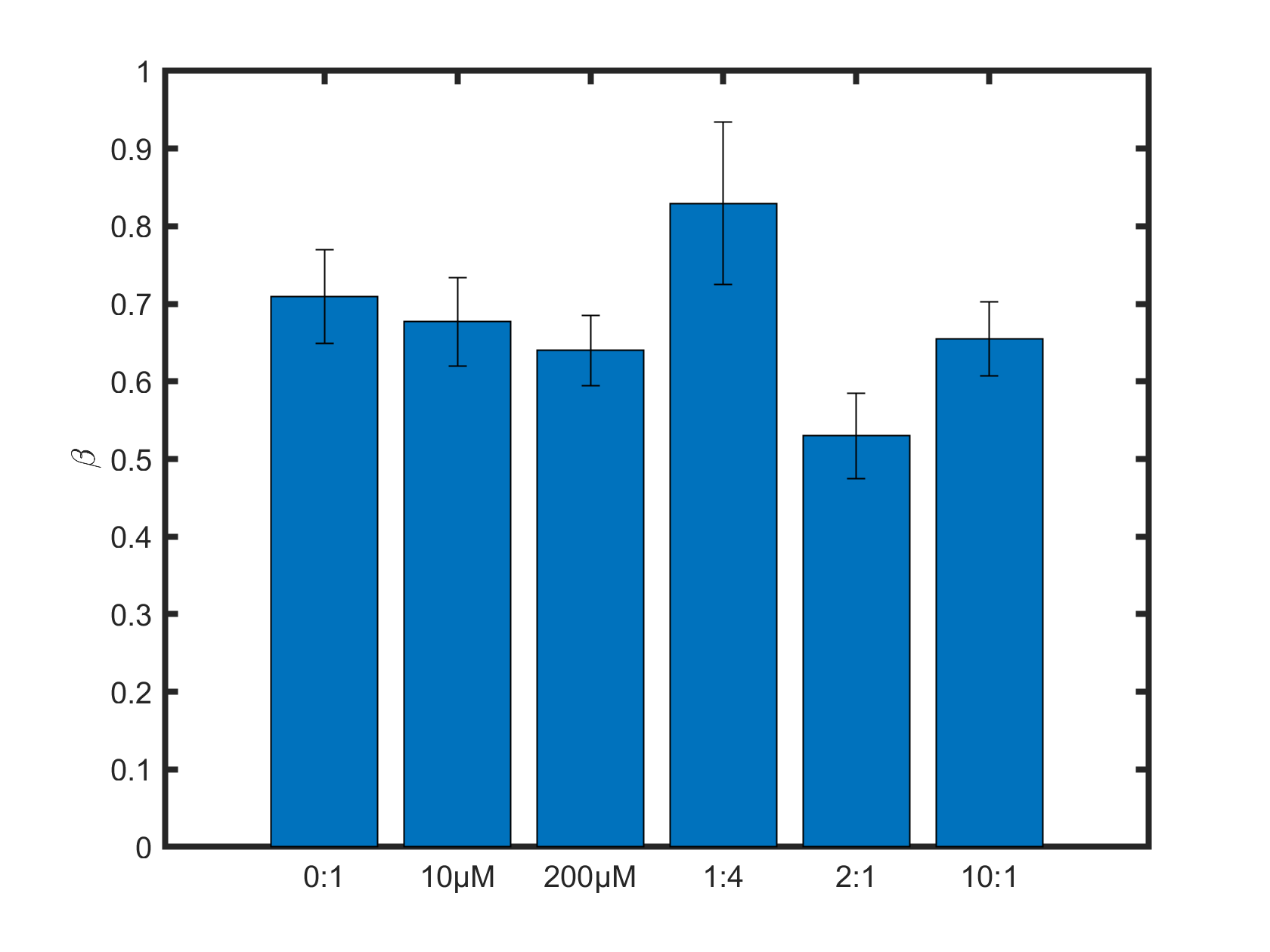}
%\centering
\caption[Free Gd$^{3+}$ Click Conjugation Control:$\beta$]{Comparison of $\beta$ for samples of EM-Hem/Chol without any Gd$^{3+}$ (0:1), with 10 $\mu$M and 200 $\mu$M concentrations of amine-terminated Gd$^{3+}$, and post dialysis of EM-Hem/Chol conjugated to DBCO-terminated Gd$^{3+}$ at initial DBCO to azide ($-N_{3}$) molar ratios of  1:4, 2:1 and 10:1 respectively.
 These ratios translate to initially added [Gd$^{3+}$] of approximately 25 $\mu$M, 200 $\mu$M, and 1000 $\mu$M, respectively. 
See Figure 6 in the Main Text for  ICP-OES measurements of remaining  Gd$^{3+}$ in the click conjugated samples. Error bars represent the best-fit confidence intervals. }
\label{fig:GDCtrlButBeta}

 \end{figure}}

\def\GDCtrlButCon{\begin{figure}[t]
\renewcommand\figurename{Figure}
%\centering
\includegraphics[width = \textwidth ]{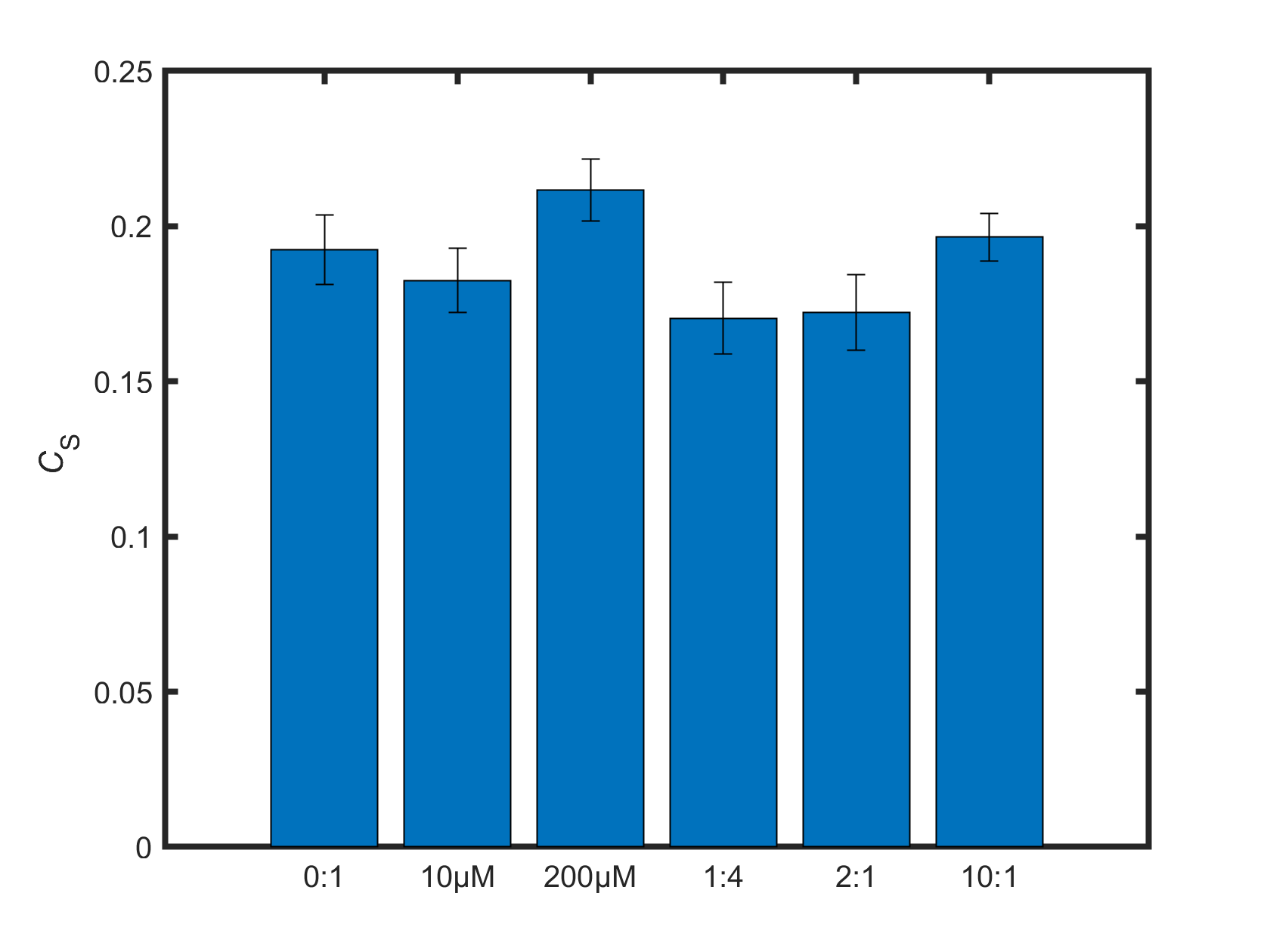}
%\centering
\caption[Free Gd$^{3+}$ Click Conjugation Control:PL Contrast]{Comparison of photoluminescence contrast ($C_S$) for samples of EM-Hem/Chol without any Gd$^{3+}$ (0:1), with 10 $\mu$M and 200 $\mu$M concentrations of amine-terminated Gd$^{3+}$, and post dialysis of EM-Hem/Chol conjugated to DBCO-terminated Gd$^{3+}$ at initial DBCO to azide ($-N_{3}$) molar ratios of  1:4, 2:1 and 10:1 respectively.
 These ratios translate to initially added [Gd$^{3+}$] of approximately 25 $\mu$M, 200 $\mu$M, and 1000 $\mu$M, respectively. 
See Figure 6 in the Main Text for  ICP-OES measurements of remaining  Gd$^{3+}$ in the click conjugated samples. Error bars represent the best-fit confidence intervals.}
\label{fig:GDCtrlButCon}

 \end{figure}}

\def\GDCtrlButTA{\begin{figure}[t]
\renewcommand\figurename{Figure}
%\centering
\includegraphics[width = \textwidth ]{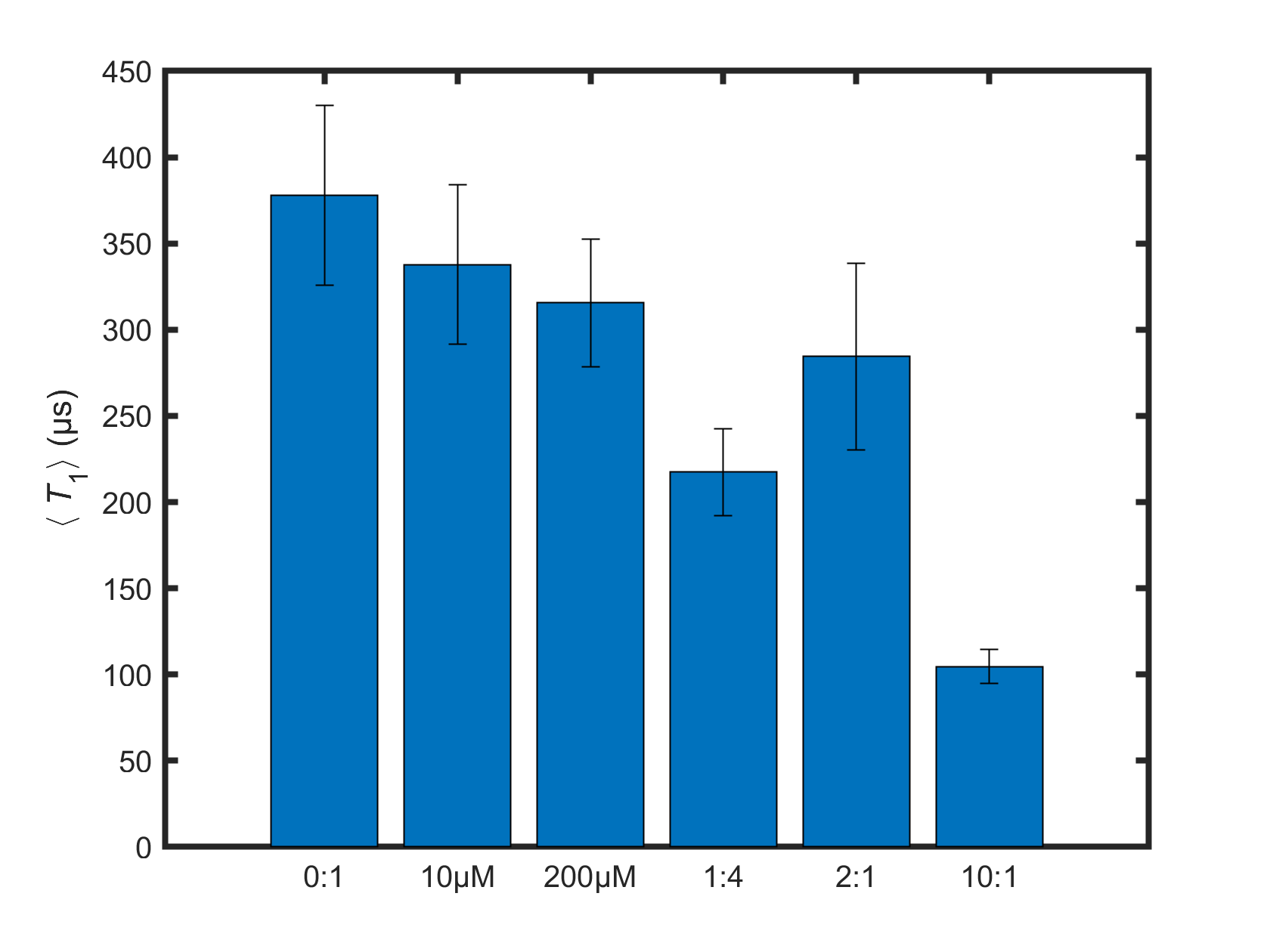}
%\centering
\caption[Free Gd$^{3+}$ Click Conjugation Control:  $\langle{T}_1\rangle$]{Comparison of $\langle{T}_1\rangle$ for samples of EM-Hem/Chol without any Gd$^{3+}$ (0:1), with 10 $\mu$M and 200 $\mu$M concentrations of amine-terminated Gd$^{3+}$, and post dialysis of EM-Hem/Chol conjugated to DBCO-terminated Gd$^{3+}$ at initial DBCO to azide ($-N_{3}$) molar ratios of  1:4, 2:1 and 10:1 respectively.
 These ratios translate to initially added [Gd$^{3+}$] of approximately 25 $\mu$M, 200 $\mu$M, and 1000 $\mu$M, respectively. 
See Figure 6 in the Main Text for  ICP-OES measurements of remaining  Gd$^{3+}$ in the click conjugated samples. Error bars represent the best-fit confidence intervals.}
\label{fig:GDCtrlButTA}

 \end{figure}}

\def\ClickGDDLS{\begin{figure}[t]
\renewcommand\figurename{Figure}
%\centering
\includegraphics[width = \textwidth ]{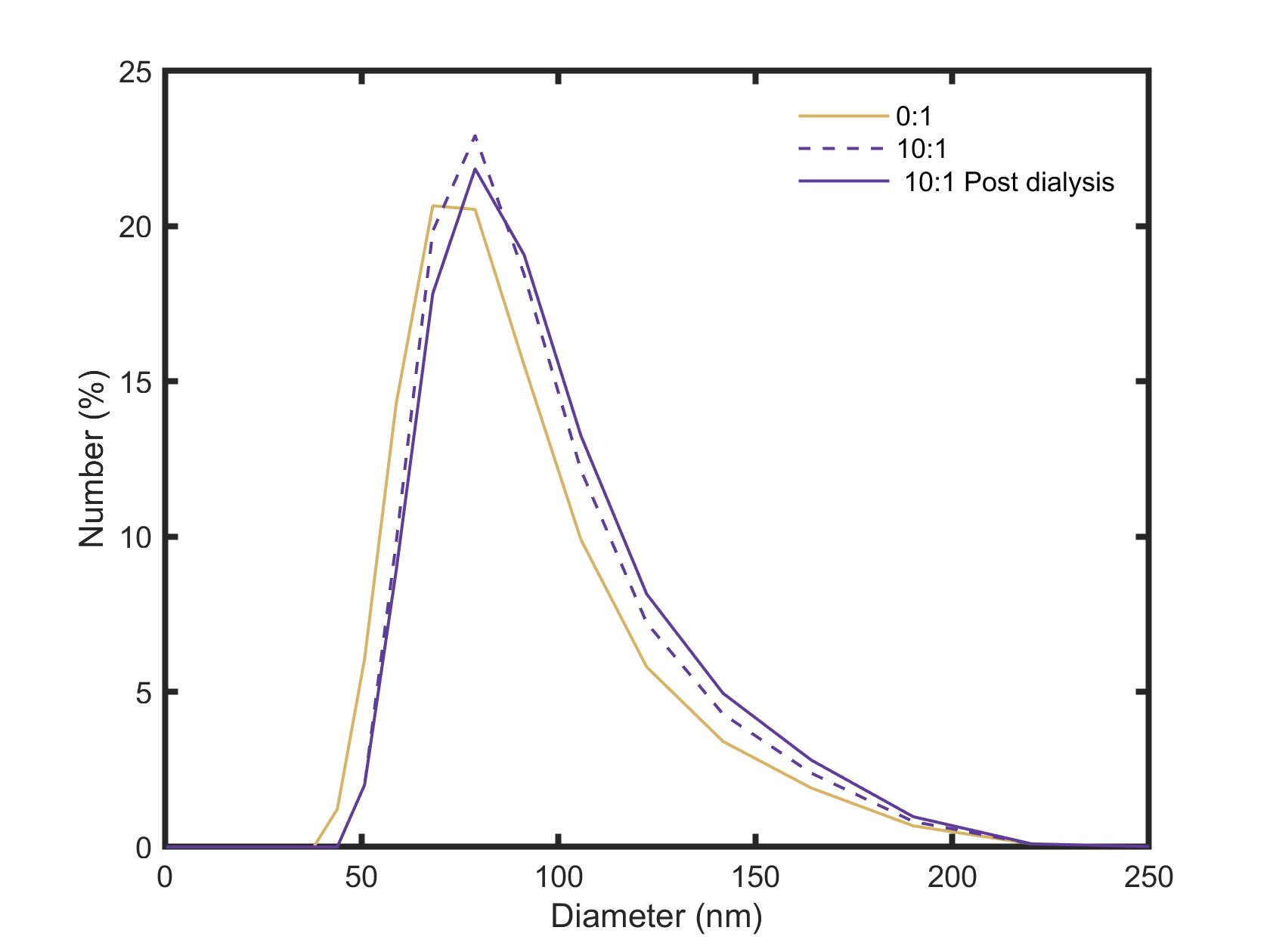}
%\centering

\caption[Post Gd$^{+3}$ Click Conjugation DLS]{DLS measurements of EM-Hem/Chol dispersions before conjugation (0:1), after click conjugation (10:1), and post dialysis for the 10:1 conjugated sample (10:1 post dialysis), where 10:1  is the DBCO to azide ($-N_{3}$) molar ratio sample. }
\label{fig:Click Gd DLS}

 \end{figure}}

\def\TSVGdPerNDSummary{\begin{figure}[t]
\renewcommand\figurename{Figure}
%\centering
\includegraphics[width = \textwidth ]{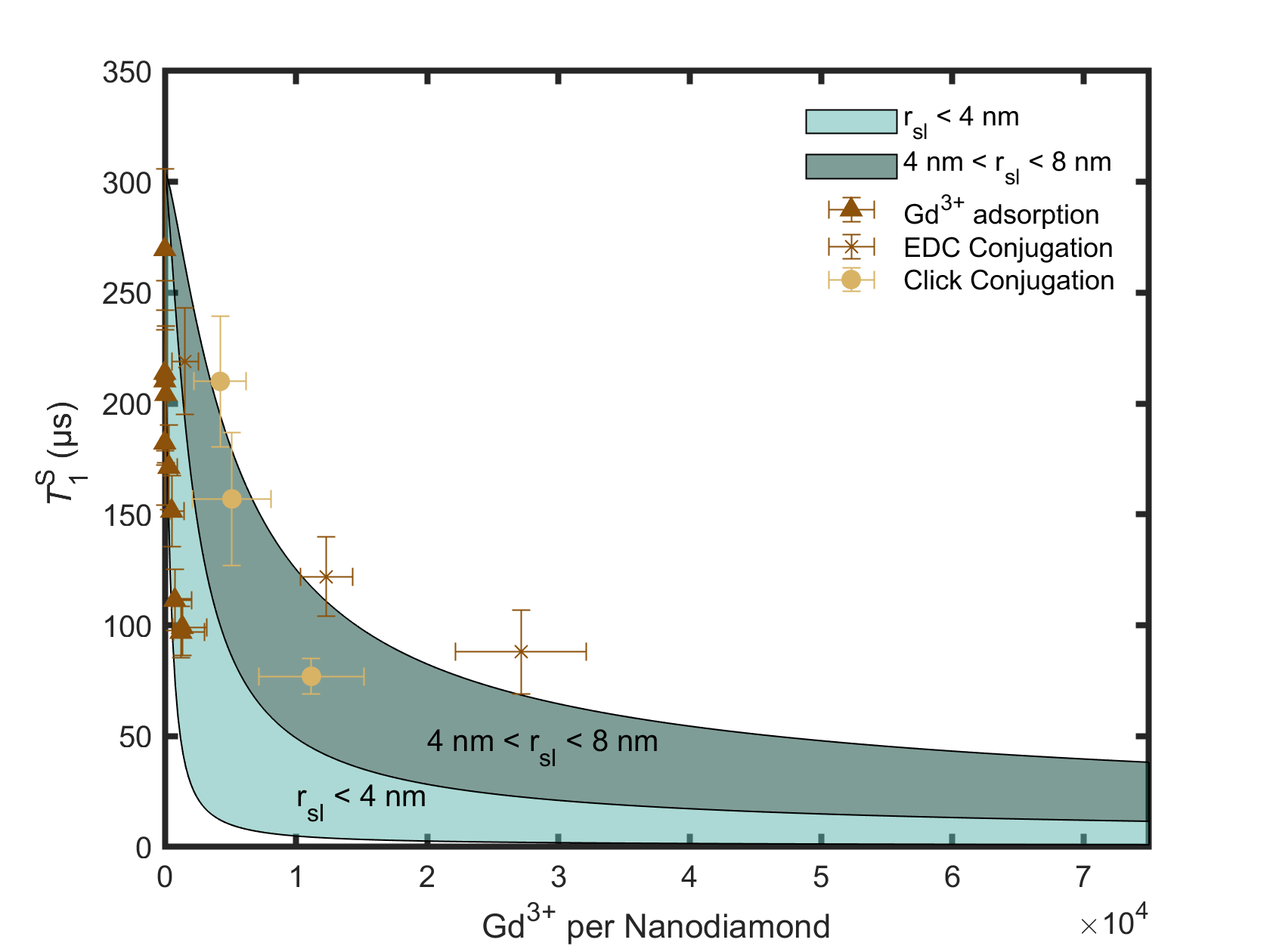}
%\centering

\caption[ ${T}_1^S$ \textit{vs} average number of Gd$^{+3}$ Per Nanodiamond ]{  ${T}_1^S$ \textit{vs} average number of Gd$^{+3}$ per nanodiamond for adsorbed chelates from solution and for both carbodiimide and click chemistry conjugation. A model of ${T}_1^S$ with $T_1^{\text{0}}$ = 269 $\pm$ 36 $\mu$s and  $R_{ND} = $34 nm and with $r_{sl} <$ 4 nm (light green region) and with $r_{sl}$ is between 4 nm and 8 nm (dark green region). Brown triangles display values of ${T}_1^S$ for the adsorbed Gd$^{3+}$ chelates on EM-Hem samples, as seen in Figure 4 of the Main Text. Error bars represent  represent 68\% confidence intervals for $\ToneS$ and the 68\% confidence intervals for Langmuir model fit described in the Main Text. 
Brown stars depict values of ${T}_1^S$ for the EDC-mediated conjugation at 0:1, 2:1 and 3:1 [EDC]:[Gd$^{3+}$] ratios respectively. Error bars represent  represent 68\% confidence intervals for $\ToneS$ and uncertainty from ICP-OES measurements, assuming a 68 nm diameter spherical nanodiamond. 
Yellow circles  depict values of ${T}_1^S$ for the click chemistry  conjugation at 1:4, 2:1 and 10:1 DBCO to  N$_{3}$ ratios.  Error bars represent  represent 68\% confidence intervals for $\ToneS$ and uncertainty from ICP-OES measurements, assuming a 68 nm diameter spherical nanodiamond. 
}
\label{fig:T1sVsGdPerNDConSummary}

 \end{figure}}

Here we compare the sensitivity of EM-Hem/Chol emulsions to  DBCO-terminated and amine-terminated Gd$^{+3}$ chelate spin labels.

\GDCtrlButTS
\GDCtrlButBeta
\GDCtrlButCon
\GDCtrlButTA
\ClickGDDLS
\clearpage

The results of the ND-Gd$^{+3}$ chelate experiments and the previously described $T_1$ model are in close agreement (Fig.~\ref{fig:T1sVsGdPerNDConSummary}). 
The values of $\ToneS$ for nanodiamonds with unconjugated Gd$^{+3}$ chelates are found in the region where $r_{sl} <$ 1 nm, consistent with surface adsorption. 
In contrast, the values of $\ToneS$ for the conjugated nanodiamond emulsions are consistent with the Gd$^{+3}$ spin labels spaced from the nanodiamond surface by the nanometer-scale thickness of the the emulsion coatings. 
\TSVGdPerNDSummary
\clearpage

%%% Add this line AFTER all your figures and tables

\bibliography{Bib_merged}